\documentclass[twocolumn]{aastex701}
\usepackage{amsmath}
\usepackage[acronym]{glossaries} % Using the acronym option

\definecolor{myGreen}{rgb}{0.1, 0.6, 0.2}


\newacronym{nircam}{NIRCam}{Near Infrared Camera}
\newacronym{alma}{ALMA}{Atacama Large Millimeter/submillimeter Array}
\newacronym{hst}{HST}{Hubble Space Telescope}
\newacronym{stis}{STIS}{Space Telescope Imaging Spectrograph}
\newacronym{nicmos}{NICMOS}{Near Infrared Camera and Multi-Object Spectrometer}
\newacronym{sphere}{SPHERE}{Spectro-Polarimetric High-contrast Exoplanet REsearch}
\newacronym{hpfrdi}{HPFRDI}{high-pass filter reference-star differential imaging}
\newacronym{mcmc}{MCMC}{Markov Chain Monte Carlo}
\newacronym{pa}{PA}{position angle}
\newacronym{psf}{PSF}{point spread function}
\newacronym{vlt}{VLT}{Very Large Telescope}
\newacronym{adi}{ADI}{angular differential imaging}
\newacronym{klip}{KLIP}{Karhunen–Lo\`{e}ve image projection}
\newacronym{sed}{SED}{spectral energy distribution}
\newacronym{iwa}{IWA}{inner working angle}
\newacronym{ki}{KI}{Keck Interferometer}
\newacronym{sem}{SEM}{standard error of the mean}
\newacronym{mast}{MAST}{Mikulski Archive for Space Telescopes}
\newacronym{miri}{MIRI}{Mid-Infrared Instrument}
\newacronym{SMSS}{SMSS}{SkyMapper Southern Survey}

\newcommand{\mjup}{$M_{\mathrm{Jup}}$}
\newcommand{\mum}{$\mu$m}

\def\addtodot#1.#2\relax{#1\rlap{.}^{\dotadd}#2}
\newcommand{\dotdeg}[1]{\def\dotadd{\circ}\addtodot#1\relax}

\shorttitle{JWST/NIRCam Imaging of the TW Hya Disk}
\shortauthors{Lin et al.}
\begin{document}

\title{JWST/NIRCam Imaging of Young Stellar Objects. IV. Detailed Imaging of the Protoplanetary Disk around TW Hya}
\correspondingauthor{Yu-Chia Lin}

\author[orcid=0000-0002-1511-310X,gname=Yu-Chia, sname='Lin']{Yu-Chia Lin}
\affiliation{Department of Astronomy and Steward Observatory, University of Arizona, USA}
\affiliation{Department of Physics, University of Arizona, USA}
\email[show]{yuchialin@arizona.edu}

\author[orcid=0000-0002-0834-6140,gname=Jarron, sname='Leisenring']{Jarron Leisenring}
\affiliation{Department of Astronomy and Steward Observatory, University of Arizona, USA}
\email{jarronl@arizona.edu}

\author[orcid=0000-0002-9977-8255,gname=Schuyler, sname='Wolff']{Schuyler G. Wolff}
\affiliation{Department of Astronomy and Steward Observatory, University of Arizona, USA}
\email{sgwolff@arizona.edu}

\author[orcid=0000-0001-9994-2142,gname=Justin, sname='Hom']{Justin Hom}
\affiliation{Department of Astronomy and Steward Observatory, University of Arizona, USA}
\email{jrhom@arizona.edu}

\author[orcid=0000-0002-6964-8732, gname=Kellen, sname='Lawson']{Kellen Lawson}
\affiliation{Center for Space Sciences and Technology, University of Maryland, Baltimore County, Baltimore, MD, USA}
\affiliation{Astrophysics Science Division, NASA Goddard Space Flight Center, Greenbelt, MD, USA}
\affiliation{Center for Research and Exploration in Space Science and Technology, NASA Goddard Space Flight Center, Greenbelt, MD, USA}
\email{kellen.d.lawson@nasa.gov}

\author[orcid=0000-0002-0813-4308,gname=Ewan, sname='Douglas']{Ewan S. Douglas}
\affiliation{Department of Astronomy and Steward Observatory, University of Arizona, USA}
\email{douglase@arizona.edu}

\author[0000-0003-2303-6519]{George Rieke}
\affiliation{Department of Astronomy and Steward Observatory, University of Arizona, USA}
\email{grieke@arizona.edu}

\author[0000-0001-7255-3251]{Gabriele Cugno}
\affiliation{Department of Astrophysics, University of Z\"urich, Winterthurerstrasse 190, 8057 Z\"urich, Switzerland}
\affiliation{Department of Astronomy, University of Michigan, Ann Arbor, MI 48109, USA}
\email{gabriele.cugno@uzh.ch}

\author[0000-0002-1783-8817]{John Debes}
\affiliation{Space Telescope Science Institute, Baltimore, MD, USA}
\email{debes@stsci.edu}

\author[0000-0001-9290-7846]{Ruobing Dong}
\affiliation{Kavli Institute for Astronomy and Astrophysics, Peking University, Beijing 100871, People's Republic of China}
\affiliation{Department of Physics and Astronomy, University of Victoria, Victoria, BC, V8P 5C2, Canada}
\email{rbdong@gmail.com}

\author[0000-0002-6773-459X]{Doug Johnstone}
\affiliation{NRC Herzberg Astronomy and Astrophysics, 5071 West Saanich Road, Victoria, BC, V9E 2E7, Canada}
\affiliation{Department of Physics and Astronomy, University of Victoria, Victoria, BC, V8P 5C2, Canada}
\email{Douglas.Johnstone@nrc-cnrc.gc.ca}

\author[0009-0007-3210-4356]{Camryn Mullin}
\affiliation{Department of Physics and Astronomy, University of Victoria, Victoria, BC, V8P 5C2, Canada}
\email{camrynmullin@uvic.ca}

\author[0000-0001-8103-5499]{Taylor L. Tobin}
\affiliation{Department of Astronomy, University of Michigan, 1085 S. University, Ann Arbor, MI 48109, USA}
\email{tltobin@umich.edu}

\author[0000-0002-4309-6343]{Kevin R. Wagner}
\affiliation{Department of Astronomy and Steward Observatory, University of Arizona, USA}
\email{kevinwagner@arizona.edu}

\author[0000-0002-8963-8056]{Thomas Greene}
\affiliation{IPAC, California Institute of Technology, Pasadena, CA 91125, USA}
\email{tgreene@caltech.edu}

\author[0000-0003-1227-3084]{Michael R. Meyer}
\affiliation{Department of Astronomy, University of Michigan, Ann Arbor, MI 48109, USA}
\email{mrmeyer@umich.edu}

\author[0000-0002-7893-6170]{Marcia Rieke}
\affiliation{Department of Astronomy and Steward Observatory, University of Arizona, USA}
\email{mrieke@arizona.edu}

\begin{abstract}
As the nearest protoplanetary disk to Earth ($d=60.14$\,pc), TW Hya is one of the most studied protoplanetary disks and a critical benchmark for testing planet formation theories. We present high-contrast coronagraphic imaging of the TW Hya disk from JWST/NIRCam across four filters (F187N, F200W, F356W, and F444W). We detect the disk's scattered-light emission in F200W, F356W, and F444W. An elliptical fit to the disk image yields an average inclination of $i = \dotdeg{8.74}^{\dotdeg{+1.03}}_{\dotdeg{-0.94}}$ and a position angle of $\mathrm{PA} = \dotdeg{75.62}^{\dotdeg{+7.86}}_{\dotdeg{-6.56}}$. We find tentative evidence for radial variations in these parameters, a trend consistent with a disk warp. Our companion search yields no new detections, placing the lowest mass limits yet on companions that might be responsible for carving out the dust gap. Assuming no local extinction and a system age of 10~Myr, the F444W data are sensitive to masses down to $\sim$0.4~\mjup\ at separations of 1\arcsec\ ($\sim$60~AU). Accounting for local disk extinction analogous to the AS 209 system, our limits reach sub-Jupiter masses beyond 2\arcsec. Furthermore, our analysis provides a detailed view of a previously detected feature in the outer disk at $\sim$120~AU, confirming its morphology as a distinct bifurcation structure. This feature may indicate the presence of complex substructures arising from dynamical planet-disk interactions. These results demonstrate JWST's ability to characterize the architecture of protoplanetary disks and constrain the properties of forming worlds.
\end{abstract}


\section{Introduction} \label{sec:intro}
Protoplanetary disks are the birthplaces of planets. Understanding their structure, evolution, and composition is key to unraveling the processes of planet formation \citep{2019SAAS...45.....A}. The TW Hya system is a valuable laboratory for such studies; it is centered on a young M0.5V T Tauri star with an age of $8 \pm 3$ ~Myr \citep{2018ApJ...853..120S} located at a distance of only $60.14 \pm 0.05$\,pc \citep{2023A&A...674A...1G}, nearly two times closer than other well-characterized protoplanetary disks. With a stellar mass of $0.87\,M_{\odot}$ and a mass accretion rate of $2.51\times 10^{-9}$ $M_{\odot}$/$\mathrm{yr}$ \citep{2023ApJ...956..102H}, the system provides a unique opportunity for high-resolution observations of planet formation around a solar analog.

The disk surrounding TW Hya is one of the best-characterized, having been observed across a broad swath of the electromagnetic spectrum. It is classified as a transition disk, featuring an inner cavity and a complex vertical structure where large grains settle to the midplane while small grains remain coupled to the gas in the upper atmosphere \citep{2006ApJ...638..314D, 2014A&A...564A..93M}. Its nearly face-on orientation ($i \approx 7^\circ$; \citealt{2016ApJ...820L..40A}) minimizes projection effects and self-obscuration, allowing a clear view of features like concentric gaps and rings. The disk was first resolved in scattered light \glstarget{hst} by \gls{hst} using the Wide Field and Planetary Camera 2 (WFPC2) \citep{2000ApJ...538..793K}. Subsequent multi-wavelength analysis confirmed the disk's vast extent to at least 230~AU and revealed initial evidence of a partially filled gap at 80~AU \citep{2013ApJ...771...45D}. Later, observations with \glstarget{alma} the \gls{alma} resolved a series of gaps in the millimeter-sized dust continuum, suggesting the potential presence of forming planets carving out orbits \citep{2016ApJ...820L..40A}. These gaps have been further characterized by high-contrast monitoring \glstarget{stis} with \gls{stis} on the \gls{hst} and ground-based instruments \glstarget{sphere} such as the \gls{sphere} at \glstarget{vlt} the \gls{vlt}, which also identified complex, evolving shadow features \citep{2017ApJ...835..205D, 2017ApJ...837..132V}. These observations establish TW Hya as a prototypical system for understanding the conditions under which planets form.

Despite these tell-tale signposts of planet-disk interactions, the embedded planets driving these disk substructures have remained elusive. TW Hya has been the subject of extensive search campaigns aimed at detecting protoplanetary companions. These efforts have included deep thermal-infrared coronagraphic imaging with the Keck/NIRC2 vortex coronagraph to search for direct thermal emission \citep{2017AJ....154...73R} as well as deep polarimetric imaging with \gls{vlt}/\gls{sphere} and Gemini Planet Imager (GPI) \citep{2015ApJ...815L..26R, 2017ApJ...837..132V} to detect indirect structural signatures. To date, no companions have been directly confirmed, suggesting that any planets responsible for the observed substructures are either lower in mass or more deeply embedded than previous instruments could detect.

In this paper, we present new high-contrast coronagraphic images of the TW Hya disk obtained with the \glstarget{nircam} JWST \gls{nircam} \citep{2023PASP..135b8001R}. While ground-based observatories with larger primary mirrors, such as the \gls{vlt}, possess a higher diffraction-limited angular resolution, their delivered image quality is fundamentally limited by the need to correct for atmospheric turbulence using adaptive optics \citep{2019A&A...631A.155B, 2020A&A...633A..63D}. In contrast, the high sensitivity and remarkable thermal and pointing stability of JWST provide \glstarget{psf} a \gls{psf} that is stable over long periods, enabling near-diffraction-limited performance that is challenging to achieve from the ground \citep{2023PASP..135d8003W}. Coupled with low background from the sky and telescope, this stability allows us to probe scattered light from the disk's surface with improved clarity, search for faint companions close to the star, and resolve fine-scale structures that were previously difficult to distinguish, highlighting the anticipated capability of JWST for coronagraphic high-contrast imaging \citep{2021MNRAS.501.1999C}. 

In Section \ref{sec:obs}, we describe the JWST/\gls{nircam} observations and detail our data reduction methods focused on revealing the faint disk structure through \gls{psf} subtraction and deconvolution.
The data analysis and results are discussed in Section \ref{sec:results}. We first present our disk photometry and spatially resolved color analysis (Sections \ref{subsec:disk_photometry} and \ref{subsec:disk_color}, respectively). We then constrain the disk's geometry (Section \ref{subsec:disk_geometry}) and perform a detailed analysis of the disk features (Section \ref{subsec:disk_features}). Subsequently, we describe the outcomes of our companion search and derived mass limits (Section \ref{subsec:mass_limit}).
We discuss the broader implications of our findings in Section \ref{sec:discussion} and provide a summary of our main conclusions in Section \ref{sec:conclusion}.

\section{Observations and Data Reduction} \label{sec:obs} 

\begin{deluxetable*}{lcccccccc}[t!]
\tablecaption{JWST/\gls{nircam} Coronagraphic Observation Log for TW Hya \label{tab:obs_log}}
\tablehead{
\colhead{Filter} & \colhead{Detector} & \colhead{Date} & \colhead{MJD Start} & \colhead{Roll Angle (PA V3)} & \colhead{N$_{\mathrm{ints}}$} & \colhead{$t_{\mathrm{int}}$} & \colhead{Total Exposure} \\
\colhead{} & \colhead{} & \colhead{(YYYY-MM-DD)} & \colhead{} & \colhead{($\degr$)} & \colhead{} & \colhead{(s)} & \colhead{(s)}
}
\startdata
F187N & NRCA2 & 2024-02-14 & 60354.41123 & 332.05 & 16 & 104.77 & 1676.3 \\
F187N & NRCA2 & 2024-02-14 & 60354.48438 & 342.04 & 16 & 104.77 & 1676.3 \\
\hline
F200W & NRCA2 & 2024-02-14 & 60354.43260 & 332.05 & 16 & 104.77 & 1676.3 \\
F200W & NRCA2 & 2024-02-14 & 60354.50567 & 342.04 & 16 & 104.77 & 1676.3 \\
\hline
F356W & NRCALONG & 2024-02-14 & 60354.41123 & 332.05 & 16 & 104.77 & 1676.3 \\
F356W & NRCALONG & 2024-02-14 & 60354.48438 & 342.04 & 16 & 104.77 & 1676.3 \\
\hline
F444W & NRCALONG & 2024-02-14 & 60354.43259 & 332.05 & 16 & 104.77 & 1676.3 \\
F444W & NRCALONG & 2024-02-14 & 60354.50567 & 342.04 & 16 & 104.77 & 1676.3 \\
\enddata
\tablecomments{All observations were taken as part of JWST Program ID 1179, and instrument settings include the \texttt{MASKRND} pupil, \texttt{MASKA335R} coronagraphic mask, and the \texttt{SUB320A335R} subarray. Each integration consisted of 10 groups using the \texttt{MEDIUM8} readout pattern.}
\end{deluxetable*}

TW Hya was observed on UT 2024-02-14 as part of the JWST Guaranteed Time Observation program “Direct Imaging of YSOs” (ID 1179; PI: J.\ Leisenring). The data products used in this analysis are available from \glstarget{mast} the \gls{mast} at the Space Telescope Science Institute under the \href{https://archive.stsci.edu/doi/resolve/resolve.html?doi=10.17909/dymv-st72}{DOI:10.17909/dymv-st72}. This program was designed to search for and characterize forming protoplanets around several young stars selected for their pre-existing evidence of planet-disk interactions. TW Hya was chosen due to its well-documented system of concentric rings and gaps observed by \gls{hst}, \gls{vlt}/\gls{sphere} and \gls{alma}  \citep{2017ApJ...837..132V, 2021A&A...648A..33M, 2023ApJ...948...36D, 2024A&A...689A.104D}. To achieve the high contrast necessary to search for forming planets within this system, the observing strategy for this target utilized the \gls{nircam} coronagraphic mode \citep{2022SPIE12180E..3QG}. The observations utilized the \texttt{MASKA335R} coronagraphic mask with the \texttt{MASKRND} pupil, which suppresses the central starlight and enhances sensitivity to faint disk structures and embedded planets.
A detailed log of these observations is presented in Table \ref{tab:obs_log}.

The observations utilized both the short-wavelength (NRCA2) and long-wavelength (NRCALONG) detectors, reading out the \texttt{SUB320A335R} subarray. Each integration was 104.77~s, configured with the MEDIUM8 frame readout pattern consisting of 10 groups. For each filter at each of the two roll angles, a sequence of 16 such integrations was acquired, resulting in a total exposure time of 1676.3~s. The observations were executed using two pairs of simultaneously observed short- and long-wavelength filters: F200W/F444W and F187N/F356W. While these pairings and their observational order were primarily driven by operational efficiency in obtaining simultaneous data in both channels, the resulting dataset supports two key scientific goals. The broadband filters (F200W, F356W, and F444W) provide useful diagnostics for the nature of any detected point sources; planetary-mass companions are expected to have very red colors (e.g., F200W--F444W) due to their low effective temperatures, which should clearly distinguish them from faint background stars and galaxies \citep{2025ApJ...987L..41C}. Additionally, the narrow-band F187N filter traces the Pa-$\alpha$ hydrogen recombination line, a key indicator of accretion shocks. Since the F187N bandpass is contained within the F200W filter, the F200W data serve as the continuum reference; a comparison of the flux ratio between these two filters allows for the identification of Pa-$\alpha$ emission line excess indicative of active accretion onto forming protoplanets \citep{2024AJ....167..183M}.

The data reduction pipeline involved three primary stages: (1) initial calibration from raw data to flux-calibrated slope images, (2) post-calibration image preparation, and (3) final \gls{psf} subtraction and deconvolution to reveal the disk structure. The initial calibration and preparation were conducted using the \texttt{spaceKLIP} package \citep[v1.1.0;][]{2022SPIE12180E..3NK, 2023ApJ...951L..20C}, which wraps the official JWST Science Calibration Pipeline \citep[version 1.17.1 with CRDS version 12.1.4;][]{bushouse_2022_7229890} with enhancements for high-contrast imaging, and the final image products were generated using the \texttt{winnie} package \citep[v1.1;][]{2023AJ....166..150L}.

\subsection{Initial Calibration and Image Preparation}
The raw uncalibrated ramp data were first converted to slope images (\texttt{rateints} files) using \texttt{spaceKLIP}'s wrapper for stage 1 of the JWST pipeline. This process included steps for group scaling, data quality initialization, saturation flagging, superbias subtraction, reference pixel correction, non-linearity correction, jump detection, 1/f noise mitigation, and ramp fitting. The subsequent stage 2 processing converted the slope images into calibrated files (\texttt{calints}), applying background subtraction, world coordinate system assignment, flat-fielding, and photometric calibration to produce images in units of MJy/sr. At this stage, the pipeline had not yet accounted for flux corrections in the regions attenuated by the coronagraphic occulting mask.

Following this standard calibration, we applied several image preparation steps using \texttt{spaceKLIP}'s \texttt{imagetools} module. These included a frame-by-frame median subtraction to reduce residual bias drifts and a multistep routine for bad-pixel correction. The stellar position was precisely determined using a model \gls{psf} generated by \texttt{STPSF} for an M0V spectral type \citep{2025zndo..15747364P}, and all frames were co-aligned using a Fourier-based registration technique. Finally, to prevent edge effects in subsequent processing, all frames were padded with NaN values to a final dimension of $650\times650$~pixels.

\subsection{PSF Subtraction and Deconvolution}
\label{subsec:psfsub_deconv}
To reveal the disk structure, we \glstarget{hpfrdi} performed \gls{hpfrdi} using the \texttt{winnie} package \citep[v1.1;][]{2023AJ....166..150L}. As our program did not include a contemporaneous reference-star observation, we built a \gls{psf} reference library from archival observations taken between November 1, 2023, and April 2, 2025. We excluded off-axis observations to ensure the reference \glspl{psf} matched the coronagraphic optical path. The size of the resulting reference library for each filter is detailed in Table \ref{tab:ref_library}.

The primary analysis in this work focuses on the F200W and F444W datasets, which benefit from large reference-star libraries, leading to robust \gls{psf} subtraction \citep{2022A&A...666A..32X, 2022AJ....163..119S, 2024AJ....168..215S}. For F187N, the archive contained only off-axis observations, preventing \gls{hpfrdi} processing. Since \glstarget{adi} standard \gls{adi} also failed to recover the disk due to self-subtraction, F187N was excluded from the disk analysis. The F356W dataset, while processed and presented for completeness in Appendix \ref{appendix:other_filters}, suffers from a small reference library and was not used for the detailed analysis.

The \gls{hpfrdi} process determines the optimal scaling coefficients for each reference \gls{psf} by performing a least-squares fit using temporary, high-pass-filtered copies of the data. This isolates the high-frequency \gls{psf} speckles from the extended disk emission, ensuring the disk flux does not bias the fit or lead to over-subtraction. The derived coefficients were then applied to the original, unfiltered reference images to model and subtract the stellar \gls{psf} from the science target, producing the \gls{hpfrdi} image.

\begin{deluxetable*}{lccl}[ht!]
  \tablecaption{Reference Star Library Composition\label{tab:ref_library}}
  \tablehead{\colhead{Filter} & \colhead{\# of Reference Stars} & \colhead{\# of Reference Images} & \colhead{Corresponding Proposal IDs}}
  \startdata
  F200W & 25 & 241 & 3337, 3840, 3947, 3973, 3989, 4014, \\
   & & & 4050, 4090, 4558, 5229, 6139\\
  \hline
  F356W & 2 & 27 & 3840, 4558\\
  \hline
  F444W & 35 & 273 & 1193, 2780, 3337, 3840, 3947, 3973, 3989, \\
   &  &  & 4050, 4090, 4558, 5229, 6012, 6139\\
  \enddata
  \tablecomments{The number of unique reference stars and the total number of individual reference images used to build the \gls{psf} library for the \gls{hpfrdi} reduction in each filter.}
\end{deluxetable*}

\begin{figure*}[ht!]
\gridline{\fig{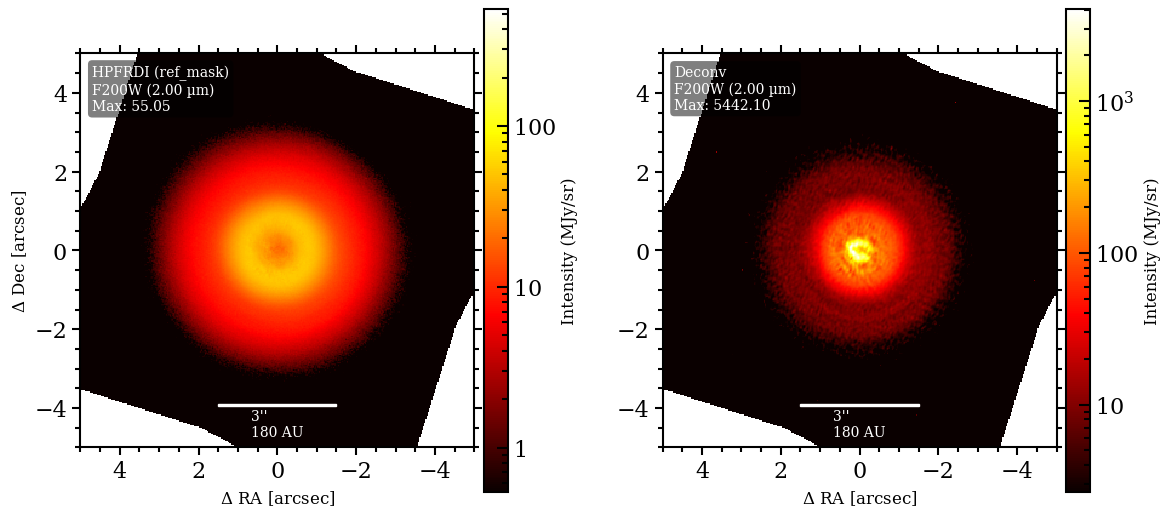}{1.0\textwidth}{(a) F200W}}
\gridline{\fig{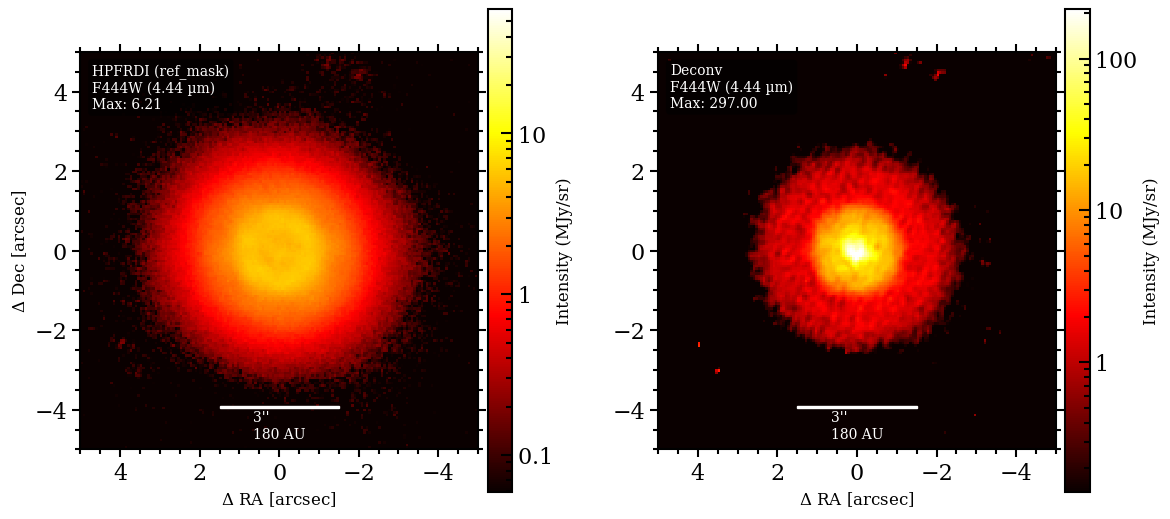}{1.0\textwidth}{(b) F444W}}
\caption{Final processed images of the TW Hya protoplanetary disk. Panel (a) shows the result in the F200W filter, and panel (b) shows the F444W result. Both left images are the product of \gls{hpfrdi}, which subtracts the stellar \gls{psf}. The right images undergo a further deconvolution process with \texttt{winnie}, utilizing 200 iterations (for F200W) and 129 iterations (for F444W) of the Richardson-Lucy algorithm. The images resolve the nearly face-on disk structure.
\label{fig:disks}}
\end{figure*}

To further sharpen the image and resolve details, the \gls{hpfrdi} results were deconvolved using the Richardson-Lucy algorithm implemented in \texttt{winnie}. The Richardson-Lucy algorithm is an iterative, maximum-likelihood technique that reconstructs the underlying source distribution by minimizing the divergence between the observed image and a model convolved with the instrumental \gls{psf}. The implementation implicitly accounts for the coronagraphic mask transmission during this process, restoring flux levels inside \glstarget{iwa} the \gls{iwa}.

The deconvolution used a grid of synthetic \glspl{psf} generated with \texttt{STPSF} for the corresponding instrument configuration and target spectral type. To identify the optimal stopping criterion and quantify algorithmic uncertainties, we implemented a forward-modeled synthetic disk injection and recovery framework (detailed in Appendix \ref{app:deconv_quality}). Based on these tests, the process was run for 200 iterations for the F200W and F356W datasets, and 129 iterations for the F444W dataset to optimize the recovery of the radial profiles, yielding the final ``Deconv" images shown in Figure \ref{fig:disks}. These images better resolve the inner structure of the nearly face-on disk; however, the bright central peak should be interpreted with caution. It arises from the amplification of noise and systematic residuals in the region of lowest mask transmission and is therefore likely a combination of real disk flux and processing artifacts. A quantitative assessment of the deconvolution's fidelity, showing the minimal residuals outside the core, is presented in Appendix \ref{app:deconv_quality}.

\section{Results and Analysis} \label{sec:results}

\subsection{Disk Photometry}
\label{subsec:disk_photometry}

To calculate the disk-to-star flux ratio, we established a \glstarget{sed} reference \gls{sed} for the host star, TW Hya. Details regarding the construction of the stellar photospheric model and the physics of the stellar and disk photometric excess are provided in Appendix \ref{app:phot_details}. The resulting modeled stellar fluxes ($F_{\star}$) are listed in Table \ref{tab:photometry}.

\begin{deluxetable*}{lcccc}[ht!]
\tablecaption{Stellar and Spatially Resolved Disk Photometry\label{tab:photometry}}
\tablehead{
\colhead{Filter} & \colhead{Stellar Flux ($F_{\star}$)\tablenotemark{a}} & \colhead{\gls{hpfrdi} Disk Flux} & \colhead{Throughput Corr. Flux} & \colhead{Deconvolved Disk Flux} \\
\colhead{} & \colhead{(mJy)} & \colhead{(mJy)} & \colhead{(mJy)} & \colhead{(mJy)}
}
\startdata
F200W & $823.27 \pm 7.22$ & $8.259 \pm 0.005$ & $23.71 \pm 0.07$ & $20.1 \pm 1.2$ \\
F356W & $351.16 \pm 5.51$ & $2.189 \pm 0.005$ & $6.45 \pm 0.13$ & $3.42 \pm 0.11$ \\
F444W & $236.43 \pm 4.26$ & $1.167 \pm 0.006$ & $4.07 \pm 0.03$ & $2.72 \pm 0.23$ \\
\enddata
\tablecomments{Comparison of the modeled stellar photosphere flux to the spatially resolved scattered-light flux from the disk. Uncertainties for the disk flux measurements (columns 3--5) represent the $1\sigma$ standard deviation from 500,000 bootstrap resamples. \tablenotetext{a}{Derived from the best-fit stellar spectral model integrated over the filter bandpass (see Appendix \ref{app:phot_details}).}}
\end{deluxetable*}

We found positive total photometric excesses in all three filters (see Appendix \ref{app:phot_details}). These excesses are found with high confidence in F444W ($20.2\sigma$) and F356W ($8.1\sigma$). While this total excess arises from a combination of scattered light and thermal emission, we refer the reader to Appendix \ref{app:phot_details} for a detailed discussion distinguishing this total system excess—which is dominated by unresolved thermal emission—from the spatially resolved scattered-light fluxes reported below.

To quantify the fraction of this total photometric excess that arises from the spatially resolved disk detected in our coronagraphic observations, we measured the total integrated flux density ($F_{\text{disk}}$) using three different methods. First, we measured the flux (1) directly from the \gls{hpfrdi} images. As expected, this yields the lowest values because the signal is attenuated by the coronagraph's mask. To account for this suppression, we utilized two reconstruction methods: (2) dividing the \gls{hpfrdi} images by the simulated coronagraphic throughput map provided by \texttt{webbpsf\_ext} (imposing a minimum transmission floor of 0.01 to limit noise amplification), and (3) measuring from the final deconvolved images, which recover the source geometry via \gls{psf} fitting. We employed a bootstrap resampling technique (detailed in Appendix \ref{appendix:bootstrap_results}), combined with a systematic deconvolution flux uncertainty derived from our synthetic injection tests (Appendix \ref{app:deconv_quality}), to estimate total uncertainties for all methods. The results are summarized in Table \ref{tab:photometry}.

\begin{figure}[ht!]
\centering
\includegraphics[width=\columnwidth]{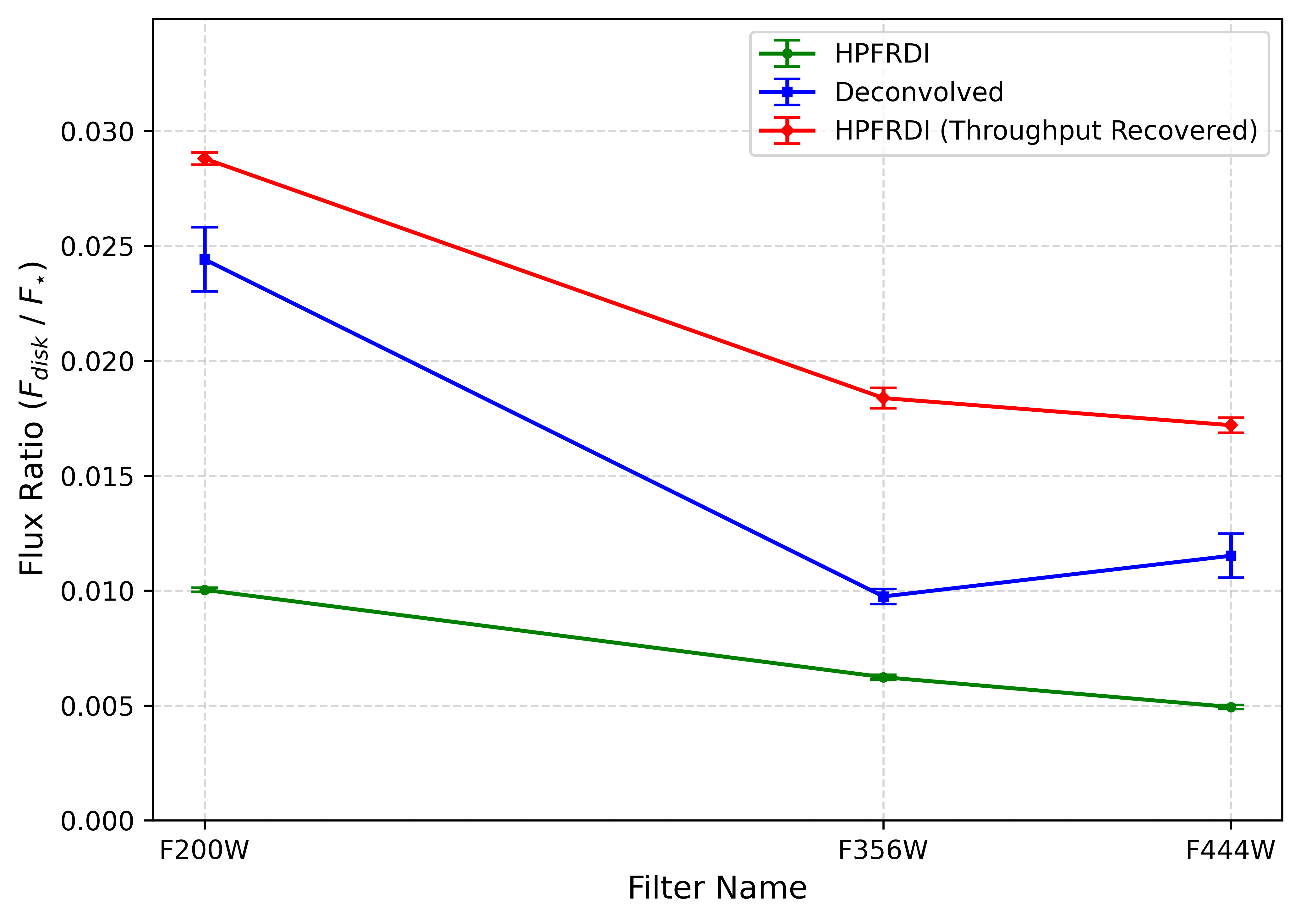}
\caption{Relative photometry of the TW Hya disk from our JWST/\gls{nircam} observations showing the ratio of the disk flux compared to that of the stellar photosphere. The green line represents the flux from the \gls{hpfrdi}-processed images, which is a lower limit due to signal attenuation from the stellar-light subtraction process and the coronagraphic mask. The red line shows the result after applying a theoretical throughput recovery to the \gls{hpfrdi} data. The blue line shows the flux from the final deconvolved images, which provides our best estimate of the disk's brightness.
}
\label{fig:flux_ratio}
\end{figure}

The ``Throughput Corrected'' fluxes provide a rough consistency check against the deconvolution results. While this simplified method gives a general sense of the flux suppression, it does not fully account for the optical propagation of a disk image convolved with the instrumental \glspl{psf} and then occulted by \gls{nircam}'s coronagraphic mask. In addition, we note that the uncertainties derived for the throughput-corrected method are likely underestimated. This simple correction involves dividing by small transmission values near the \gls{iwa}, which amplifies noise and systematic residuals that the bootstrap resampling of the total flux does not fully capture. Consequently, we adopt the deconvolved values as our best estimate of the spatially resolved disk flux and use them for the disk-to-star flux ratios ($F_{\text{disk}}/F_{\star}$) plotted in Figure \ref{fig:flux_ratio}.

Our measured disk-to-star flux ratios in the F200W filter, from both the \gls{hpfrdi} image ($1.003\% \pm 0.009\%$) and the final deconvolved image ($2.44\% \pm 0.14\%$), are higher than the disk reflectance of $\sim0.5\%$ measured at 2.04~\mum{}\ by \citet{2013ApJ...771...45D} \glstarget{nicmos} using the \gls{nicmos} on the \gls{hst}. This difference is driven by two key factors. First, our deconvolution provides a more complete flux measurement by recovering the signal that is suppressed by the coronagraph. This factor accounts for the significant flux increase from our intermediate \gls{hpfrdi} result to our final deconvolved image value. Second, the nature of the \gls{iwa} differs between the two instruments. The \gls{nicmos} observations used for the \citet{2013ApJ...771...45D} measurement employed a hard-edged occulting hole \citep{1998ApJ...492L..95T}, resulting in a complete loss of flux inside this region. In contrast, the \gls{nircam} coronagraph utilizes a mask with a graded transmission profile. While the nominal size of the mask is comparable to the \gls{nicmos} hole, the graded edges allow for signal recovery at small separations during the deconvolution process. This enables us to account for scattered light from the bright, innermost regions of the disk that were fully occulted by the hard stop in the \gls{hst} data.

\subsection{Spatially Resolved Disk Color}
\label{subsec:disk_color}

A color measurement on scattered light requires accounting for both the filter transmission curves and the TW Hya spectrum. First, to establish an accurate stellar color zero-point, we calculated the effective flux of TW Hya in the F200W and F444W filters. We utilized the fitted two-component stellar model, derived using the process described in Section \ref{subsec:disk_photometry}, and numerically integrated it over the respective transmission curve with the \texttt{webbpsf\_ext} package. This process yields an effective stellar flux that accounts for the filter's bandpass shape, providing a more robust reference than a single-wavelength interpolation.

\begin{figure*}[ht!]
\centering
\includegraphics[width=0.9\textwidth]{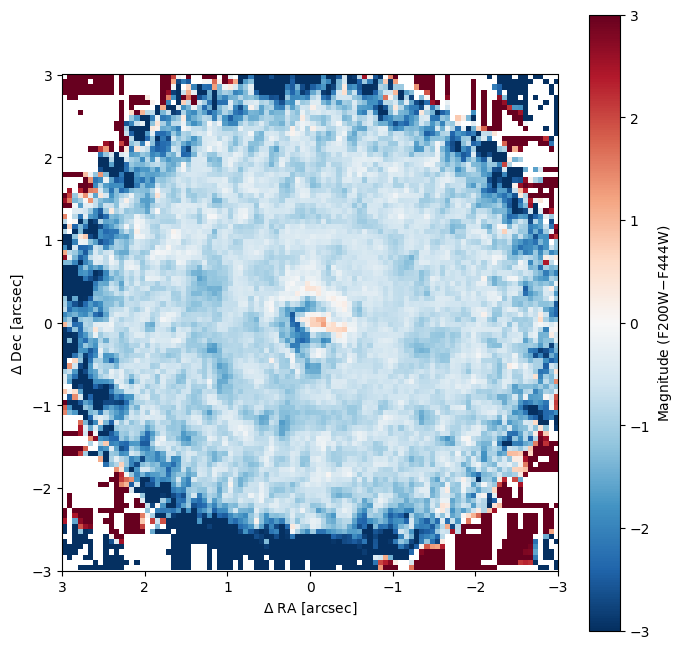}
\caption{The F200W -- F444W scattered-light color map of the TW Hya disk, derived from the final deconvolved images after resampling the F200W data to the F444W pixel grid. The color is predominantly blue (negative values), indicating that the disk is brighter at F200W than at F444W, a signature of scattering by small dust grains. A small, slightly redder region is visible near the center, which may indicate a change in dust properties or be an artifact of the image processing in this high-contrast region.
\label{fig:color_map}}
\end{figure*}

Second, because the native pixel scales of the \gls{nircam} short- (0.031\arcsec/pixel) and long-wavelength (0.063\arcsec/pixel) channels differ, the F200W deconvolved image was resampled to match the pixel grid of the F444W image using a bicubic interpolation. The final (F200W -- F444W) color for each pixel was then calculated in the Vega magnitude system, referencing the effective stellar fluxes to establish the color zero-point. The resulting map is shown in Figure \ref{fig:color_map}.

The disk exhibits a predominantly blue color with typical (F200W -- F444W) values between 0 and -2~mag. This color indicates that the scattering surface is dominated by small, micron- to sub-micron-sized dust grains. This finding is consistent with previous optical and near-infrared scattered-light studies from both \gls{hst} \citep{2013ApJ...771...45D} and \gls{vlt}/\gls{sphere} \citep{2017ApJ...837..132V} and supports the well-established model of a vertically stratified disk. Additionally, a bluer ring is visible at a radius of approximately 1.4\arcsec ($\sim$84~AU). This feature corresponds exactly to the physical radial shift of Gap 2 between the two wavelengths. As detailed in Section \ref{subsec:disk_features}, the gap is located at 91~AU in F200W, but shifts inward to 85~AU in F444W. This spatial misalignment of the scattered-light deficit creates the prominent color gradient seen at this radius.

A notable exception is a small, slightly redder (color $\gtrsim 0$) region located near the center of the image. This feature could have a physical origin, perhaps indicating the presence of larger grains or a contribution from the thermal emission in the innermost part of the disk. However, given its location in the region most affected by \gls{psf} subtraction residuals and deconvolution artifacts (as discussed in Section \ref{subsec:psfsub_deconv}), this feature should be interpreted with caution.

\subsection{Geometric Modeling of the Disk}
\label{subsec:disk_geometry}

The processed images (Figure \ref{fig:disks}) resolve the scattered light from the nearly face-on disk. To determine the disk's orientation, we fit a geometric model to the deconvolved F200W image. Our model describes the disk as a circular, flat structure viewed at an arbitrary orientation. This orientation is defined by four geometric parameters: the center of the disk in the image plane ($x_0, y_0$), the disk's inclination ($i$, where $i=0\degr$ corresponds to a face-on view), and \glstarget{pa} the \gls{pa} of the disk's projected major axis on the sky (measured East of North). Note that for our $650\times650$ pixel images, the geometric center of the frame lies at coordinates (x, y)=(324.5, 324.5) in a 0-indexed system. The model's assumption is that for the correct set of these geometric parameters, the deprojected disk image should be azimuthally uniform. In other words, all pixels at the same deprojected radius are assumed to have the same intrinsic brightness. Any observed azimuthal variance is therefore treated as a deviation from this ideal model, and the goal of the fit is to find the geometry that minimizes this variance.

\begin{figure*}[ht!]
\centering
\gridline{
    \fig{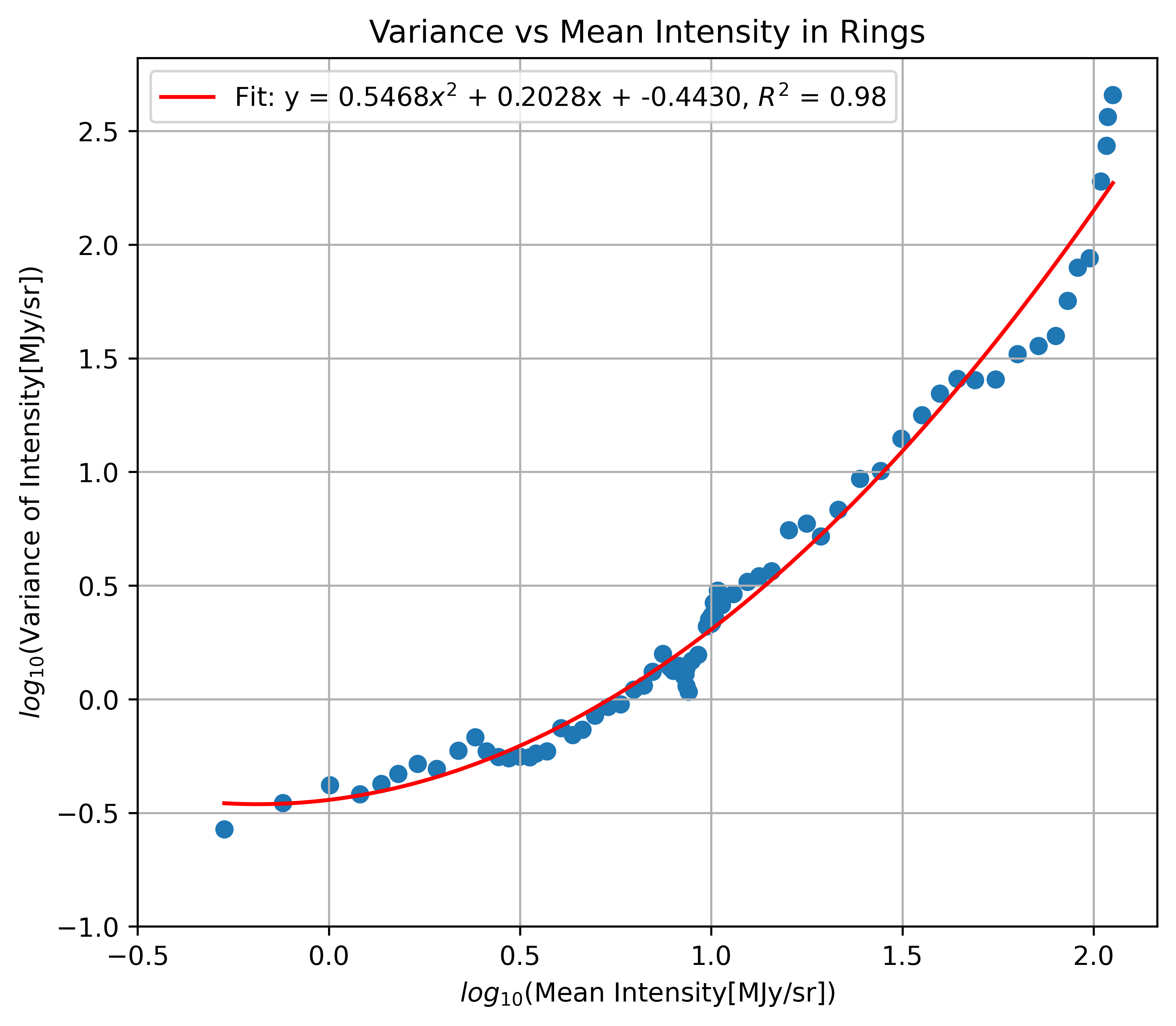}{0.54\textwidth}{(a) Empirical Variance Model}
    \fig{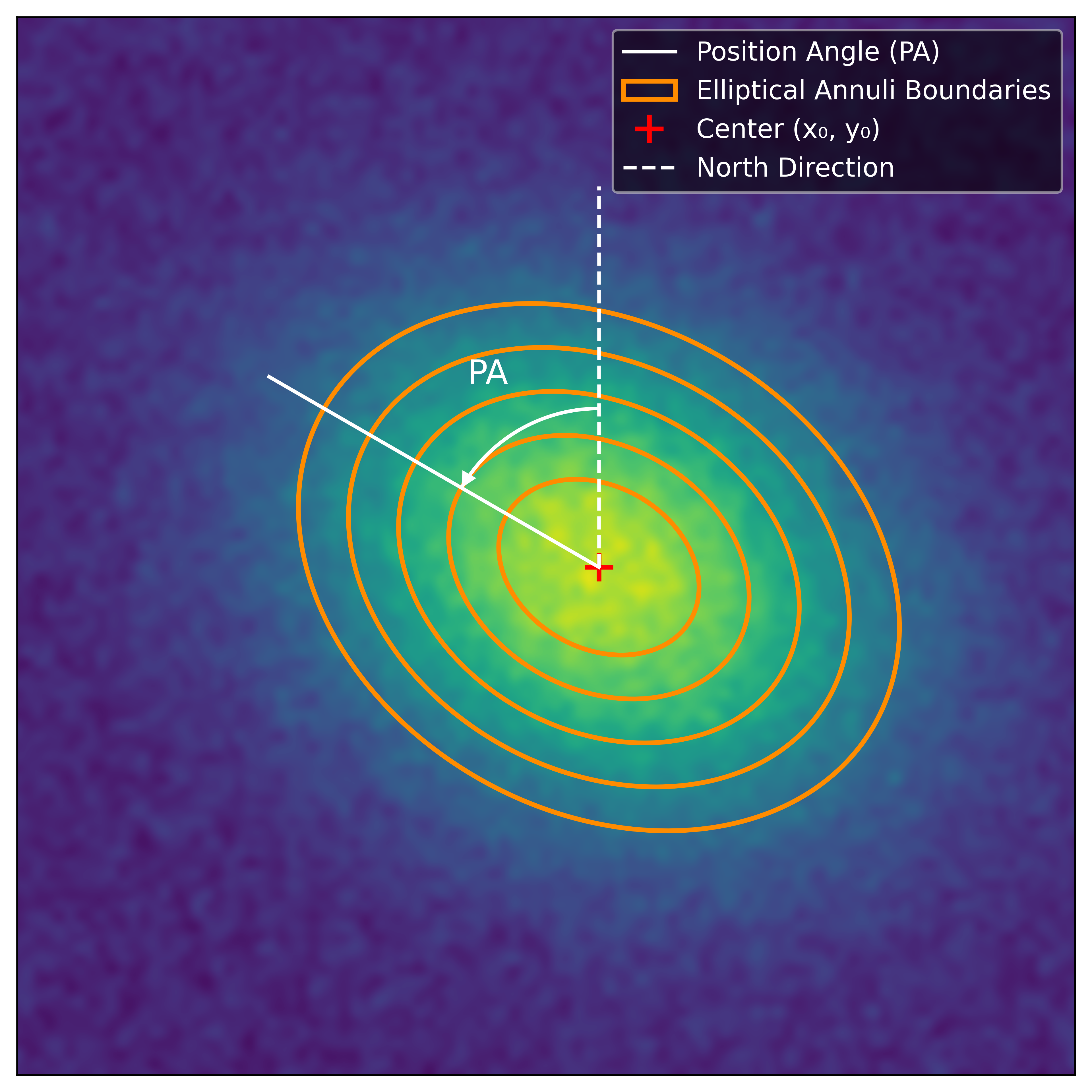}{0.46\textwidth}{(b) Geometric Fitting Annuli}
}
\caption{Methodology for the geometric fitting of the disk. \textbf{Panel (a)} shows the diagnostic plot establishing the empirical variance model. The relationship between $\log_{10}$(Variance) and $\log_{10}$(Mean Intensity) is well-described by a second-order polynomial (red line, $R^2 = 0.98$), which accounts for the non-linear combination of statistical noise and unresolved astrophysical structure. \textbf{Panel (b)} visualizes the annuli used in the fitting process (orange ellipses) overlaid on the simulated inclined disk data. \glstarget{mcmc} The \gls{mcmc} algorithm seeks the geometric parameters—center ($x_0, y_0$), inclination, and \gls{pa}—that minimize the observed brightness variance within these elliptical annuli, weighted by the model established in panel (a).
\label{fig:geo_method}}
\end{figure*}

We employed an \gls{mcmc} approach with the \texttt{emcee} package \citep{2013PASP..125..306F} to find the parameters that best satisfy this assumption. The log-likelihood function is designed to quantify how closely the observed pixel statistics within elliptical annuli match an empirical variance model. For each annulus $k$, we use an expected variance ($\sigma_k^2$) that is a function of the mean brightness of the annulus ($M_k$). This relationship was determined empirically by fitting a polynomial to the observed variance as a function of mean intensity in log-log space (Figure \ref{fig:geo_method}):
\begin{equation}
\log_{10}(\sigma_k^2) = \alpha \cdot \log_{10}(M_k)^2 + \beta \cdot \log_{10}(M_k) + \kappa
\end{equation}
The coefficients $\alpha$, $\beta$, and $\kappa$ are the best-fit parameters from the second-order polynomial regression shown in Figure \ref{fig:geo_method}. This fit was performed on statistics gathered in circular annuli, assuming a temporary face-on geometry, to establish a general phenomenological variance model before the main \gls{mcmc} fit. This approach, which temporarily assumes a face-on geometry, allows for a characterization of the relationship between signal and variance across the image. For a disk known to have a low inclination like TW Hya, the use of circular annuli is a robust approximation for this purpose, as the on-sky projection is very close to circular. The consistency of this approach is supported by our final fitting results, which are consistent with the disk being nearly face-on (Table \ref{tab:disk_fit}).

\begin{deluxetable*}{lcccc|cc}[ht!]
\tablecaption{Disk Geometry Fitting Prior and Posterior\label{tab:disk_fit}}
\tablehead{
\colhead{Parameter} & \colhead{Symbol} & \colhead{Units} & \colhead{Prior} & \colhead{Posterior (Whole Disk)} & \colhead{Posterior (Inner Disk)} & \colhead{Posterior (Outer Disk)}}
\startdata
x-center & $x_0$ & pixels & Uniform(300, 350) & $325.17^{+0.13}_{-0.08}$ & $325.17^{+0.13}_{-0.11}$ & $325.19^{+0.23}_{-0.30}$ \\
y-center & $y_0$ & pixels & Uniform(300, 350) & $324.57^{+0.08}_{-0.09}$ & $324.58^{+0.11}_{-0.12}$ & $324.44^{+0.20}_{-0.18}$ \\
Inclination & $i$ & degrees & Uniform(0, 15) & $\dotdeg{8.74}^{\dotdeg{+1.03}}_{\dotdeg{-0.94}}$ & $\dotdeg{10.62}^{\dotdeg{+2.45}}_{\dotdeg{-1.90}}$ & $<\dotdeg{9.75} (2\sigma)$\tablenotemark{a} \\
Position Angle & PA & degrees & Uniform(40, 160) & $\dotdeg{75.62}^{\dotdeg{+7.86}}_{\dotdeg{-6.56}}$ & $\dotdeg{82.80}^{\dotdeg{+10.19}}_{\dotdeg{-9.63}}$ & $\dotdeg{76.69}^{\dotdeg{+18.86}}_{\dotdeg{-14.32}}$ \\
\hline
\enddata
\tablecomments{The posterior values represent the median of the marginalized distribution, with uncertainties corresponding to the 16th and 84th percentiles. Fits were performed on the inner disk (0.675\arcsec--1.44\arcsec), outer disk (1.44\arcsec--2.98\arcsec), and whole disk (0.675\arcsec--2.98\arcsec) regions. \tablenotetext{a}{The posterior for the outer-disk inclination is not well-constrained; we report the 2$\sigma$ upper limit.}}
\end{deluxetable*}

The log-likelihood is constructed from the sum of contributions from each annulus, which are visualized for our best-fit model in Figure \ref{fig:geo_method}. Our method assumes the pixel intensities in an annulus follow a Gaussian distribution, but corrects for the correlated noise introduced by the instrumental \gls{psf} by defining an \textit{effective number of independent samples}, $N^{\text{eff}}_k = N_k / N_{\text{beam}}$ (where $N_{\text{beam}} \approx 46.5$~pixels, see Appendix \ref{app:beam_size}). The full derivation of the log-likelihood function is detailed in Appendix \ref{app:likelihood}, yielding the final expression:
\begin{equation}
\ln(\mathcal{L}) = -\frac{1}{2} \sum_{k} N^{\text{eff}}_k \left[ \frac{\mathrm{var}_k}{\sigma_k^2} + \ln(\sigma_k^2)  \right]
\label{eq:logL_final}
\end{equation}
where $\mathrm{var}_k$ is the variance of the pixel intensities within annulus $k$ and $\sigma_k^2$ is the expected variance from our empirical model. This formulation, therefore, seeks the geometry that minimizes the intrinsic azimuthal variance of the disk. The empirical variance model, $\sigma_k^2$, serves to appropriately weight the contribution of each annulus, ensuring that the fit is not dominated by brighter regions where higher variance is expected.

To validate these assumptions, we performed three consistency checks. First, re-evaluating the empirical variance model using the best-fit inclined elliptical annuli produced a variance-mean relationship nearly identical to the initial circular assumption, confirming the variance is mainly driven by intrinsic disk structure and photon noise rather than geometric projection effects. Second, we confirmed that the residual pixel intensities within each annulus follow nearly a Gaussian distribution. Third, mapping the intensity outliers ($> ~3\sigma$) revealed no systematic azimuthal clustering (e.g., along the semi-minor axis), indicating that a more complex scattering phase-function model is not required for this dataset.

To survey potential radial variations in the disk's structure, we performed three independent fits on distinct regions of the F200W image. The central region ($<$ 0.675\arcsec) was excluded to avoid biasing the fit with the significant artifacts and high-amplitude residuals that remain after the subtraction of the stellar \gls{psf} near the coronagraphic mask. Similarly, the region beyond 2.98\arcsec was excluded due to the low signal-to-noise ratio of the disk. First, a `whole-disk' fit was conducted over the resulting radial range of 0.675\arcsec--2.98\arcsec to determine the average geometry across this entire well-detected region of scattered light. To test for systematic changes in orientation with radius, such as those caused by a disk warp, we then divided this area into two zones for separate analyses. An `inner-disk' fit (0.675\arcsec--1.44\arcsec) was defined to cover the bright, primary rings of the disk, while an `outer-disk' fit (1.44\arcsec--2.98\arcsec) probes the fainter structures at larger separations. This division allows us to probe for differences in the derived geometric parameters as a function of radius. The uniform priors and resulting marginalized posterior values for all three fits are detailed in Table \ref{tab:disk_fit}.

\begin{figure*}[ht!]
\centering
\includegraphics[width=\textwidth]{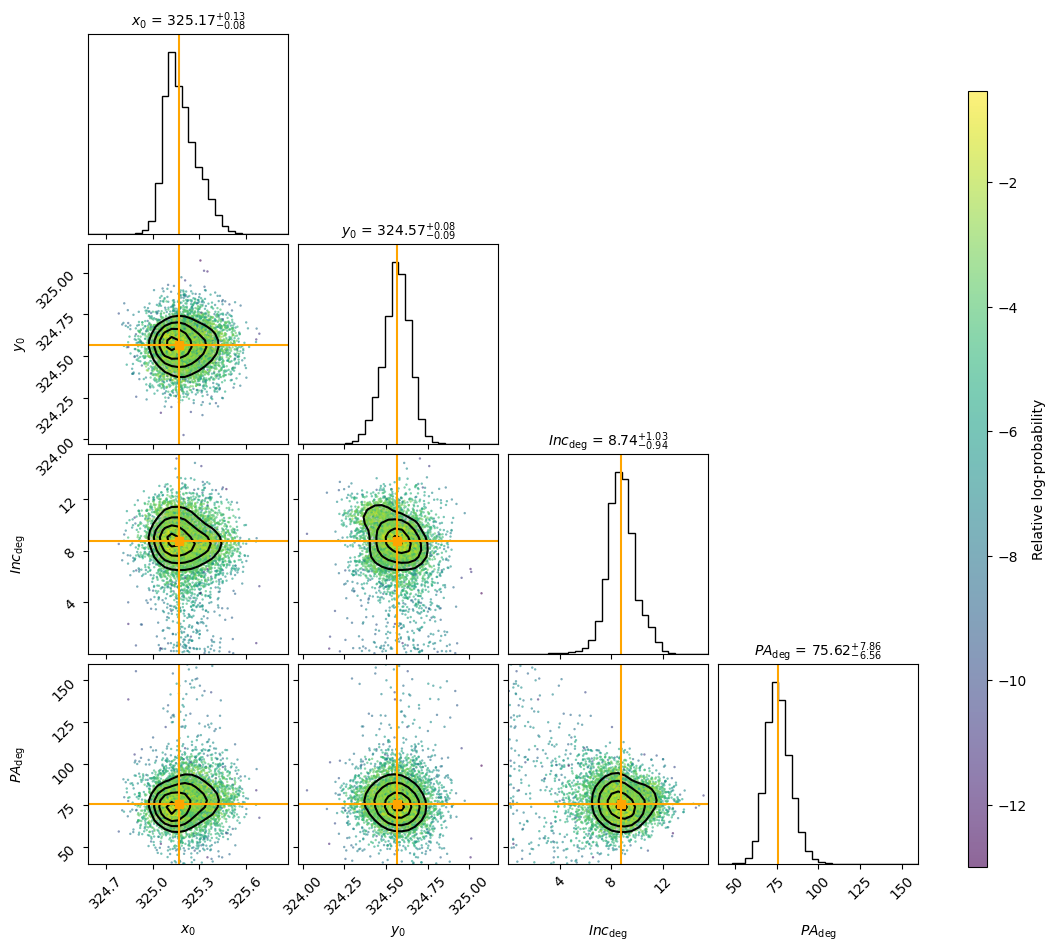}
\caption{\gls{mcmc} posterior distributions from the fit to the whole disk (0.675\arcsec--2.98\arcsec). This fit provides the averaged geometry for the system. Each panel shows the one- and two-dimensional projections of the posterior probability distributions for the geometric parameters. The histograms show the marginalized distribution for each parameter, with the median value and 1$\sigma$ uncertainties quoted; the solid orange lines indicate the median values.
\label{fig:diskgeo_whole}}
\end{figure*}

\begin{figure*}[ht!]
\centering
\includegraphics[width=\textwidth]{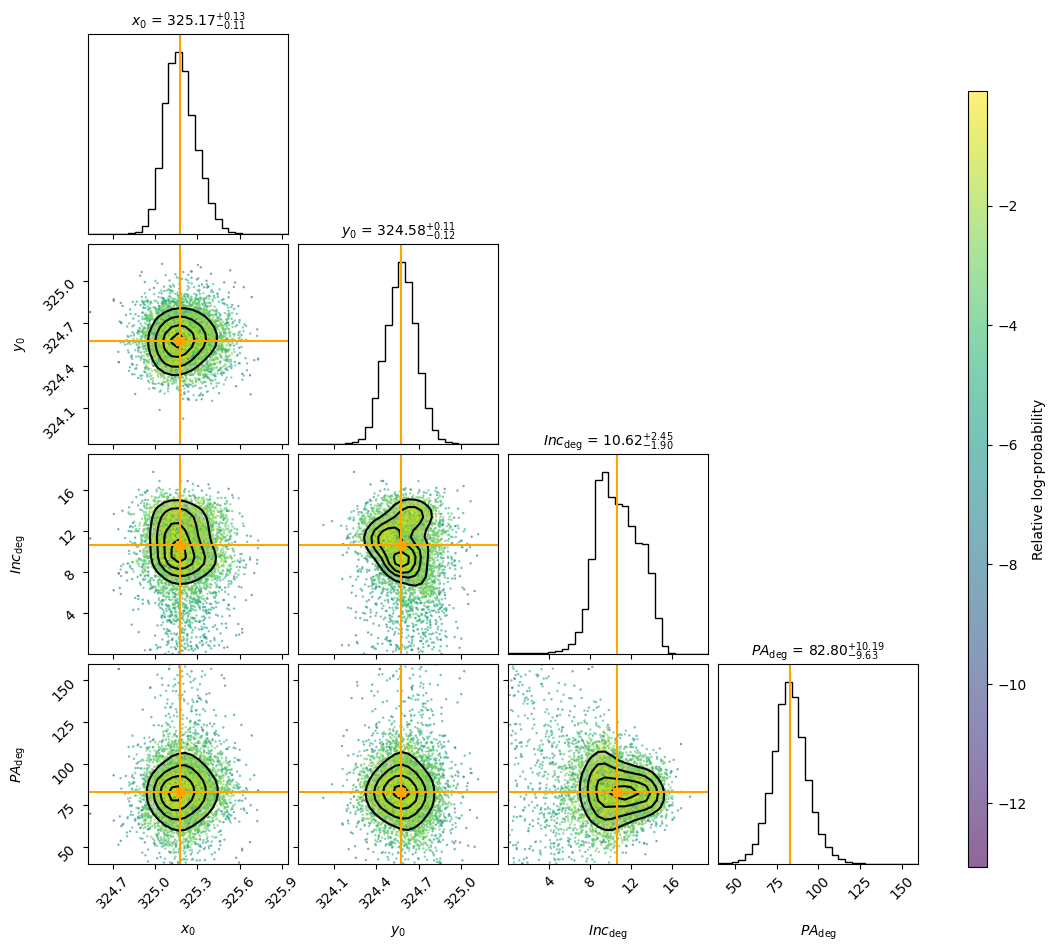}
\caption{\gls{mcmc} posterior distributions for the geometric parameters from the fit to the inner disk (0.675\arcsec--1.44\arcsec). In contrast to the outer disk, the posteriors for the inner-disk parameters are single-peaked and
comparatively well constrained.
\label{fig:diskgeo_inner}}
\end{figure*}

\begin{figure*}[ht!]
\centering
\includegraphics[width=\textwidth]{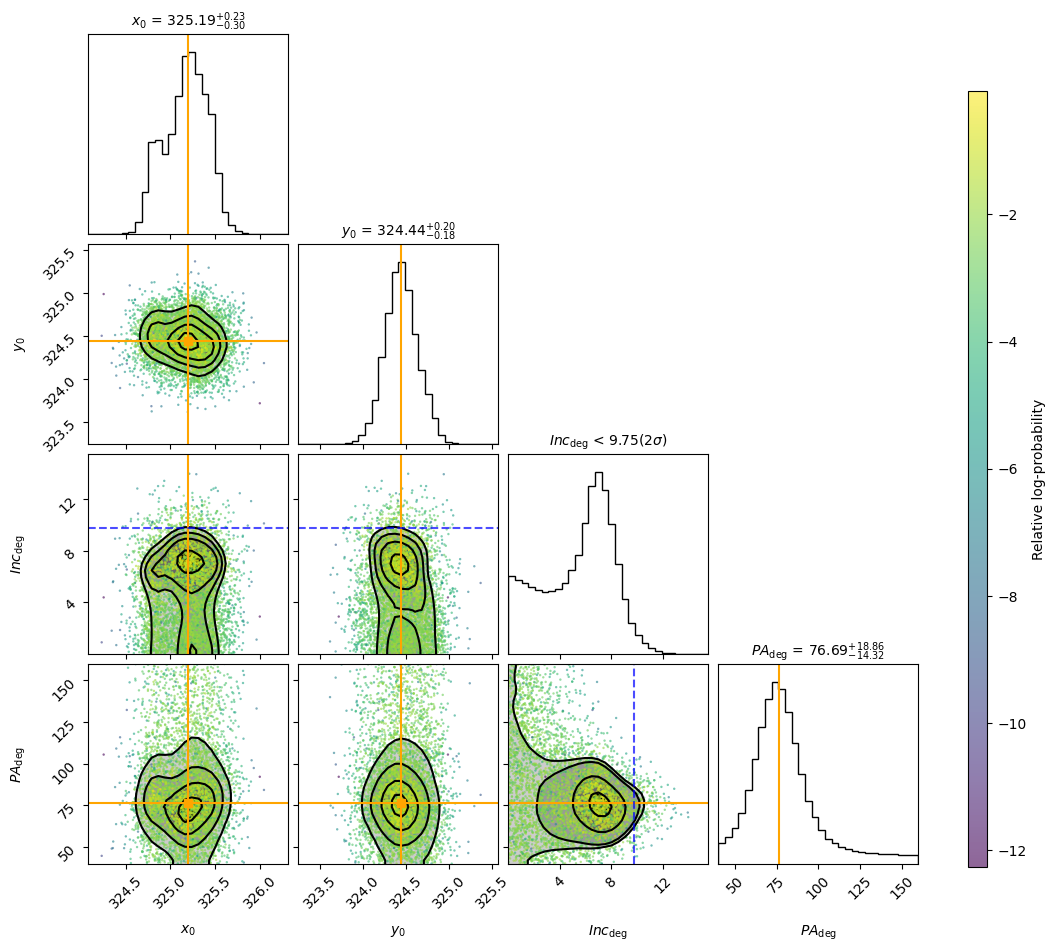}
\caption{\gls{mcmc} posterior distributions from the fit to the outer disk (1.44\arcsec--2.98\arcsec). The inclination posterior is broad and asymmetric; we therefore report a $2\sigma$ upper limit of $\dotdeg{9.75}$, marked by the blue dashed line. Because its allowed inclination range overlaps those for the whole- and inner-disk fits, this result does not by itself establish a statistically significant radial change in geometry, although it remains consistent with a warped geometry.
\label{fig:diskgeo_outer}}
\end{figure*}

The posterior distributions from the whole-disk fit, which provide our measurement of the average disk geometry, are shown in Figure \ref{fig:diskgeo_whole}. This fit yields an inclination of $i = \dotdeg{8.74}^{\dotdeg{+1.03}}_{\dotdeg{-0.94}}$ and a position angle of $\mathrm{PA} = \dotdeg{75.62}^{\dotdeg{+7.86}}_{\dotdeg{-6.56}}$. The analysis of separate disk regions, however, reveals tentative evidence for radial variation. The inner-disk fit (Figure \ref{fig:diskgeo_inner}) favors a higher inclination, yielding $i = \dotdeg{10.62}^{\dotdeg{+2.45}}_{\dotdeg{-1.90}}$ with a well-constrained, single-peaked posterior. In contrast, the fit to the outer disk (Figure \ref{fig:diskgeo_outer}) produces a broad and asymmetric inclination posterior that is skewed toward lower values. We therefore report a 2$\sigma$ upper limit of $\dotdeg{9.75}$. Although the inner-disk fit favors a higher inclination, the allowed
inclination ranges for the whole-, inner-, and outer-disk fits overlap. Consequently, the geometric fitting alone does not establish a statistically significant radial variation in inclination, although the results remain consistent with the presence of a warp.

\clearpage

\begin{figure*}[ht!]
\gridline{\fig{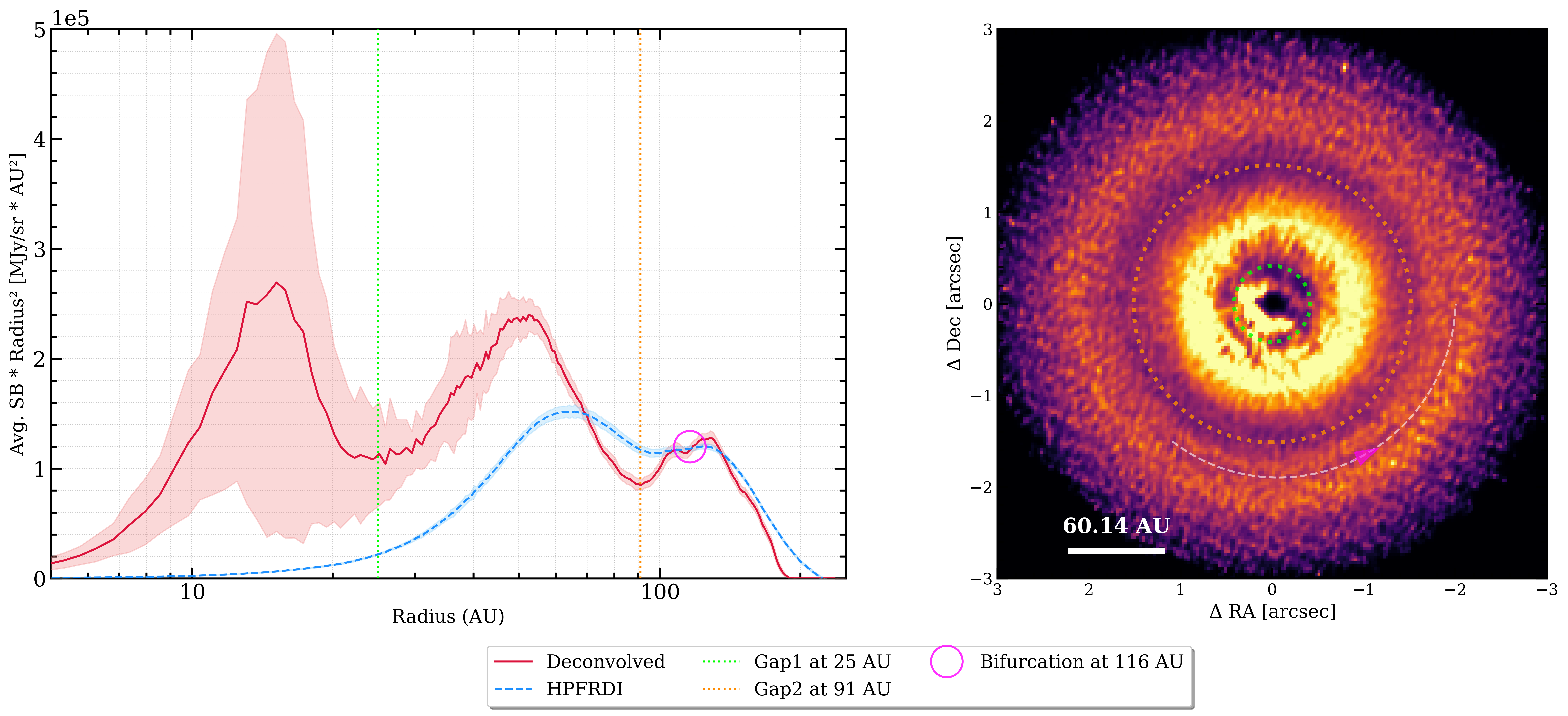}{\textwidth}{(a) F200W}}
\gridline{\fig{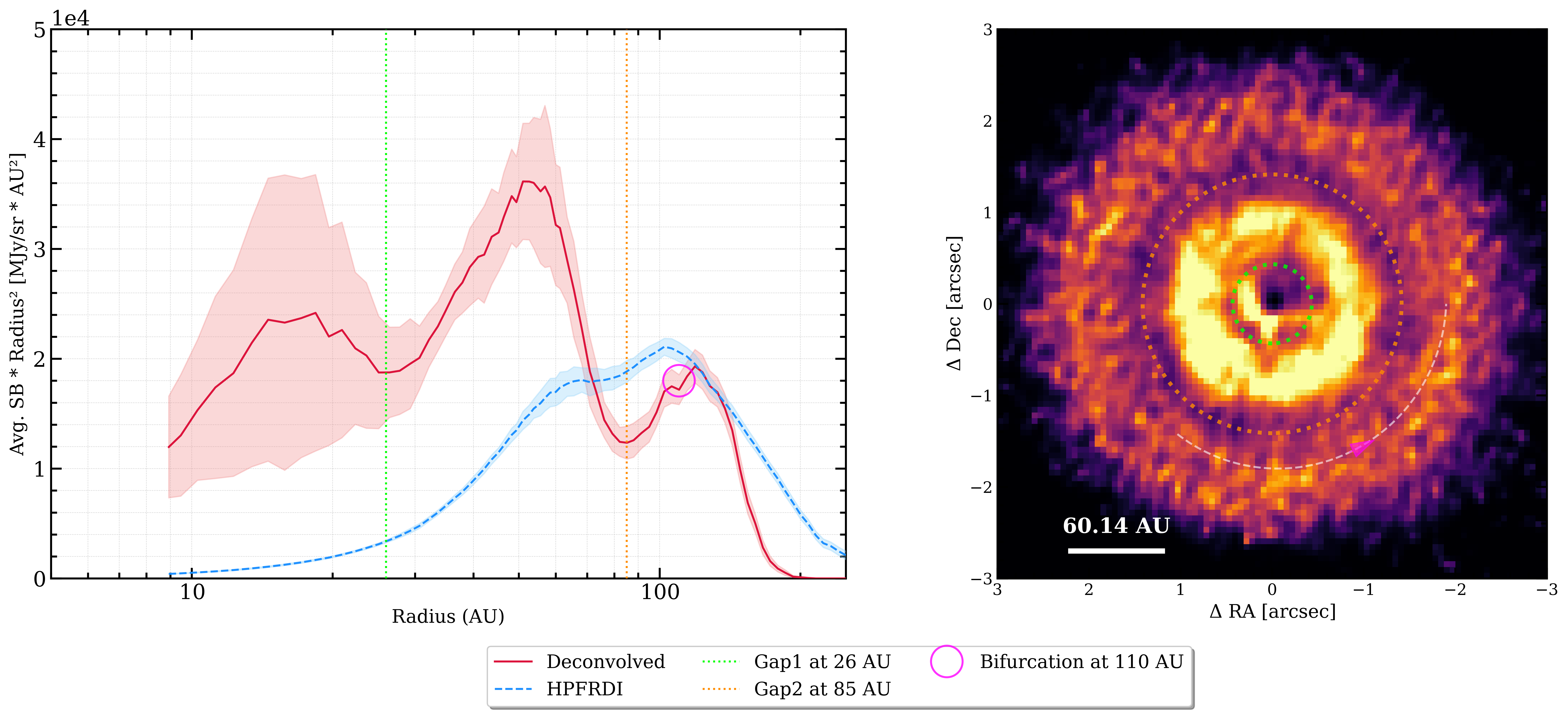}{\textwidth}{(b) F444W}}

\caption{Radial surface brightness profiles and corresponding $r^2$-scaled images of the TW Hya disk in the (a) F200W and (b) F444W filters. \textbf{Left panels:} The y-axis shows the average surface brightness multiplied by the radius squared ($r^2$) to compensate for the geometric fall-off of starlight. The solid red line shows the profile from the deconvolved image, with the shaded region indicating the \glstarget{sem} 1$\sigma$ \gls{sem}—corrected for the effective beam size to account for correlated noise—across the radial bins. The dashed blue line is the profile from the \gls{hpfrdi}-processed image. Vertical dotted lines highlight radial gaps at 25~AU (green) and 91~AU (orange) in F200W, and 26~AU (green) and 85~AU (orange) in F444W, while the open magenta circles mark the bifurcation structure at 116~AU in F200W and 110~AU in F444W. \textbf{Right panels:} The 2D images of the disk, similarly scaled by $r^2$ and normalized to reveal faint outer disk structures. Dotted rings spatially trace the locations of the Gap 1 and Gap 2 structures. A dashed white curve with a magenta arrow traces the morphology of the spiral/bifurcation feature extending outward.
\label{fig:radial_profile}}
\end{figure*}

\begin{figure*}[ht!]
\centering
\gridline{
\fig{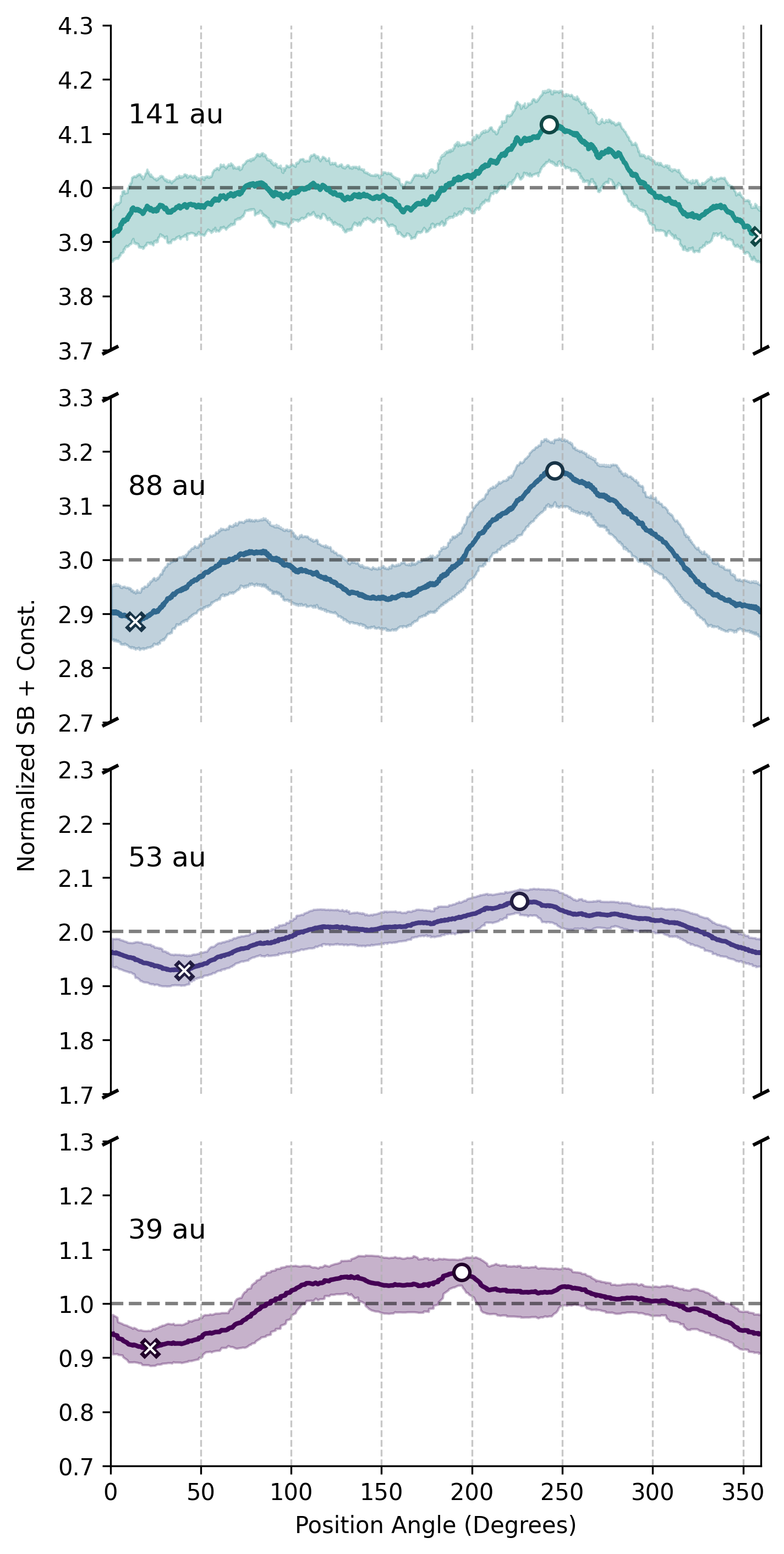}{0.5\textwidth}{(a) F200W Filter}   
\fig{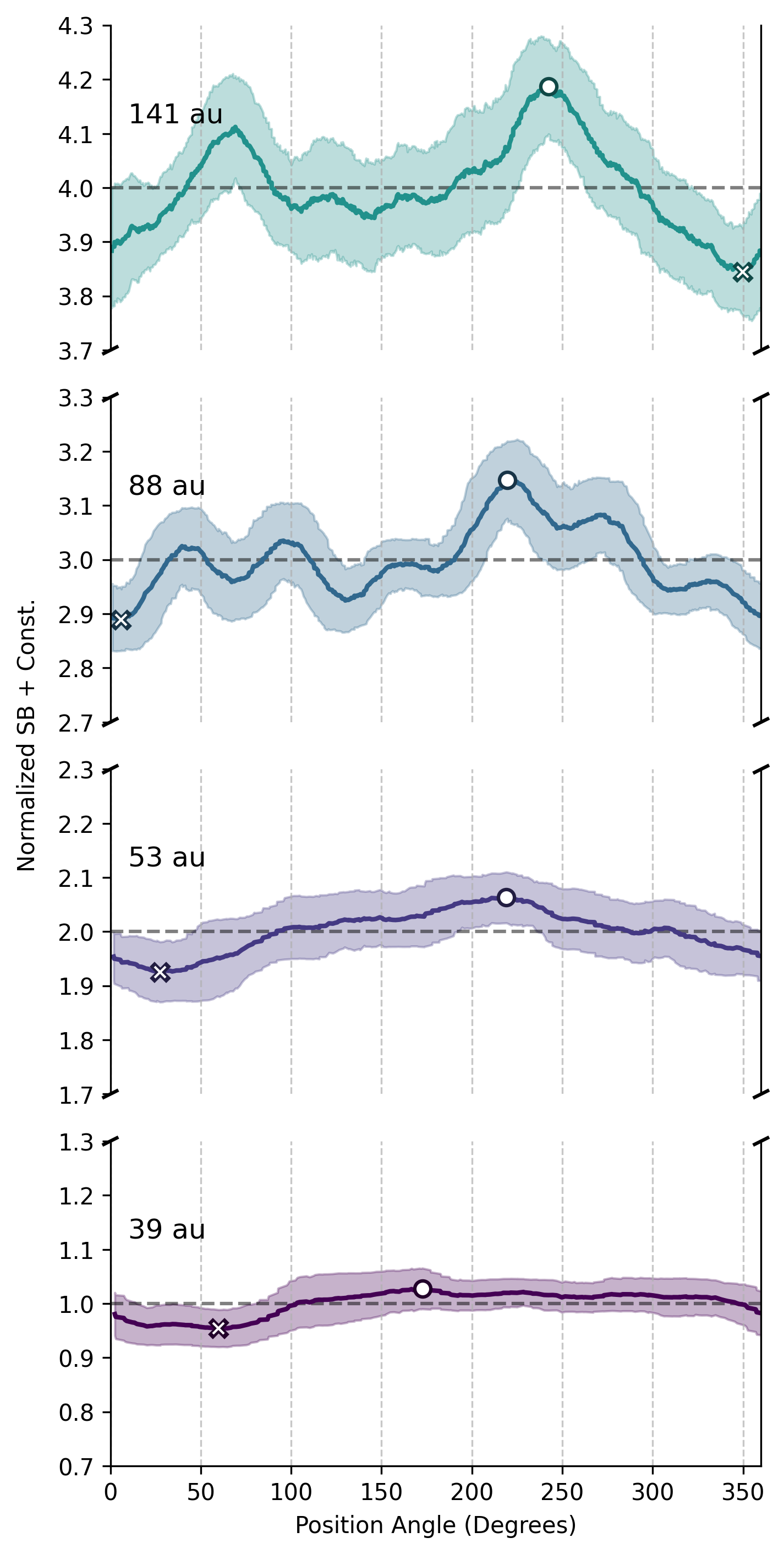}{0.5\textwidth}{(b) F444W Filter}}
\caption{Azimuthal surface brightness profiles from the JWST/\gls{nircam} observations, derived from the \gls{hpfrdi} images. This approach was chosen to analyze the on-sky brightness distribution before deconvolution, avoiding potential artifacts that could affect the measurement of azimuthal structures. Panel (a) shows the F200W data, and panel (b) shows the F444W data. Each profile is calculated using a pixel-by-pixel sliding window (3~pixels radially by 15~pixels azimuthally) to ensure the extraction area approximates the instrumental beam size. The profiles are normalized by their respective means and offset vertically for clarity. The solid line represents the normalized mean, the shaded region indicates the \gls{sem} corrected for the effective sample size ($N_{\text{eff}}$), and the measured brightness minima (crosses) and maxima (circles) are marked. The systematic radial shift in the positions of these features across both filters is consistent with a shadow being projected onto a warped outer disk.}
\label{fig:azimuthal_shadow}
\end{figure*}

\subsection{Disk Features}
\label{subsec:disk_features}
\begin{figure*}[ht]
\centering
\includegraphics[width=0.85\textwidth]{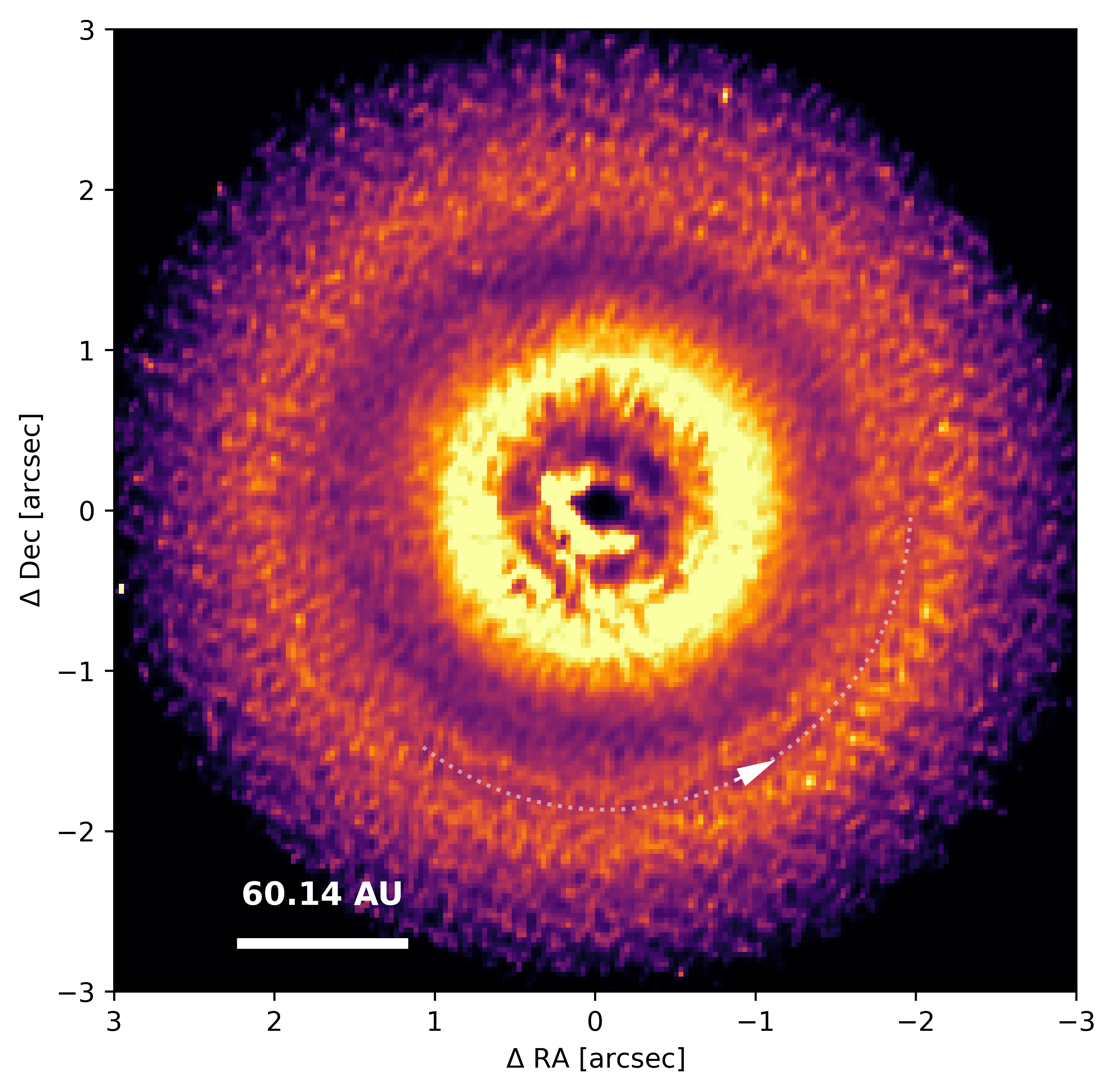}
\caption{An $R^2$-scaled view of the deconvolved F200W image, enhancing the outer disk structures. This visualization shows the bifurcation feature at approximately 120~AU, traced by a dotted logarithmic spiral with a 2\degr pitch angle. A white arrow points to the spiral track to aid identification; note that the arrow's direction is for visual guidance only and carries no physical significance. The image is normalized, and a scale bar indicates the physical size of 60.1~AU (1\arcsec) on the sky.\label{fig:bifurcation}}
\end{figure*}

\subsubsection{Radial and Azimuthal Profiles}

To quantitatively analyze the disk's annular features, we extracted both radial and azimuthal surface brightness profiles. As demonstrated by our synthetic injection tests (Appendix \ref{app:deconv_quality}), the Richardson-Lucy algorithm requires a higher number of iterations to accurately recover the global radial profile, which inevitably amplifies pixel-to-pixel grain noise. Therefore, we extracted the radial profiles from the final deconvolved images to accurately locate the gaps and bifurcation structure. Conversely, because azimuthal profiles are highly sensitive to this amplified grain noise, we extracted them from the non-deconvolved \gls{hpfrdi} images. This choice also ensures a direct, consistent comparison with the non-deconvolved \gls{hst}/\gls{stis} azimuthal analysis presented in \citet{2023ApJ...948...36D}.

The radial profiles (Figure \ref{fig:radial_profile}) were calculated using a hybrid extraction method: a sliding window for the inner profile and logarithmic binning for the outer profile. Given the system's low inclination, we treat the disk as effectively face-on for the extraction of these features. First, to enhance the visibility of faint structures, the surface brightness at each pixel was scaled by the square of its radial distance from the star ($R^2$). This technique compensates for the geometric dilution of starlight and makes features at large radii more prominent. Second, to minimize the impact of noise and artifacts, a 6\% outlier clip was applied within each sliding window (for the inner profile) and within each logarithmic bin (for the outer profile), removing the brightest and faintest 3\% of pixels before statistical calculation. To preserve spatial resolution, the inner profile utilizes a sliding window whose size matches the \gls{psf} beam size. To reduce computational overhead, this inner window advances outward with a progressively increasing pixel step size. Beyond this densely sampled inner region (which spans the first $N_{\text{beam}}$ dynamically spaced sliding windows), the remaining outer profile is divided into logarithmically spaced bins (200 bins for F200W and 80 bins for F444W). The uncertainty for each sliding window and radial bin, represented by the shaded error bands in Figure \ref{fig:radial_profile}, is the \gls{sem}, which we have corrected for correlated noise by dividing the pixel count by the effective beam size ($N_{\text{eff}} = N_{\text{pixels}} / N_{\text{beam}}$). These profiles reveal distinct substructures: in F200W, we identify Gap 1 at 25~AU, Gap 2 at 91~AU, and a bifurcation at 116~AU; in F444W, these features are located at 26~AU, 85~AU, and 110~AU, respectively.

For the azimuthal profiles (Figure \ref{fig:azimuthal_shadow}), we applied a pixel-by-pixel sliding window along the azimuthal direction. We extracted surface brightness profiles at radii of 39, 53, 88, and 141~AU. These specific radii were selected to enable a direct comparison with the analysis of historical \gls{hst} observations presented in Figure 8 of \citet{2023ApJ...948...36D}. To properly account for correlated noise and ensure a statistically robust effective sample size ($N_{\text{eff}} \approx 1$), the sliding window was sized to roughly match the area of one instrumental beam (3~pixels radially by 15~pixels azimuthally). Within each window, the mean brightness and beam-size-corrected \gls{sem} were calculated. To accurately trace the relative brightness variation as a function of \gls{pa}, each profile was then normalized by the overall mean brightness of its respective radial annulus and offset vertically for clarity.

\subsubsection{Disk Substructures}
\label{subsubsec:substructures}
The radial surface brightness profile, shown in Figure \ref{fig:radial_profile}, reveals a complex structure. Most notably, our observations resolve the structure at approximately 110--116~AU, revealing a complex morphology that appears to bifurcate or split into two components, a feature that is visualized in Figure \ref{fig:bifurcation}. This corresponds to the complex feature first identified in $H$-band scattered-light images from \gls{vlt}/\gls{sphere} \citep{2017ApJ...837..132V} \footnote{\citet{2017ApJ...837..132V} noted this feature at $\sim$100~AU based on the smaller assumed distance of TW Hya (54 pc, versus the 60.14 pc used in this work)}. While \citet{2017ApJ...837..132V} suggested the feature could be traced by a spiral with a $\dotdeg{1.5}$ pitch angle, we find that a logarithmic spiral with a slightly larger pitch angle of $2\degr$ provides a better match to our high-resolution deconvolved data assuming a face-on disk geometry.

Complementary to the radial profile, the azimuthal brightness profiles in Figure \ref{fig:azimuthal_shadow} trace the brightness variations at these specific radii. Both the F200W and F444W data reveal significant, and in some cases complex, azimuthal brightness modulations, suggesting a departure from simple axisymmetric structures. However, careful consideration of instrumental effects is required when comparing the two filters. For instance, the multiple-peaked, ``wiggly" structure seen at 88~AU in the F444W profile is an instrumental artifact arising from the hexagonal diffraction pattern of the JWST primary mirror, which is broader and more prominent at longer wavelengths. The F200W profile, therefore, provides a more reliable representation of the disk at this separation.

At larger separations (e.g., 141~AU), the pairwise morphological differences are potentially physical. The F444W profile exhibits a distinct double-peak structure, whereas F200W shows a broader, flatter plateau preceding the main peak. Notably, the secondary peak at 141~AU in F444W aligns more closely with the shape of features seen further inward at 88~AU in F200W. One possible explanation for these azimuthal discrepancies is 3D projection effects. Because the F444W bandpass probes deeper into the disk at a lower optical depth, it traces the same physical structures at different apparent projected radii compared to the smaller grains traced by F200W. However, this is only one possible interpretation; localized variations in dust properties or density could also contribute. Detailed 3D radiative transfer modeling is required to accurately distinguish between these mechanisms and confirm the origin of these morphological differences.
 
These wavelength-dependent differences also manifest in the radial profiles. While the overall disk morphology is broadly consistent between filters, there are notable differences in the radial locations of specific features. Our deconvolved radial profiles (Figure \ref{fig:radial_profile}) reveal a physical shift in the scattering surface with wavelength. Gap 2 and the bifurcation structure in F444W are located 6~AU inward compared to F200W. This radial offset may be driven by several distinct physical mechanisms. First, it may simply be a geometric consequence of the disk's well-established vertical stratification \citep{2006ApJ...638..314D, 2014A&A...564A..93M}. Because dust opacity generally decreases at longer near- and mid-infrared wavelengths \citep{2003ARA&A..41..241D}, the F444W bandpass could probe deeper optical depths than the F200W bandpass. If the gap wall possesses a tapered vertical profile, tracing this scattering surface at varying depths would naturally yield a smaller apparent radius at longer wavelengths. Alternatively, this discrepancy might reflect a true radial variation in the dust distribution driven by aerodynamic filtration across the gap edge \citep{2006MNRAS.373.1619R, 2012ApJ...755....6Z, 2015ApJ...809...93D}. Despite the small difference in probed grain sizes, a sharp pressure gradient may still trap the slightly larger grains at a smaller radius. Localized variations in dust properties, such as composition and scattering albedo near the gap edge, could also contribute to this wavelength-dependent spatial shift.

Separately, the comparison between the \gls{hpfrdi} and deconvolved data highlights the impact of instrumental broadening on identifying these structures. The diffraction-limited \gls{psf} of JWST is more than twice as broad at F444W as it is at F200W. In the \gls{hpfrdi} profiles (dashed lines in Figure \ref{fig:radial_profile}), this larger \gls{psf} acts as a smoothing kernel, blending the bifurcated structures resolved at shorter wavelengths into broader features. The deconvolution process effectively removes this blurring, recovering the split structure and allowing for the physical comparison described above.

\subsection{Companion Search and Detection Limits}
\label{subsec:mass_limit}
To search for faint companions, we leveraged the two distinct spacecraft roll angles to perform a separate \gls{adi} reduction. We employed \glstarget{klip} the \gls{klip} algorithm \citep{2012ApJ...755L..28S}, implemented within the \texttt{spaceKLIP} pipeline, to model and subtract the residual stellar \gls{psf}. Because TW Hya's disk is nearly face-on, \gls{klip} \gls{adi} subtraction removes the vast majority of the disk signal, allowing the analysis to focus on faint point sources. We concentrated our sensitivity analysis on the F444W dataset because this long-wavelength filter offers optimal sensitivity to the thermal emission of cool, planetary-mass companions \citep{2021MNRAS.501.1999C}. Because the observational sequence contains only two discrete roll angles, the principal-component basis for the \gls{klip} reduction contains only one component; consequently, exploring a larger number of KL modes is not applicable. No statistically significant point sources were detected in the final processed image. We therefore derived detection limits to quantify the constraints on unseen companions. The methodology for deriving these limits is detailed below, and the resulting sensitivity curves are presented in Figure \ref{fig:mass_limit}. Detection limits and mass-sensitivity curves for the other three filters are presented in Appendix \ref{app:add_limits}.

\begin{figure*}[ht]
\includegraphics[width=1.0\textwidth]{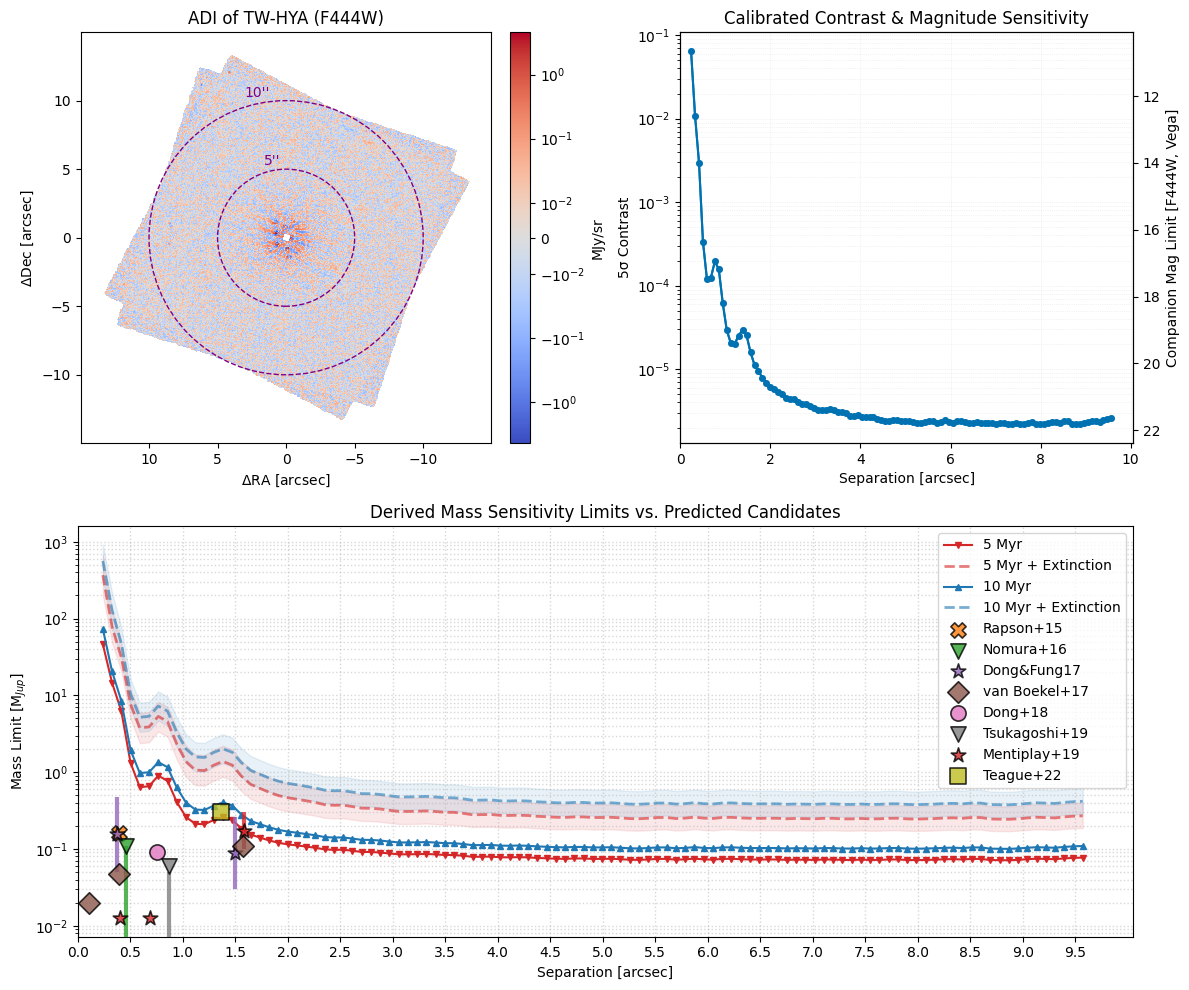}
\caption{Sensitivity limits from our companion search in the F444W filter. The top left panel shows the \gls{adi}-processed image. The top right panel presents the 5$\sigma$ calibrated contrast curve (left axis) and corresponding apparent magnitude limit (right axis). The bottom panel displays the derived upper mass limits using \citet{2019A&A...623A..85L} evolutionary models for system ages of 5 (red) and 10 (blue) Myr. 
Solid lines represent limits assuming clear conditions, while dashed lines with shaded error regions indicate limits adjusted for local extinction ($A_{F444W} \approx 2.7 \pm 0.7$~mag). 
The various markers plot the estimated masses and separations of candidate protoplanets proposed in the literature (detailed in Table \ref{tab:planet_predictions}) to explain previously observed disk substructures.
\label{fig:mass_limit}}
\end{figure*}

\begin{deluxetable*}{lcccc}
\tablecaption{Predicted Protoplanet Candidates in the TW Hya Disk \label{tab:planet_predictions}}
\tablehead{
\colhead{} & \colhead{Assumed} & \colhead{Modeled} & \colhead{Projected} & \colhead{Predicted} \\
\colhead{Reference} & \colhead{Distance} & \colhead{Radius} & \colhead{Separation\tablenotemark{a}} & \colhead{Mass} \\
\colhead{} & \colhead{(pc)} & \colhead{(AU)} & \colhead{(arcsec)} & \colhead{($M_{\rm Jup}$)}
}
\startdata
\citet{2015ApJ...815L..26R} & 54 & 21 & 0.39 & $\sim$0.16 \\
\citet{2016ApJ...819L...7N} & 54 & 25 & 0.46 & $\lesssim$0.11\tablenotemark{b} \\
\citet{2017ApJ...835..146D} & 54 & 20 & 0.37 & 0.05--0.48 \\
 & 54 & 81 & 1.5 & 0.03--0.26 \\
\citet{2017ApJ...837..132V} & 54 & 6 & 0.11 & $\sim$0.020 \\
 & 54 & 21 & 0.39 & $\sim$0.047 \\
 & 54 & 85 & 1.57 & $\sim$0.11 \\
\citet{2018ApJ...866..110D} & 60 & 45 & 0.75 & $\sim$0.09 \\
\citet{2019ApJ...878L...8T} & 59.5 & 52 & 0.87 & $\lesssim$0.06\tablenotemark{c} \\
\citet{2019MNRAS.484L.130M} & 59.5 & 24 & 0.40 & $\sim$0.0126\tablenotemark{d} \\
 & 59.5 & 41 & 0.69 & $\sim$0.0126\tablenotemark{d} \\
 & 59.5 & 94 & 1.58 & 0.1--0.3\\
\citet{2022ApJ...936..163T} & 60.1 & 82 & 1.36 & $\sim$0.30 \\
\enddata
\tablecomments{Estimates of planet masses and locations from the literature proposed to explain various disk substructures.}
\tablenotetext{a}{Calculated as $r (\text{AU}) / d (\text{pc})$ using the specific distance adopted by each reference.}
\tablenotetext{b}{Upper limit of $2\,M_{\rm Neptune}$ converted to Jupiter masses.}
\tablenotetext{c}{Upper limit of $1\,M_{\rm Neptune}$ converted to Jupiter masses.}
\tablenotetext{d}{$\sim 0.4\,M_{\oplus}$ converted to Jupiter masses.}
\end{deluxetable*}

\subsubsection{Derivation of the Calibrated Contrast Curve}
The foundation of our sensitivity analysis is the 5$\sigma$ calibrated contrast curve. The raw contrast was first determined by measuring the standard deviation of flux in concentric annuli in the final \gls{klip}-processed image. However, the \gls{klip} algorithm can partially suppress the flux from a real astrophysical source due to self-subtraction of a companion signal, meaning the raw noise level does not represent the true detection limit.

To correct for this algorithmic throughput loss, we performed synthetic planet injection tests using \texttt{spaceKLIP}. Synthetic planet \glspl{psf}, generated using \texttt{STPSF}, were injected into the calibrated data prior to \gls{klip} processing at a range of separations and \glspl{pa}. We then processed this injected dataset using the identical \gls{klip} reduction pipeline. The throughput at a given separation was calculated as the ratio of the recovered flux of the synthetic planet to its known injected flux. The raw contrast curve was then divided by this empirically measured throughput curve to produce the final, calibrated 5$\sigma$ contrast curve.

\subsubsection{Mass Sensitivity and Protoplanet Candidates}
The calibrated contrast curve was converted into a limit on the apparent Vega magnitude of any potential companion using the stellar magnitude of TW Hya in the F444W filter. Subsequently, we translated these apparent magnitude limits into planet mass limits. This final conversion was achieved by interpolating the \citet{2019A&A...623A..85L} planetary evolution models, assuming the system distance of 60.14 pc and two representative ages of 5 and 10 Myr.

For the conservative 10~Myr case, our F444W observations rule out the presence of non-embedded companions with masses greater than $\sim0.4$~\mjup beyond 60~AU ($1\arcsec$) and $\sim2.0$~\mjup beyond 30~AU ($0.5\arcsec$). These limits are consistent with pre-launch predictions of JWST's performance, which forecasted sensitivity to sub-Jupiter-mass objects beyond 30~AU \citep{2021MNRAS.501.1999C}.

However, such limits assume non-embedded companions. Planets forming within gaps could be subject to significant local extinction from circumstellar material \citep{2020MNRAS.492.3440S}, as demonstrated by \citet{2025AJ....170..317C} for the system AS 209, where a background star's signal within a gap was found to be attenuated by $2.7 \pm 0.7$~mag at 4~\mum{}. To account for this scenario, we computed the extinction-corrected mass sensitivity curves (dashed lines in Figure \ref{fig:mass_limit}) by adopting this attenuation value for our F444W data. The shaded region represents the uncertainty in the mass limit derived from the $\pm 0.7$~mag error in the extinction estimate. While AS 209 and TW Hya are different systems, this adjustment provides a more realistic constraint for embedded protoplanets than the dust-free assumption. Even with this significant extinction correction, our observations reach sub-Jupiter masses ($\lesssim 1\,M_{\rm Jup}$) beyond $\sim2$\arcsec.

We compare these limits to specific planet candidates proposed in the literature to explain the diverse substructures observed in TW Hya (Table \ref{tab:planet_predictions}). Our F444W sensitivity has a chance to detect some of the predicted high-mass candidates if they were not obscured, particularly those proposed to drive the large-scale spiral features \citep{2022ApJ...936..163T}. The non-detection of these sources suggests that if they exist, they are likely lower in mass or subject to similar extinction as the AS 209 proxy applied here.

\section{Discussion} \label{sec:discussion}

\subsection{An Evolved Understanding of the TW Hya Shadow}

The two-ring precession model from \citet{2023ApJ...948...36D} provided a promising explanation for the state of the TW Hya shadow in 2021, locating the shadowing structures at 5--7~AU. This geometry implies differential rotation with Keplerian periods ranging from 13--17~yr. Our new JWST observations from February 2024 allow for a quantitative test of this system's evolution. The time elapsed between the \gls{hst} observation (June 7, 2021) and our JWST observation (February 14, 2024) is $\sim$2.69~yr. To provide a baseline for comparison, we test the projection of the canonical 15.9~yr period \citep{2017ApJ...835..205D} that described the shadow's bulk motion prior to 2016. Based on this period, the features are predicted to have rotated counter-clockwise by:
\[ \Delta\theta = 2.69 \text{~yr}/15.9 \text{~yr} \times 360^{\circ} \approx \dotdeg{60.9} \] Applying this rotation to the 2021 positions yields the following predictions for February 2024: shadow local minima at \glspl{pa} of $49^{\circ} + 60.9^{\circ} \approx 110^{\circ}$ and $175^{\circ} + 60.9^{\circ} \approx 236^{\circ}$, with corresponding brightness local maxima at the midpoints, \glspl{pa} $\approx 173^{\circ}$ and $353^{\circ}$. Table \ref{tab:shadow_locations} compares these predictions with the features observed in our JWST \gls{hpfrdi} images (quantified in Figure \ref{fig:azimuthal_shadow}), revealing a clear departure from the steady precession forecast.

\begin{deluxetable*}{l|cccc}
\tablecaption{Predicted vs. Observed Shadow Locations \label{tab:shadow_locations}}
\tablewidth{0pt}
\tablehead{
\colhead{Feature} & \colhead{Observed 2021 (\gls{hst})} & \colhead{Model Prediction} & \colhead{Observed (F200W)} & \colhead{Observed (F444W)}
}
\startdata
\multicolumn{5}{c}{\textbf{Radius: 88~AU}} \\
\hline
Minimum \gls{pa} & $\sim49^{\circ}$, $\sim175^{\circ}$ & $\sim110^{\circ}$, $\sim236^{\circ}$ &
$\dotdeg{13.9}$ & $\dotdeg{6.1}$ \\
Maximum \gls{pa} & $\sim112^{\circ}$, $\sim292^{\circ}$ & $\sim173^{\circ}$, $\sim353^{\circ}$ &
$\dotdeg{245.9}$ & $\dotdeg{219.6}$ \\
\hline
\multicolumn{5}{c}{\textbf{Radius: 141~AU}} \\
\hline
Minimum \gls{pa} & $\sim49^{\circ}$, $\sim175^{\circ}$ & $\sim110^{\circ}$, $\sim236^{\circ}$ &
$\dotdeg{359.6}$ & $\dotdeg{349.9}$ \\
Maximum \gls{pa} & $\sim112^{\circ}$, $\sim292^{\circ}$ & $\sim173^{\circ}$, $\sim353^{\circ}$ & $\dotdeg{242.6}$ & $\dotdeg{242.4}$ \\
\enddata
\tablecomments{Historical \gls{hst} locations and 2024 model predictions are based on the two-ring precession model from \citet{2023ApJ...948...36D}. The observed JWST locations represent the absolute minimum and maximum surface brightness extracted from the azimuthal profiles.}
\end{deluxetable*}

The observed positions of both minima and maxima in both filters do not align with the model's predictions. The quantitative discrepancy is now confirmed across two \gls{nircam} filters using the post-processed, pre-deconvolution data. Notably, the observed maxima at $219^{\circ}$--$246^{\circ}$ are located where one of the shadow local minima was predicted to be ($\sim$236$^{\circ}$), which is inconsistent with the model. The observation of this single, broad shadow in 2024---more reminiscent of the pre-2016 morphology previously characterized by \citet{2013ApJ...771...45D} and \citet{2023ApJ...948...36D} ---implies the ``two-shadow'' state of 2021 may have been a transient phase in a non-linear progression.

While the exact \glspl{pa} show minor variations between the F200W and F444W filters, the overall behavior is consistent. Both datasets (Figure \ref{fig:azimuthal_shadow}) reveal a systematic shift in the shadow's \gls{pa} with increasing radius. This behavior is consistent with historical trends, as \cite{2023ApJ...948...36D} noted that in their observations, the shadow's \gls{pa} at $r<50$~AU did not always align with that of the outer disk. The most direct explanation for such a radial twist in a shadow's projected position is a warp in the surface of the outer disk itself. Additionally, the color map (Figure \ref{fig:color_map}) does not show significantly large-scale azimuthal color variations that correlate with the shadow locations seen in either the historical \gls{hst}/\gls{stis} data or our new JWST images. This suggests that the shadowing mechanism primarily reduces the intensity of scattered light without altering its color, implying that the dust properties on the disk surface are relatively uniform, even within the shadowed regions. Other complex physical effects, such as azimuthally and radially dependent light scattering from a non-uniform dust distribution, could potentially mimic this signature \citep{2016ApJ...821...82O, 2017A&A...605A..34P}.

Disentangling the effects of scattering from the disk's physical structure is a task for future, detailed radiative transfer modeling. In conclusion, our analysis demonstrates that the shadow's behavior is more complex than predicted by steady precession. The combination of its non-linear temporal evolution and the observed radial twist in its \gls{pa} suggests that dynamic processes likely govern the system within a probable warped disk architecture. These features provide constraints for future models of TW Hya; future work will focus on detailed radiative transfer modeling to reproduce these observed structures and on combining these results with existing multi-wavelength data from \gls{alma}, \gls{hst}, and \gls{sphere}.

\subsection{Origin of the Outer Disk Bifurcation}

We have presented JWST/\gls{nircam} coronagraphic observations that resolve the fine-scale architecture of the TW Hya protoplanetary disk. The principal result of this imaging is the detailed characterization of the bifurcation feature in the outer disk. As visualized in Figure \ref{fig:bifurcation}, this structure manifests as a distinct splitting of the outer disk brightness into two components (peaking at projected separations of 108 and 128~AU). The feature is not azimuthally uniform; the splitting is prominent on one side of the disk but appears to merge on the opposite side. 

This complex morphology corresponds to the feature previously identified in \gls{sphere} scattered-light images by \citet{2017ApJ...837..132V}. We considered the temporal evolution of this feature by comparing our 2024 observation to the \gls{sphere} epoch from February 2015. For a stellar mass of $0.87\,M_{\odot}$, local Keplerian motion at $\sim$116~AU would result in a shift of only $2.4\degr$ over this 9~yr  baseline. A visual comparison might suggest a much larger apparent rotation ($\sim 30 \degr$, which would require co-rotation with an inner disk structure at $\sim$22~AU); however, the low signal-to-noise ratio at this structure barely allows for the identification of the bifurcation itself. Consequently, any attempt to ﻿ track its precise azimuthal phase across different instruments and wavelengths is highly susceptible to morphological artifacts and noise. We therefore caution that the current data quality does not allow us to robustly differentiate between local Keplerian motion, a co-rotating inner disk shadow, and observational artifacts. \citet{2017ApJ...837..132V} originally interpreted the feature as a ``dark spiral" (finding a $1.5\degr$ pitch angle), and our own logarithmic spiral fit yields a similarly shallow pitch angle of $2\degr$ assuming a face-on geometry. Nevertheless, we cannot definitively distinguish between a tightly wound spiral arm and an azimuthally asymmetric or eccentric double-ring structure based solely on these observations. The feature is most clearly resolved in our deconvolved F200W imagery; in the F444W data, the larger \gls{psf} blurs the feature, though the structure remains detectable in the radial profiles (Figure \ref{fig:radial_profile}).

Regardless of whether the feature is classified as a spiral or a split ring, its physical origin is likely dynamical. We first considered thermal condensation mechanisms, as dust pile-ups at snowlines can theoretically produce ring and gap morphologies \citep{2016ApJ...821...82O, 2018ApJ...867L..14V}. However, interferometric observations of CO and N$_2$H$^+$ emission in TW Hya have firmly constrained the CO snowline to a radius of $\sim$30~AU \citep{2013Sci...341..630Q}. The significant spatial discrepancy between this thermal front and the bifurcation at $\sim$120~AU suggests that volatile condensation is unlikely to be the primary driver.

With this thermal mechanism disfavored, a dynamical interaction with an embedded planetary companion emerges as the most compelling alternative. Hydrodynamical simulations indicate that a planet embedded in the disk can open a gap in the gas and dust, creating pressure maxima at the gap edges that trap particles and manifest as a double-ring structure \citep{2015ApJ...809...93D, 2017ApJ...843..127D}. Alternatively, such a planet can launch spiral density waves that would also account for the non-axisymmetric nature of the observed bifurcation feature \citep{2016MNRAS.459.2790R, 2017ApJ...837..132V}. If this bifurcated structure indeed represents a spiral density wave, its morphology depends heavily on the ratio of the planet's mass to the local disk thermal mass \citep{2002ApJ...569..997R}. Adopting a midplane temperature of $12.5$\,K at a radius of 115\,AU---consistent with recent thermal structure models for the TW Hya disk \citep{2021AJ....161...38O, 2021ApJ...908....8C}---we estimate a hydrostatic aspect ratio of $h/r \approx 0.081$\footnote{The hydrostatic aspect ratio is calculated as $h/r = c_s / v_K = \left[ \frac{k_B T r}{\mu m_H G M_{\star}} \right]^{1/2}$, where $c_s$ is the isothermal sound speed, $v_K$ is the Keplerian orbital velocity, $k_B$ is the Boltzmann constant, $m_H$ is the mass of hydrogen, and we assume a mean molecular weight of $\mu = 2.34$.}. For a stellar mass of 0.87 $M_{\odot}$, this yields a thermal mass of $M_{\mathrm{th}} = M_{\star}(h/r)^3 \approx 0.49$ \mjup. Linear spiral wave theory is particularly viable if the structure is excited by a perturber well below this thermal mass threshold. By synthesizing this structural evidence with the sensitivity limits derived in Section \ref{subsec:mass_limit}, we can place constraints on the nature of such a perturber. Our F444W observations yield a non-detection of point sources, ruling out non-embedded companions more massive than $\sim$0.2 \mjup at these separations. This indicates that any potential perturber driving this spiral must either be a sub-thermal-mass object or be significantly obscured by local disk extinction. While a full kinematic spiral-fitting analysis could theoretically constrain the perturber's exact mass and location, the current lack of specific predicted planet candidates beyond 100\,AU in the literature (see Table \ref{tab:planet_predictions}) leaves such parameter space highly unconstrained. We therefore defer a detailed modeling of the spiral's exact origin to future work.

\subsection{Constraints on Planet Mass}

Our non-detection of companions down to Jupiter-mass levels beyond $\sim$30~AU places constraints on the bodies responsible for sculpting the disk's complex architecture. The disk exhibits numerous substructures, including gaps, rings, a possible warp, and a bifurcated outer ring, which are often interpreted as signposts of planet-disk interactions. Our deep imaging limits demonstrate that massive, long-period, non-embedded companions are not the primary drivers of these features. This suggests that the observed morphology may be sculpted by planets---potentially sub-thermal-mass perturbers---that have so far eluded detection. These could be either lower-mass planets that fall below our current detection thresholds or more massive companions embedded within the disk and obscured by dust extinction. Follow-up observations with \glstarget{miri} JWST's \gls{miri} could push detection limits to even cooler, lower-mass companions, providing a more comprehensive census of planet formation in this benchmark system \citep{2025ApJ...987L..41C, 2025Natur.642..905L}.

\section{Conclusions}  \label{sec:conclusion}
In this paper, we presented high-contrast JWST/\gls{nircam} coronagraphic imaging of the TW Hya protoplanetary disk. Our analysis of the disk's geometry, photometry, substructures, and limits on planetary companions yields the following primary conclusions:

\begin{enumerate}
    \item \textbf{We find geometric evidence consistent with a disk warp.} While an elliptical fit to the full detected extent of the disk yields an average inclination of $i = \dotdeg{8.74}^{\dotdeg{+1.03}}_{\dotdeg{-0.94}}$ and position angle of $\mathrm{PA} = \dotdeg{75.62}^{\dotdeg{+7.86}}_{\dotdeg{-6.56}}$, spatially stratified modeling suggests radial variations. The inner-disk fit favors a higher inclination ($i = \dotdeg{10.62}^{\dotdeg{+2.45}}_{\dotdeg{-1.90}}$), while the outer-disk fit yields a $2\sigma$ upper limit of $i < \dotdeg{9.75}$. Although the statistical uncertainties allow for a uniform geometry, the trend is consistent with a warped architecture.
    
    \item \textbf{The disk surface is brighter than seen in historical data and remains dominated by small grains.} We measure a disk-to-star flux ratio in the F200W filter of $2.44\% \pm 0.14\%$, significantly higher than previous \gls{hst} measurements due to JWST's ability to recover more flux inside the classical coronagraphic \gls{iwa}. The scattered-light color is predominantly blue ($F200W-F444W<0$~mag) across most of the disk, indicating that the scattering surface is dominated by micron- to sub-micron-sized dust grains.
    
    \item \textbf{The disk shadow has evolved into a state that contradicts steady precession models.} Our observations characterize the variable shadowing on the outer disk, revealing a return to a single, broad shadow morphology similar to that observed prior to 2016. The position of the shadow contradicts the predictions of the two-ring linear precession model proposed to explain the 2021 epoch, suggesting the system is governed by complex, non-linear dynamics potentially linked to a disk warp.
    
    \item \textbf{High-resolution imaging resolves the outer disk bifurcation.} We spatially resolve the feature at $\sim120$~AU, confirming a distinct bifurcation whose morphology can be traced by a logarithmic spiral with a pitch angle of $\sim2^\circ$, although an asymmetric double-ring interpretation cannot be excluded. This morphology supports a dynamical origin, possibly a spiral density wave driven by interactions with unseen sub-thermal-mass planetary companions.

    \item \textbf{Deep imaging places stringent mass limits on potential companions.} These data provide the deepest constraints yet on point-source companions within the TW Hya disk. For the conservative 10~Myr case, our F444W observations rule out the presence of non-embedded companions with masses greater than $\sim0.4$~\mjup beyond 60~AU ($1\arcsec$) and $\sim2.0$~\mjup beyond 30~AU ($0.5\arcsec$). These limits suggest that the planets responsible for the observed gaps and spiral features are either sub-Jupiter-mass objects or are deeply embedded and obscured by circumstellar dust.
\end{enumerate}

\begin{acknowledgments}
This work is based on observations made with the NASA/ESA/CSA James Webb Space Telescope.
The authors are grateful for support from NASA for the \textit{JWST}/NIRCam project under contract number NAS5-02105 (M. Rieke, University of Arizona, PI).
The data were obtained from the \gls{mast} at the Space Telescope Science Institute, which is operated by the Association of Universities for Research in Astronomy, Inc., under NASA contract NAS 5-03127 for JWST. The data products used in this analysis are available under the \href{https://archive.stsci.edu/doi/resolve/resolve.html?doi=10.17909/dymv-st72}{DOI:10.17909/dymv-st72}. 
The work of GHR was supported in part by grant 80NSSC18K0555 from NASA Goddard Space Flight Center to the University of Arizona. 
D.J.\ is supported by NRC Canada and by an NSERC Discovery Grant. 
This work has been carried out within the framework of the NCCR PlanetS supported by the Swiss National Science Foundation under grant 51NF40\_205606.
\end{acknowledgments}

\facilities{JWST(NIRCam)}

\software{\texttt{Astropy} \citep{2013A&A...558A..33A, 2018AJ....156..123A, 2022ApJ...935..167A}, \texttt{emcee} \citep{2013PASP..125..306F}, JWST Science Calibration Pipeline \citep{bushouse_2022_7229890}, \texttt{spaceKLIP}\citep{2025ascl.soft02014C}, \texttt{STPSF} \citep{2025zndo..15747364P}, \texttt{winnie} \citep{2022ApJ...935L..25L, 2023AJ....166..150L}}
\section*{Data Availability}
The data reduction pipelines, analysis codes, and supplemental figures used in this work are available on Zenodo under an open-source Creative Commons Attribution license: \dataset[doi:10.5281/zenodo.21294173]{https://doi.org/10.5281/zenodo.21294173}.

\appendix
\section{Images in F187N and F356W Filters}
\label{appendix:other_filters}
\begin{figure*}[ht!]
\centering
\includegraphics[width=0.85\textwidth]{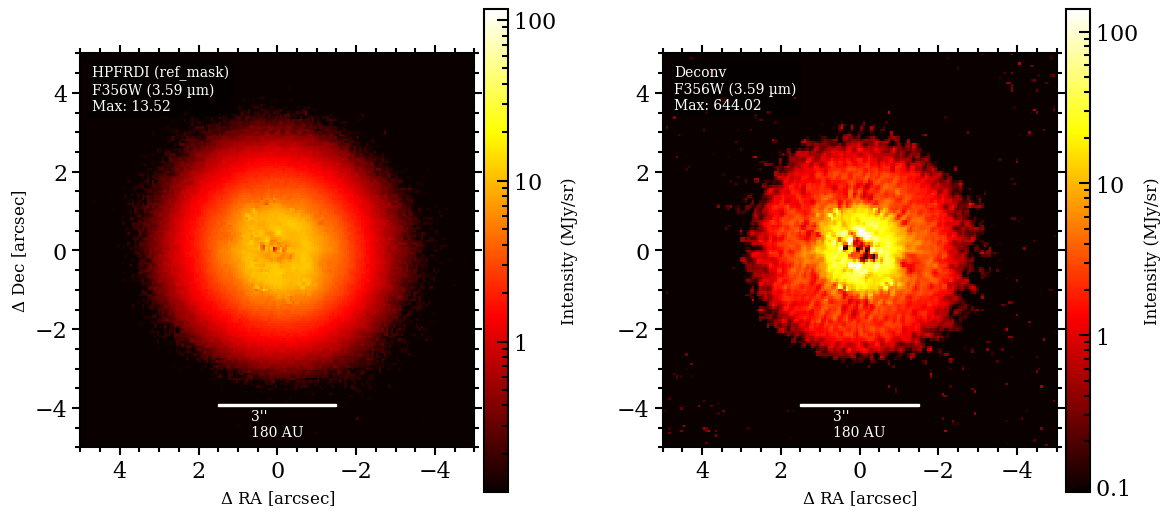}
\caption{Final processed images of the TW Hya protoplanetary disk in the F356W filter. Similar to Figure \ref{fig:disks}, the right panel shows the deconvolved image produced with \texttt{winnie}, while the left panel shows the \gls{hpfrdi}-processed image. 
\label{fig:356}}
\end{figure*}

In addition to the F200W and F444W filters presented in the main text, observations were also carried out in the F187N and F356W filters. The data were processed using the same pipeline described in Section \ref{sec:obs}. However, as discussed in Section \ref{subsec:psfsub_deconv}, the F187N data could not be processed via \gls{hpfrdi} due to the lack of a valid reference star. The final image for F356W is shown in Figure \ref{fig:356}. As discussed in Section \ref{subsec:psfsub_deconv}, the small reference library for F356W led to a less optimal \gls{psf} subtraction, resulting in a final image with more residual noise and artifacts compared to the primary F200W and F444W datasets.

\section{Deconvolution Quality Assessment}
\label{app:deconv_quality}

To validate the fidelity of the Richardson-Lucy deconvolution presented in Section \ref{subsec:psfsub_deconv}, we performed a forward-modeling consistency check. We utilized the \texttt{winnie} package to convolve the final deconvolved source distribution with the instrumental \glspl{psf}. This generates a ``Reconvolved'' model that can be directly compared to the input \gls{hpfrdi} data.

\begin{figure*}[ht!]
\centering
\gridline{\fig{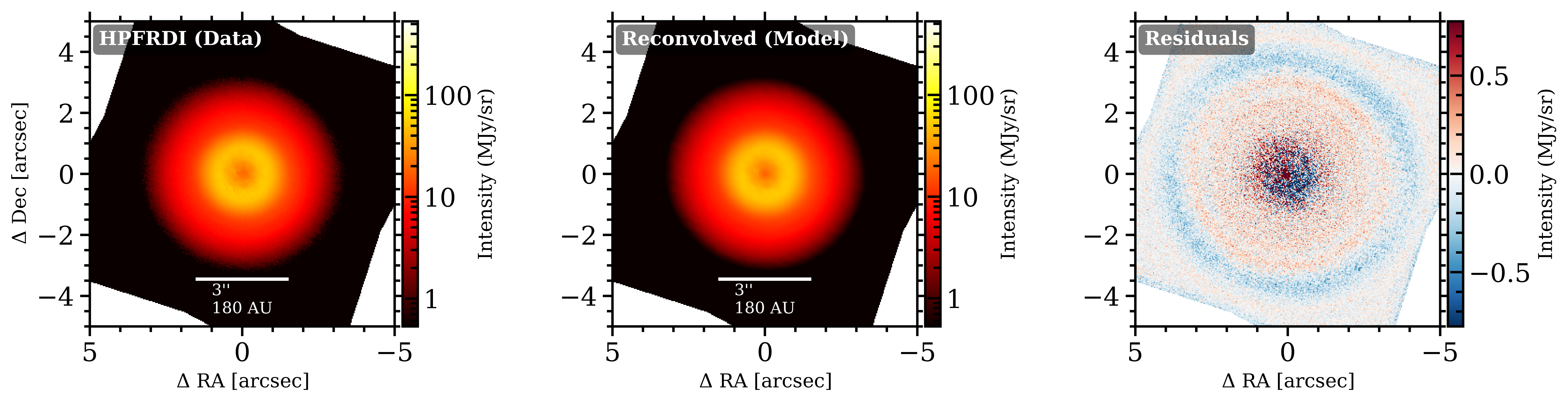}{1.0\textwidth}{(a) F200W}}
\gridline{\fig{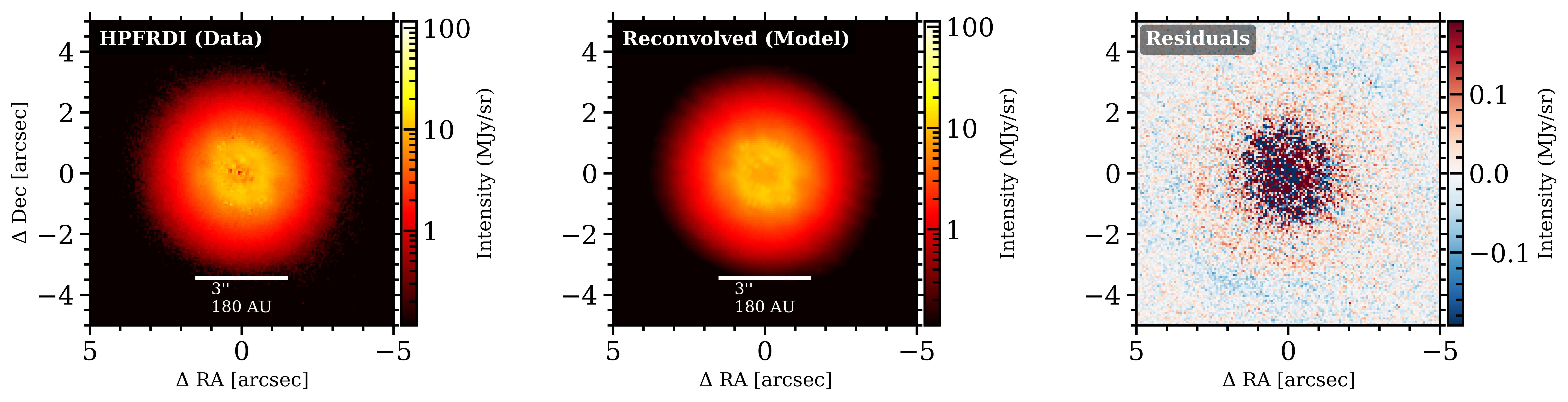}{1.0\textwidth}{(b) F356W}}
\gridline{\fig{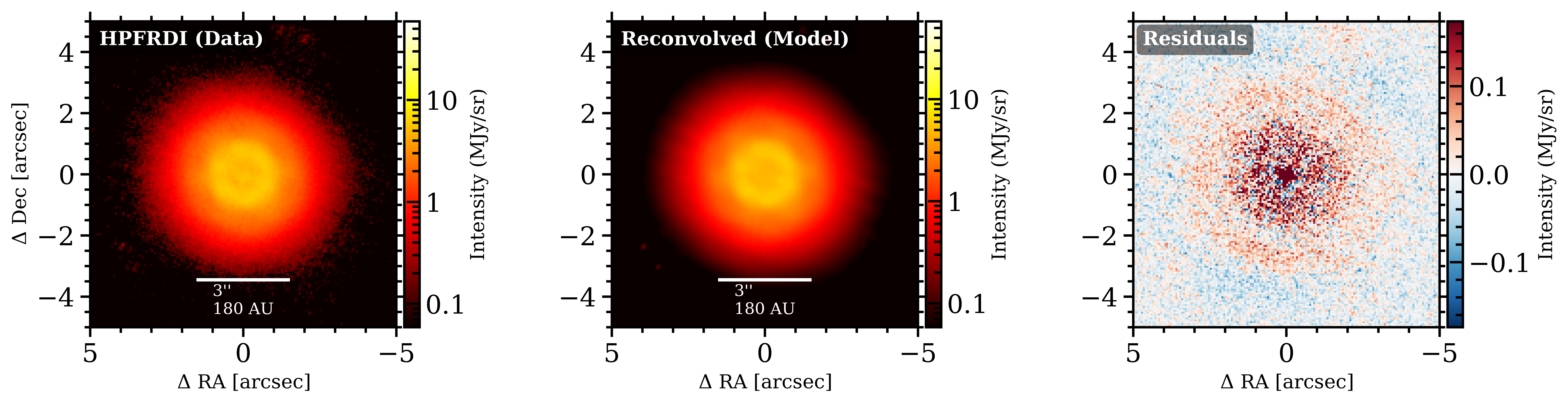}{1.0\textwidth}{(c) F444W}}
\caption{Quality assessment for the deconvolution process. Each row presents the results for a specific filter: (a) F200W, (b) F356W, and (c) F444W. In each row, the left panel shows the input \gls{hpfrdi} data; the middle panel shows the deconvolved source distribution forward-modeled (reconvolved) with the instrumental \glspl{psf}; and the right panel shows the residuals (Data $-$ Model). The residuals show a pattern that correlates with the disk's surface brightness, as expected for intensity-dependent noise and deconvolution residuals, but reveal no significant unmodeled astrophysical structures.
\label{fig:qa_all}}
\end{figure*}

The results of this comparison for the F200W, F356W, and F444W filters are presented in Figure \ref{fig:qa_all}. The residual maps (Data minus Model) demonstrate the consistency of the deconvolution. The residual flux levels are negligible compared to the source signal, indicating that the model accounts for the vast majority of the disk flux. The residuals do not exhibit spatially coherent astrophysical structures, such as distinct spiral arms or point sources, that were missed by the model. Instead, they display a concentric annular pattern that spatially correlates with the disk's brightness profile. This behavior is expected, as both statistical noise and potential deconvolution artifacts scale with source intensity. The absence of systematic, large-scale deviations confirms that the algorithm has successfully recovered the overall disk geometry. Minor structured residuals persist within the center region, which could be attributed to the amplification of noise in the region where the coronagraphic mask transmission is lowest.

To further quantify the structural limitations, optimal stopping criterion, and algorithmic uncertainties of our deconvolution approach, we implemented a forward-modeled synthetic injection and recovery framework. We constructed a purely morphological synthetic model consisting of three concentric annuli, including a sharp bifurcation feature in the outermost ring. Assuming a face-on orientation, the optimal parameters for these structures---including their specific radii and relative fluxes---were determined by convolving the model with the \gls{nircam} \glspl{psf}, passing it through our \gls{hpfrdi} forward-modeling process, and minimizing the residuals against the actual \gls{hpfrdi} observations. To accurately evaluate how the deconvolution algorithm handles noise amplification, we injected synthetic noise matching the statistical properties of the real data into these forward-modeled images. We note that this model was designed solely to stress-test our algorithmic ability to recover spatial features and flux distributions under realistic observational conditions, rather than to serve as a rigorous physical exploration of the disk's intrinsic properties. Once established, this optimized and noisy synthetic model was subsequently deconvolved using our pipeline. These tests revealed that optimizing the algorithm to recover the accurate radial locations of the gaps and the bifurcation structure required higher iteration counts (200 for F200W/F356W and 129 for F444W). While a lower number of iterations ($\sim$30) minimized absolute pixel-wise grain noise, it failed to fully resolve the structural depths. Crucially, the recovery tests demonstrated that the radial locations of the gaps and the bifurcation structure do not depend on the specific number of iterations and are accurately recovered with no systematic spatial offset introduced by the deconvolution process. We also utilized the residual flux from these tests to incorporate a systematic deconvolution uncertainty into our total disk flux measurements.

\section{Stellar Spectrum and SED Excess Analysis}
\label{app:phot_details}

To accurately identify and quantify the \gls{sed} excess in the near-infrared bands, we first established a reference \gls{sed} for the host star, TW Hya. As a T Tauri star, TW Hya exhibits significant accretion luminosity with equivalent blackbody temperatures of 10,000\,K or higher that generate accretion excess emission above the photospheric continuum at optical and UV wavelengths \citep{1998ApJ...509..802C}. In addition, the \gls{sed} of such an object typically shows near- and mid-IR photometric excesses resulting from circumstellar material re-radiating the incident UV and optical emission \citep{1987ApJ...323..714K}. 

To construct the photospheric reference \gls{sed}, we selected archival
photometry within the near-infrared fitting interval
0.7 $< \lambda <$ 1.8~\mum{}. The available measurements within this
interval span approximately 0.76--1.65~\mum{}. This selection avoids the UV and optical regimes, which are heavily affected by variable accretion luminosity and flaring events \citep{2018MNRAS.478..758S}, as well as the mid- to far-infrared wavelengths dominated by thermal emission from the circumstellar disk. We compiled data points for TW Hya from the VizieR Photometry viewer service \citep{vizier}. During this compilation, we excluded any photometric measurements lacking reported uncertainties. We also excluded two anomalously \glstarget{SMSS} low $z$-band \gls{SMSS} measurements from both the composite-model fit and the LOWESS smoothing of the observed \gls{sed}. These measurements are likely affected by detector saturation for a target as bright as TW Hya, as documented in the survey’s release notes \citep{2018PASA...35...10W}.

\begin{figure*}[ht]
\centering
\includegraphics[width=1.0\textwidth]{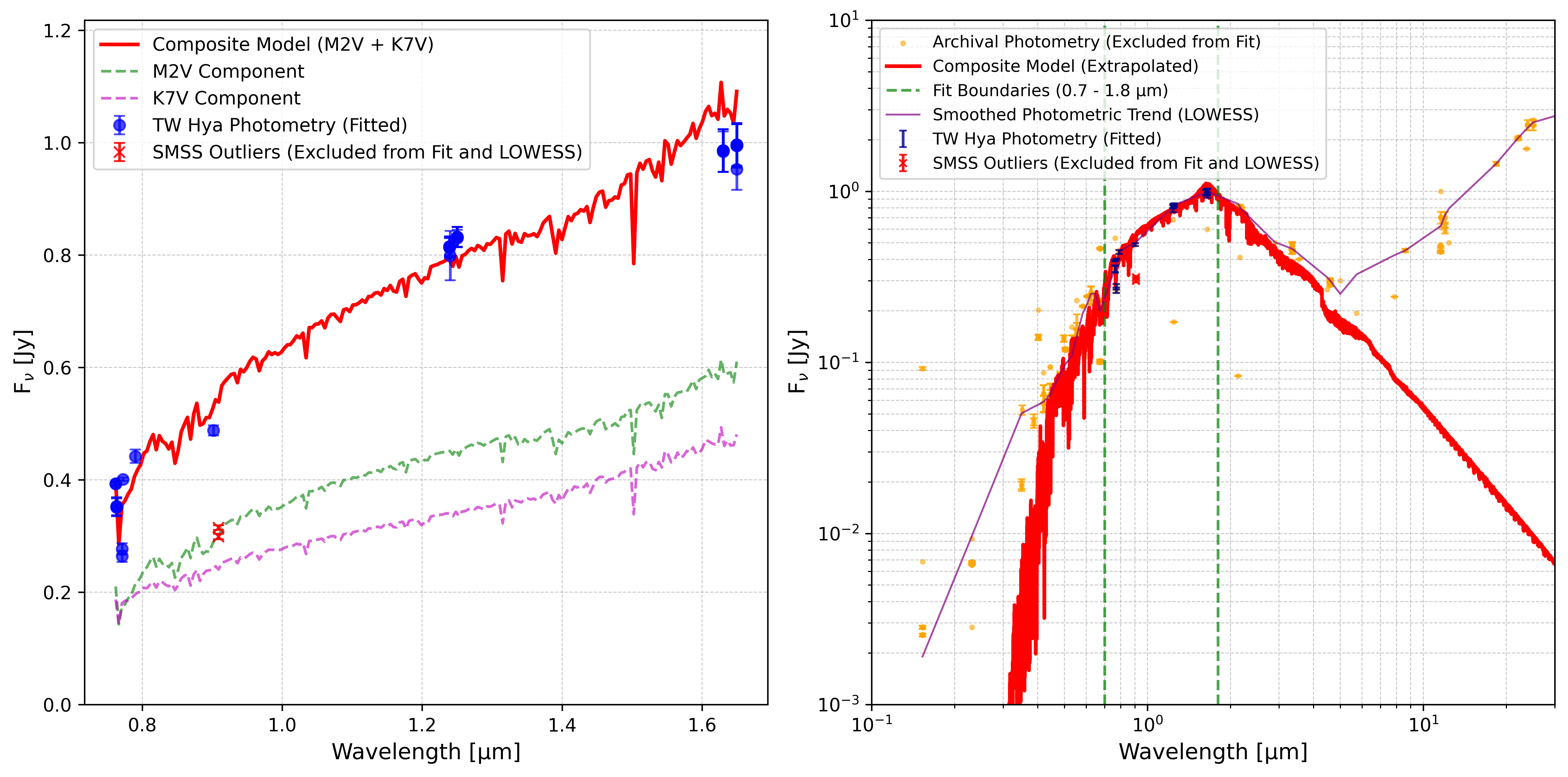}
\caption{Determination of the TW Hya stellar reference spectrum. 
\textbf{Left:} A two-component composite spectrum (red solid line) fitted to the selected near-infrared photometry of TW Hya (blue points). The individual components of the model---an M2V component (green dashed line) and a K7V component (magenta dashed line)---are shown separately. Two anomalously low \gls{SMSS} measurements are marked with red crosses and were excluded from both the composite-model fit and the LOWESS smoothing. 
\textbf{Right:} The full observed \gls{sed} of TW Hya. Archival photometry not used in the stellar fit is shown as orange points, while the fitted photometry and excluded \gls{SMSS} measurements are shown as blue points and red crosses, respectively. The vertical green dashed lines indicate the 0.7--1.8~\mum{} fitting interval. The best-fit composite spectrum is extrapolated over the full wavelength range (red solid line), and the purple line shows the LOWESS-smoothed trend derived from the archival photometry after excluding the two \gls{SMSS} outliers. The ultraviolet and optical excess is associated primarily with accretion, whereas the infrared excess arises from thermal emission and scattered light from circumstellar material.}
\label{fig:sed_fit}
\end{figure*}

To account for the complex, spotted nature of the TW Hya photosphere—which \citet{2013ApJ...771...45D} demonstrated is better characterized by a combination of a cooler underlying star with hotter accretion spots, or conversely, a hotter star with cool magnetic spots—we fit a two-component composite spectral model to the filtered near-infrared data. We utilized the ``BOSZ" Synthetic Stellar Spectral Library\footnote{\url{https://archive.stsci.edu/hlsp/bosz}} \citep{2017AJ....153..234B, 2024A&A...688A.197M} to generate the stellar templates. Specifically, we generated an M2V template ($T_{\text{eff}} = 3600$\,K, $\log g = 4.2$, $[\text{M/H}] = 0.0$) and a K7V template ($T_{\text{eff}} = 3990$\,K, $\log g = 4.2$, $[\text{M/H}] = 0.0$). These templates were combined and scaled to the TW Hya data using a weighted least-squares optimization routine implemented with the \texttt{scipy} Python library\footnote{\texttt{scipy.optimize.curve\_fit}}. The resulting best-fit scaled composite model (Figure \ref{fig:sed_fit}) serves as our reference spectrum for the central star.

To quantify the total photometric infrared excess in the JWST bands, we performed synthetic photometry on both the observed \gls{sed} and the modeled stellar spectrum. We utilized the \texttt{synphot} and \texttt{webbpsf\_ext} packages to integrate the spectra over the specific system throughput curves for the F200W, F356W, and F444W filters. The effective flux densities were calculated using the \texttt{effstim} method in \texttt{synphot}. Uncertainties were determined by propagating multiple sources of error in quadrature. For the observed TW Hya flux, we calculated the \gls{sem} of the archival photometric data points falling within the rectangular width of each filter's bandpass. For the stellar model flux, the uncertainty was derived from the statistical error of the scaling factor determined during the \gls{sed} fitting process.

The resulting photometry is presented in Table \ref{tab:sed_photometry}. A positive total photometric excess (Archive minus Model) is detected in all three filters. This excess is primarily due to thermal emission from the inner disk. This is shown quantitatively by \citet{2006ApJ...637L.133E}, who report $K$-band interferometry with \glstarget{ki} the \gls{ki}. They find an extended flux within the \gls{ki} primary beam (diameter 45~mas) consistent with the $K$-band spectral veiling of $\sim$ 7\% of the stellar flux. This agrees well with the values for F200W in Table~\ref{tab:sed_photometry}. The excesses in the three filters are consistent within their uncertainties with a single-temperature blackbody of approximately 1200~K. This temperature is consistent with the expected sublimation temperature of silicate grains, suggesting that the thermal emission arises from dust transported inward to 
near the sublimation radius.

\begin{deluxetable}{lcccc}[ht!]
\tablecaption{Total System Photometry and Total Photometric Excess \label{tab:sed_photometry}}
\tablehead{
\colhead{Filter} & \colhead{TW Hya (Archive)} & \colhead{Stellar (Model)} & \colhead{Total Photometric Excess} & \colhead{Significance} \\
\colhead{} & \colhead{(mJy)} & \colhead{(mJy)} & \colhead{(mJy)} & \colhead{($\sigma$)}
}
\startdata
F200W & $894 \pm 25$ & $823 \pm 8$ & $71 \pm 26$ & 2.8 \\
F356W & $434 \pm 9$ & $351 \pm 6$ & $83 \pm 11$ & 8.1 \\
F444W & $337 \pm 3$ & $236 \pm 5$ & $100 \pm 5$ & 20.2 \\
\enddata
\tablecomments{Photometry derived by integrating the LOWESS-smoothed observed \gls{sed}, after excluding the two anomalous \gls{SMSS} measurements, and the two-component stellar model over the \gls{nircam} bandpasses. The calculated residual in each filter represents the total photometric excess from the circumstellar environment (inner thermal emission + outer scattered light). The displayed values are rounded independently; the total photometric excesses and detection significances were calculated using the unrounded fluxes and uncertainties.}
\end{deluxetable}

Our coronagraphic images isolate extended structures by suppressing the central point source. Because the thermal emission from the inner disk is spatially unresolved from the star, the \gls{psf} subtraction process does not distinguish between the two, removing this inner flux along with the stellar photosphere. Consequently, the disk fluxes reported in Section \ref{subsec:disk_photometry} represent only the extended, scattered-light component of the disk, which is significantly fainter than the emission by the hot dust near the star.

\section{Disk Flux Distributions}
\label{appendix:bootstrap_results}

We utilized a spatially stratified bootstrap resampling technique to determine the total integrated disk flux and its uncertainty. To account for the intrinsic morphology of the disk—specifically its nearly face-on orientation (see Section \ref{subsec:disk_geometry}) and concentric ring structure (see Section \ref{subsec:disk_features})—pixels across the entire image array were first grouped into concentric annuli based on their integer pixel radius. This approach assumes that pixels at the same radial distance sample a similar underlying flux distribution.

For each bootstrap iteration, we independently resampled the pixel intensities within each annulus with replacement, preserving the total number of pixels per annulus. The total disk flux for a given iteration was then calculated by summing these resampled values across all annuli. To ensure high statistical precision, we generated 500,000 bootstrap samples. This computationally intensive procedure was accelerated using a custom routine implemented in the Rust programming language, utilizing the \texttt{rayon} library for parallel processing.

The resulting probability density functions (histograms) for the total disk flux derived from the \gls{hpfrdi} images, the throughput-corrected data, and the final deconvolved images for all filters (F200W, F356W, and F444W) are available on Zenodo under an open-source Creative Commons Attribution license: \dataset[doi:10.5281/zenodo.21294173]{https://doi.org/10.5281/zenodo.21294173}. The corresponding filenames in the repository are:

\begin{itemize}
    \item \textbf{\gls{hpfrdi} (Raw):} \texttt{HPFRDI\_flux\_[FILTER].png}
    \item \textbf{Throughput Corrected:} \texttt{HPFRDI\_throughput\_flux\_[FILTER].png}
    \item \textbf{Deconvolved:} \texttt{Deconv\_flux\_[FILTER].png}
\end{itemize}

The mean values and $1\sigma$ uncertainties derived from these distributions are reported in Table \ref{tab:photometry}.

\section{Effective Beam Size}
\label{app:beam_size}

The log-likelihood function used in our \gls{mcmc} analysis (Section \ref{subsec:disk_geometry}) requires an estimate of the effective number of independent samples within each elliptical annulus. The instrumental \gls{psf} creates correlations between adjacent pixels, which fundamentally persist through the deconvolution process. Simply using the total number of pixels in an annulus, $N_k$, would therefore overestimate the statistical weight of that annulus. To mitigate this, we correct for the number of pixels within one resolution element, which we term the effective beam size, $N_{\text{beam}}$.

We calculated $N_{\text{beam}}$ by simulating the \gls{psf} for our specific observational setup using the \texttt{webbpsf\_ext} package. The simulation was configured for an \gls{nircam} observation with the F200W filter, the MASK335R coronagraphic image mask, and the CIRCLYOT pupil mask. To ensure the simulated \gls{psf} was representative of the location of the circumstellar disk, the source was offset by 1.5\arcsec\ at a \gls{pa} of 0\degr, corresponding to a shift along the instrument's +Y axis.

The \gls{psf} was calculated on a grid with an oversampling factor of 10 relative to the native detector pixel scale. The effective beam size in detector pixels can be derived from the oversampled \gls{psf} array by calculating the ratio of the integrated flux to the peak flux value and correcting for the oversampling factor. This is conceptually equivalent to calculating the area of a uniform beam that has the same total flux and peak brightness as the simulated \gls{psf}. The formula is given by:

$$ N_{\text{beam}} = \frac{\sum I_{ij}}{\max(I_{ij})} \cdot \frac{1}{(\text{oversample factor})^2} $$

where the sum and maximum are taken over all pixels $ij$ of the oversampled \gls{psf} array. For our simulation with an oversampling factor of 10, this calculation yielded an effective beam size of $N_{\text{beam}} \simeq 46.5$~pixels. This value was then used to determine the effective number of independent pixels, $N^{\text{eff}}_k = N_k / N_{\text{beam}}$, for each annulus in our log-likelihood function.

\section{Derivation of the Log-Likelihood Function}
\label{app:likelihood}

The log-likelihood function used in our \gls{mcmc} analysis is derived from the principle of maximum likelihood, assuming that the underlying pixel noise is Gaussian. This appendix outlines the derivation, including the correction for correlated noise between pixels.

We begin by assuming that the intensity values, $x_i$, of the $N_k$ pixels within a given elliptical annulus $k$ are independent draws from a Gaussian distribution. The probability density for a single pixel is:
\begin{equation}
    P(x_i | M_k, \sigma_k^2) = \frac{1}{\sqrt{2\pi\sigma_k^2}} \exp\left(-\frac{(x_i - M_k)^2}{2\sigma_k^2}\right)
\end{equation}
where $M_k$ is the mean intensity and $\sigma_k^2$ is the expected variance from our empirical model (Figure \ref{fig:geo_method}). For $N_k$ independent pixels, the total log-likelihood for the annulus, $\ln(\mathcal{L}_k)$, is the sum of the individual log-probabilities. After dropping constant terms that do not affect the maximization, this is:
\begin{equation}
    \ln(\mathcal{L}_k) = -\frac{1}{2} \left( N_k\ln(\sigma_k^2) + \frac{\mathrm{SSR}_k}{\sigma_k^2} \right)
\end{equation}
where $\mathrm{SSR}_k = \sum_{i=1}^{N_k} (x_i - M_k)^2$ is the sum of squared residuals. We can rewrite this by factoring out $N_k$ and substituting the variance of the pixel intensities, $\mathrm{var}_k = \mathrm{SSR}_k/N_k$:
\begin{equation}
    \ln(\mathcal{L}_k) = -\frac{N_k}{2} \left( \ln(\sigma_k^2) + \frac{\mathrm{var}_k}{\sigma_k^2} \right)
\end{equation}

The assumption of pixel independence is invalidated by the instrumental \gls{psf}, which creates correlations between adjacent pixels. To account for this, we replace the total number of pixels, $N_k$, which acts as the statistical weight of the annulus, with an \textit{effective number of independent samples}, $N^{\text{eff}}_k$. This is defined as:
\begin{equation}
    N^{\text{eff}}_k = \frac{N_k}{N_{\text{beam}}}
\end{equation}
where $N_{\text{beam}}$ is the effective beam size in pixels, representing the area of correlated data (derived in Appendix \ref{app:beam_size}). This correction ensures that annuli with many correlated pixels are not overweighted in the fit.

The total log-likelihood for the model is then the sum of these corrected log-likelihood contributions from each annulus. By substituting the leading statistical weight term ($N_k$) with $N^{\text{eff}}_k$, we arrive at the final expression:
\begin{equation}
    \ln(\mathcal{L}) = \sum_k \ln(\mathcal{L}_k)_{\text{corrected}} = -\frac{1}{2} \sum_k N^{\text{eff}}_k \left( \frac{\mathrm{var}_k}{\sigma_k^2} + \ln(\sigma_k^2) \right)
\end{equation}
This is the final expression (equivalent to Equation \ref{eq:logL_final}) that our \gls{mcmc} algorithm maximizes to find the best-fit geometric parameters for the disk.

\section{Detection Limits for the F187N, F200W, and F356W Filters}
\label{app:add_limits}

While the F444W filter provides the deepest sensitivity to Jovian-mass companions in the thermal infrared, the other filters provide complementary constraints. The F187N filter is sensitive to Pa-$\alpha$ emission, a tracer of active accretion. The F200W data offer the highest spatial resolution ($\sim$0.031\arcsec/pixel), allowing us to probe the smallest \gls{iwa}. The F356W limits provide an intermediate baseline between the short- and long-wavelength channels.

We processed these datasets using the same \gls{adi} and \gls{klip} reduction pipeline described in Section \ref{subsec:mass_limit}. No statistically significant point sources were resolved in any of the final processed images. 

The final \gls{adi}-processed images, calibrated $5\sigma$ contrast curves, and derived mass sensitivity limits for these three filters are available on Zenodo under an open-source Creative Commons Attribution license: \dataset[doi:10.5281/zenodo.21294173]{https://doi.org/10.5281/zenodo.21294173}. The corresponding figures are:

\begin{itemize}
    \item \textbf{F187N:} \texttt{mass\_limit\_f187n.png}
    \item \textbf{F200W:} \texttt{mass\_limit\_f200w.png}
    \item \textbf{F356W:} \texttt{mass\_limit\_f356w.png}
\end{itemize}

\clearpage % Ensures the bibliography starts on a new page
\bibliography{main}{}

@ARTICLE{2020MNRAS.492.3440S,
       author = {{Sanchis}, E. and {Picogna}, G. and {Ercolano}, B. and {Testi}, L. and {Rosotti}, G.},
        title = "{Detectability of embedded protoplanets from hydrodynamical simulations}",
      journal = {\mnras},
         year = 2020,
        month = mar,
       volume = {492},
       number = {3},
        pages = {3440-3458},
          doi = {10.1093/mnras/staa074},
archivePrefix = {arXiv},
       eprint = {2001.03565},
 primaryClass = {astro-ph.EP},
       adsurl = {https://ui.adsabs.harvard.edu/abs/2020MNRAS.492.3440S}
}

@ARTICLE{2017ApJ...837..132V,
       author = {{van Boekel}, R. and {Henning}, Th. and {Menu}, J. and {de Boer}, J. and {Langlois}, M. and {M{\"u}ller}, A. and {Avenhaus}, H. and {Boccaletti}, A. and {Schmid}, H.~M. and {Thalmann}, Ch. and {Benisty}, M. and {Dominik}, C. and {Ginski}, Ch. and {Girard}, J.~H. and {Gisler}, D. and {Lobo Gomes}, A. and {Menard}, F. and {Min}, M. and {Pavlov}, A. and {Pohl}, A. and {Quanz}, S.~P. and {Rabou}, P. and {Roelfsema}, R. and {Sauvage}, J. -F. and {Teague}, R. and {Wildi}, F. and {Zurlo}, A.},
        title = "{Three Radial Gaps in the Disk of TW Hydrae Imaged with SPHERE}",
      journal = {\apj},
         year = 2017,
        month = mar,
       volume = {837},
       number = {2},
          eid = {132},
        pages = {132},
          doi = {10.3847/1538-4357/aa5d68},
archivePrefix = {arXiv},
       eprint = {1610.08939},
 primaryClass = {astro-ph.EP},
       adsurl = {https://ui.adsabs.harvard.edu/abs/2017ApJ...837..132V}
}

@ARTICLE{2016ApJ...820L..40A,
       author = {{Andrews}, Sean M. and {Wilner}, David J. and {Zhu}, Zhaohuan and {Birnstiel}, Tilman and {Carpenter}, John M. and {P{\'e}rez}, Laura M. and {Bai}, Xue-Ning and {{\"O}berg}, Karin I. and {Hughes}, A. Meredith and {Isella}, Andrea and {Ricci}, Luca},
        title = "{Ringed Substructure and a Gap at 1 au in the Nearest Protoplanetary Disk}",
      journal = {\apjl},
         year = 2016,
        month = apr,
       volume = {820},
       number = {2},
          eid = {L40},
        pages = {L40},
          doi = {10.3847/2041-8205/820/2/L40},
archivePrefix = {arXiv},
       eprint = {1603.09352},
 primaryClass = {astro-ph.EP},
       adsurl = {https://ui.adsabs.harvard.edu/abs/2016ApJ...820L..40A}
}

@ARTICLE{2013A&A...558A..33A,
       author = {{Astropy Collaboration} and {Robitaille}, Thomas P. and {Tollerud}, Erik J. and {Greenfield}, Perry and {Droettboom}, Michael and {Bray}, Erik and {Aldcroft}, Tom and {Davis}, Matt and {Ginsburg}, Adam and {Price-Whelan}, Adrian M. and {Kerzendorf}, Wolfgang E. and {Conley}, Alexander and {Crighton}, Neil and {Barbary}, Kyle and {Muna}, Demitri and {Ferguson}, Henry and {Grollier}, Fr{\'e}d{\'e}ric and {Parikh}, Madhura M. and {Nair}, Prasanth H. and {Unther}, Hans M. and {Deil}, Christoph and {Woillez}, Julien and {Conseil}, Simon and {Kramer}, Roban and {Turner}, James E.~H. and {Singer}, Leo and {Fox}, Ryan and {Weaver}, Benjamin A. and {Zabalza}, Victor and {Edwards}, Zachary I. and {Azalee Bostroem}, K. and {Burke}, D.~J. and {Casey}, Andrew R. and {Crawford}, Steven M. and {Dencheva}, Nadia and {Ely}, Justin and {Jenness}, Tim and {Labrie}, Kathleen and {Lim}, Pey Lian and {Pierfederici}, Francesco and {Pontzen}, Andrew and {Ptak}, Andy and {Refsdal}, Brian and {Servillat}, Mathieu and {Streicher}, Ole},
        title = "{Astropy: A community Python package for astronomy}",
      journal = {\aap},
         year = 2013,
        month = oct,
       volume = {558},
          eid = {A33},
        pages = {A33},
          doi = {10.1051/0004-6361/201322068},
archivePrefix = {arXiv},
       eprint = {1307.6212},
 primaryClass = {astro-ph.IM},
       adsurl = {https://ui.adsabs.harvard.edu/abs/2013A&A...558A..33A}
}

@ARTICLE{2018AJ....156..123A,
       author = {{Astropy Collaboration} and {Price-Whelan}, A.~M. and {Sip{\H{o}}cz}, B.~M. and {G{\"u}nther}, H.~M. and {Lim}, P.~L. and {Crawford}, S.~M. and {Conseil}, S. and {Shupe}, D.~L. and {Craig}, M.~W. and {Dencheva}, N. and {Ginsburg}, A. and {VanderPlas}, J.~T. and {Bradley}, L.~D. and {P{\'e}rez-Su{\'a}rez}, D. and {de Val-Borro}, M. and {Aldcroft}, T.~L. and {Cruz}, K.~L. and {Robitaille}, T.~P. and {Tollerud}, E.~J. and {Ardelean}, C. and {Babej}, T. and {Bach}, Y.~P. and {Bachetti}, M. and {Bakanov}, A.~V. and {Bamford}, S.~P. and {Barentsen}, G. and {Barmby}, P. and {Baumbach}, A. and {Berry}, K.~L. and {Biscani}, F. and {Boquien}, M. and {Bostroem}, K.~A. and {Bouma}, L.~G. and {Brammer}, G.~B. and {Bray}, E.~M. and {Breytenbach}, H. and {Buddelmeijer}, H. and {Burke}, D.~J. and {Calderone}, G. and {Cano Rodr{\'\i}guez}, J.~L. and {Cara}, M. and {Cardoso}, J.~V.~M. and {Cheedella}, S. and {Copin}, Y. and {Corrales}, L. and {Crichton}, D. and {D'Avella}, D. and {Deil}, C. and {Depagne}, {\'E}. and {Dietrich}, J.~P. and {Donath}, A. and {Droettboom}, M. and {Earl}, N. and {Erben}, T. and {Fabbro}, S. and {Ferreira}, L.~A. and {Finethy}, T. and {Fox}, R.~T. and {Garrison}, L.~H. and {Gibbons}, S.~L.~J. and {Goldstein}, D.~A. and {Gommers}, R. and {Greco}, J.~P. and {Greenfield}, P. and {Groener}, A.~M. and {Grollier}, F. and {Hagen}, A. and {Hirst}, P. and {Homeier}, D. and {Horton}, A.~J. and {Hosseinzadeh}, G. and {Hu}, L. and {Hunkeler}, J.~S. and {Ivezi{\'c}}, {\v{Z}}. and {Jain}, A. and {Jenness}, T. and {Kanarek}, G. and {Kendrew}, S. and {Kern}, N.~S. and {Kerzendorf}, W.~E. and {Khvalko}, A. and {King}, J. and {Kirkby}, D. and {Kulkarni}, A.~M. and {Kumar}, A. and {Lee}, A. and {Lenz}, D. and {Littlefair}, S.~P. and {Ma}, Z. and {Macleod}, D.~M. and {Mastropietro}, M. and {McCully}, C. and {Montagnac}, S. and {Morris}, B.~M. and {Mueller}, M. and {Mumford}, S.~J. and {Muna}, D. and {Murphy}, N.~A. and {Nelson}, S. and {Nguyen}, G.~H. and {Ninan}, J.~P. and {N{\"o}the}, M. and {Ogaz}, S. and {Oh}, S. and {Parejko}, J.~K. and {Parley}, N. and {Pascual}, S. and {Patil}, R. and {Patil}, A.~A. and {Plunkett}, A.~L. and {Prochaska}, J.~X. and {Rastogi}, T. and {Reddy Janga}, V. and {Sabater}, J. and {Sakurikar}, P. and {Seifert}, M. and {Sherbert}, L.~E. and {Sherwood-Taylor}, H. and {Shih}, A.~Y. and {Sick}, J. and {Silbiger}, M.~T. and {Singanamalla}, S. and {Singer}, L.~P. and {Sladen}, P.~H. and {Sooley}, K.~A. and {Sornarajah}, S. and {Streicher}, O. and {Teuben}, P. and {Thomas}, S.~W. and {Tremblay}, G.~R. and {Turner}, J.~E.~H. and {Terr{\'o}n}, V. and {van Kerkwijk}, M.~H. and {de la Vega}, A. and {Watkins}, L.~L. and {Weaver}, B.~A. and {Whitmore}, J.~B. and {Woillez}, J. and {Zabalza}, V. and {Astropy Contributors}},
        title = "{The Astropy Project: Building an Open-science Project and Status of the v2.0 Core Package}",
      journal = {\aj},
         year = 2018,
        month = sep,
       volume = {156},
       number = {3},
          eid = {123},
        pages = {123},
          doi = {10.3847/1538-3881/aabc4f},
archivePrefix = {arXiv},
       eprint = {1801.02634},
 primaryClass = {astro-ph.IM},
       adsurl = {https://ui.adsabs.harvard.edu/abs/2018AJ....156..123A}
}

@ARTICLE{2022ApJ...935..167A,
       author = {{Astropy Collaboration} and {Price-Whelan}, Adrian M. and {Lim}, Pey Lian and {Earl}, Nicholas and {Starkman}, Nathaniel and {Bradley}, Larry and {Shupe}, David L. and {Patil}, Aarya A. and {Corrales}, Lia and {Brasseur}, C.~E. and {N{\"o}the}, Maximilian and {Donath}, Axel and {Tollerud}, Erik and {Morris}, Brett M. and {Ginsburg}, Adam and {Vaher}, Eero and {Weaver}, Benjamin A. and {Tocknell}, James and {Jamieson}, William and {van Kerkwijk}, Marten H. and {Robitaille}, Thomas P. and {Merry}, Bruce and {Bachetti}, Matteo and {G{\"u}nther}, H. Moritz and {Aldcroft}, Thomas L. and {Alvarado-Montes}, Jaime A. and {Archibald}, Anne M. and {B{\'o}di}, Attila and {Bapat}, Shreyas and {Barentsen}, Geert and {Baz{\'a}n}, Juanjo and {Biswas}, Manish and {Boquien}, M{\'e}d{\'e}ric and {Burke}, D.~J. and {Cara}, Daria and {Cara}, Mihai and {Conroy}, Kyle E. and {Conseil}, Simon and {Craig}, Matthew W. and {Cross}, Robert M. and {Cruz}, Kelle L. and {D'Eugenio}, Francesco and {Dencheva}, Nadia and {Devillepoix}, Hadrien A.~R. and {Dietrich}, J{\"o}rg P. and {Eigenbrot}, Arthur Davis and {Erben}, Thomas and {Ferreira}, Leonardo and {Foreman-Mackey}, Daniel and {Fox}, Ryan and {Freij}, Nabil and {Garg}, Suyog and {Geda}, Robel and {Glattly}, Lauren and {Gondhalekar}, Yash and {Gordon}, Karl D. and {Grant}, David and {Greenfield}, Perry and {Groener}, Austen M. and {Guest}, Steve and {Gurovich}, Sebastian and {Handberg}, Rasmus and {Hart}, Akeem and {Hatfield-Dodds}, Zac and {Homeier}, Derek and {Hosseinzadeh}, Griffin and {Jenness}, Tim and {Jones}, Craig K. and {Joseph}, Prajwel and {Kalmbach}, J. Bryce and {Karamehmetoglu}, Emir and {Ka{\l}uszy{\'n}ski}, Miko{\l}aj and {Kelley}, Michael S.~P. and {Kern}, Nicholas and {Kerzendorf}, Wolfgang E. and {Koch}, Eric W. and {Kulumani}, Shankar and {Lee}, Antony and {Ly}, Chun and {Ma}, Zhiyuan and {MacBride}, Conor and {Maljaars}, Jakob M. and {Muna}, Demitri and {Murphy}, N.~A. and {Norman}, Henrik and {O'Steen}, Richard and {Oman}, Kyle A. and {Pacifici}, Camilla and {Pascual}, Sergio and {Pascual-Granado}, J. and {Patil}, Rohit R. and {Perren}, Gabriel I. and {Pickering}, Timothy E. and {Rastogi}, Tanuj and {Roulston}, Benjamin R. and {Ryan}, Daniel F. and {Rykoff}, Eli S. and {Sabater}, Jose and {Sakurikar}, Parikshit and {Salgado}, Jes{\'u}s and {Sanghi}, Aniket and {Saunders}, Nicholas and {Savchenko}, Volodymyr and {Schwardt}, Ludwig and {Seifert-Eckert}, Michael and {Shih}, Albert Y. and {Jain}, Anany Shrey and {Shukla}, Gyanendra and {Sick}, Jonathan and {Simpson}, Chris and {Singanamalla}, Sudheesh and {Singer}, Leo P. and {Singhal}, Jaladh and {Sinha}, Manodeep and {Sip{\H{o}}cz}, Brigitta M. and {Spitler}, Lee R. and {Stansby}, David and {Streicher}, Ole and {{\v{S}}umak}, Jani and {Swinbank}, John D. and {Taranu}, Dan S. and {Tewary}, Nikita and {Tremblay}, Grant R. and {de Val-Borro}, Miguel and {Van Kooten}, Samuel J. and {Vasovi{\'c}}, Zlatan and {Verma}, Shresth and {de Miranda Cardoso}, Jos{\'e} Vin{\'\i}cius and {Williams}, Peter K.~G. and {Wilson}, Tom J. and {Winkel}, Benjamin and {Wood-Vasey}, W.~M. and {Xue}, Rui and {Yoachim}, Peter and {Zhang}, Chen and {Zonca}, Andrea and {Astropy Project Contributors}},
        title = "{The Astropy Project: Sustaining and Growing a Community-oriented Open-source Project and the Latest Major Release (v5.0) of the Core Package}",
      journal = {\apj},
         year = 2022,
        month = aug,
       volume = {935},
       number = {2},
          eid = {167},
        pages = {167},
          doi = {10.3847/1538-4357/ac7c74},
archivePrefix = {arXiv},
       eprint = {2206.14220},
 primaryClass = {astro-ph.IM},
       adsurl = {https://ui.adsabs.harvard.edu/abs/2022ApJ...935..167A}
}

@ARTICLE{2017ApJ...835..205D,
       author = {{Debes}, John H. and {Poteet}, Charles A. and {Jang-Condell}, Hannah and {Gaspar}, Andras and {Hines}, Dean and {Kastner}, Joel H. and {Pueyo}, Laurent and {Rapson}, Valerie and {Roberge}, Aki and {Schneider}, Glenn and {Weinberger}, Alycia J.},
        title = "{Chasing Shadows: Rotation of the Azimuthal Asymmetry in the TW Hya Disk}",
      journal = {\apj},
         year = 2017,
        month = feb,
       volume = {835},
       number = {2},
          eid = {205},
        pages = {205},
          doi = {10.3847/1538-4357/835/2/205},
archivePrefix = {arXiv},
       eprint = {1701.03152},
 primaryClass = {astro-ph.SR},
       adsurl = {https://ui.adsabs.harvard.edu/abs/2017ApJ...835..205D}
}

@ARTICLE{2013PASP..125..306F,
       author = {{Foreman-Mackey}, Daniel and {Hogg}, David W. and {Lang}, Dustin and {Goodman}, Jonathan},
        title = "{emcee: The MCMC Hammer}",
      journal = {\pasp},
         year = 2013,
        month = mar,
       volume = {125},
       number = {925},
        pages = {306},
          doi = {10.1086/670067},
archivePrefix = {arXiv},
       eprint = {1202.3665},
 primaryClass = {astro-ph.IM},
       adsurl = {https://ui.adsabs.harvard.edu/abs/2013PASP..125..306F}
}

@ARTICLE{2023A&A...674A...1G,
       author = {{Gaia Collaboration} and {Vallenari}, A. and {Brown}, A.~G.~A. and {Prusti}, T. and {de Bruijne}, J.~H.~J. and {Arenou}, F. and {Babusiaux}, C. and {Biermann}, M. and {Creevey}, O.~L. and {Ducourant}, C. and {Evans}, D.~W. and {Eyer}, L. and {Guerra}, R. and {Hutton}, A. and {Jordi}, C. and {Klioner}, S.~A. and {Lammers}, U.~L. and {Lindegren}, L. and {Luri}, X. and {Mignard}, F. and {Panem}, C. and {Pourbaix}, D. and {Randich}, S. and {Sartoretti}, P. and {Soubiran}, C. and {Tanga}, P. and {Walton}, N.~A. and {Bailer-Jones}, C.~A.~L. and {Bastian}, U. and {Drimmel}, R. and {Jansen}, F. and {Katz}, D. and {Lattanzi}, M.~G. and {van Leeuwen}, F. and {Bakker}, J. and {Cacciari}, C. and {Casta{\~n}eda}, J. and {De Angeli}, F. and {Fabricius}, C. and {Fouesneau}, M. and {Fr{\'e}mat}, Y. and {Galluccio}, L. and {Guerrier}, A. and {Heiter}, U. and {Masana}, E. and {Messineo}, R. and {Mowlavi}, N. and {Nicolas}, C. and {Nienartowicz}, K. and {Pailler}, F. and {Panuzzo}, P. and {Riclet}, F. and {Roux}, W. and {Seabroke}, G.~M. and {Sordo}, R. and {Th{\'e}venin}, F. and {Gracia-Abril}, G. and {Portell}, J. and {Teyssier}, D. and {Altmann}, M. and {Andrae}, R. and {Audard}, M. and {Bellas-Velidis}, I. and {Benson}, K. and {Berthier}, J. and {Blomme}, R. and {Burgess}, P.~W. and {Busonero}, D. and {Busso}, G. and {C{\'a}novas}, H. and {Carry}, B. and {Cellino}, A. and {Cheek}, N. and {Clementini}, G. and {Damerdji}, Y. and {Davidson}, M. and {de Teodoro}, P. and {Nu{\~n}ez Campos}, M. and {Delchambre}, L. and {Dell'Oro}, A. and {Esquej}, P. and {Fern{\'a}ndez-Hern{\'a}ndez}, J. and {Fraile}, E. and {Garabato}, D. and {Garc{\'\i}a-Lario}, P. and {Gosset}, E. and {Haigron}, R. and {Halbwachs}, J. -L. and {Hambly}, N.~C. and {Harrison}, D.~L. and {Hern{\'a}ndez}, J. and {Hestroffer}, D. and {Hodgkin}, S.~T. and {Holl}, B. and {Jan{\ss}en}, K. and {Jevardat de Fombelle}, G. and {Jordan}, S. and {Krone-Martins}, A. and {Lanzafame}, A.~C. and {L{\"o}ffler}, W. and {Marchal}, O. and {Marrese}, P.~M. and {Moitinho}, A. and {Muinonen}, K. and {Osborne}, P. and {Pancino}, E. and {Pauwels}, T. and {Recio-Blanco}, A. and {Reyl{\'e}}, C. and {Riello}, M. and {Rimoldini}, L. and {Roegiers}, T. and {Rybizki}, J. and {Sarro}, L.~M. and {Siopis}, C. and {Smith}, M. and {Sozzetti}, A. and {Utrilla}, E. and {van Leeuwen}, M. and {Abbas}, U. and {{\'A}brah{\'a}m}, P. and {Abreu Aramburu}, A. and {Aerts}, C. and {Aguado}, J.~J. and {Ajaj}, M. and {Aldea-Montero}, F. and {Altavilla}, G. and {{\'A}lvarez}, M.~A. and {Alves}, J. and {Anders}, F. and {Anderson}, R.~I. and {Anglada Varela}, E. and {Antoja}, T. and {Baines}, D. and {Baker}, S.~G. and {Balaguer-N{\'u}{\~n}ez}, L. and {Balbinot}, E. and {Balog}, Z. and {Barache}, C. and {Barbato}, D. and {Barros}, M. and {Barstow}, M.~A. and {Bartolom{\'e}}, S. and {Bassilana}, J. -L. and {Bauchet}, N. and {Becciani}, U. and {Bellazzini}, M. and {Berihuete}, A. and {Bernet}, M. and {Bertone}, S. and {Bianchi}, L. and {Binnenfeld}, A. and {Blanco-Cuaresma}, S. and {Blazere}, A. and {Boch}, T. and {Bombrun}, A. and {Bossini}, D. and {Bouquillon}, S. and {Bragaglia}, A. and {Bramante}, L. and {Breedt}, E. and {Bressan}, A. and {Brouillet}, N. and {Brugaletta}, E. and {Bucciarelli}, B. and {Burlacu}, A. and {Butkevich}, A.~G. and {Buzzi}, R. and {Caffau}, E. and {Cancelliere}, R. and {Cantat-Gaudin}, T. and {Carballo}, R. and {Carlucci}, T. and {Carnerero}, M.~I. and {Carrasco}, J.~M. and {Casamiquela}, L. and {Castellani}, M. and {Castro-Ginard}, A. and {Chaoul}, L. and {Charlot}, P. and {Chemin}, L. and {Chiaramida}, V. and {Chiavassa}, A. and {Chornay}, N. and {Comoretto}, G. and {Contursi}, G. and {Cooper}, W.~J. and {Cornez}, T. and {Cowell}, S. and {Crifo}, F. and {Cropper}, M. and {Crosta}, M. and {Crowley}, C. and {Dafonte}, C. and {Dapergolas}, A. and {David}, M. and {David}, P. and {de Laverny}, P. and {De Luise}, F. and {De March}, R.},
        title = "{Gaia Data Release 3. Summary of the content and survey properties}",
      journal = {\aap},
         year = 2023,
        month = jun,
       volume = {674},
          eid = {A1},
        pages = {A1},
          doi = {10.1051/0004-6361/202243940},
archivePrefix = {arXiv},
       eprint = {2208.00211},
 primaryClass = {astro-ph.GA},
       adsurl = {https://ui.adsabs.harvard.edu/abs/2023A&A...674A...1G}
}

@ARTICLE{2018ApJ...853..120S,
       author = {{Sokal}, Kimberly R. and {Deen}, Casey P. and {Mace}, Gregory N. and {Lee}, Jae-Joon and {Oh}, Heeyoung and {Kim}, Hwihyun and {Kidder}, Benjamin T. and {Jaffe}, Daniel T.},
        title = "{Characterizing TW Hydra}",
      journal = {\apj},
         year = 2018,
        month = feb,
       volume = {853},
       number = {2},
          eid = {120},
        pages = {120},
          doi = {10.3847/1538-4357/aaa1e4},
archivePrefix = {arXiv},
       eprint = {1712.04785},
 primaryClass = {astro-ph.SR},
       adsurl = {https://ui.adsabs.harvard.edu/abs/2018ApJ...853..120S}
}

@misc{2025ascl.soft02014C,
       author = {{Carter}, Aarynn and {Kammerer}, Jens and {Leisenring}, Jarron and {Perrin}, Marshall and {Balmer}, William O. and {Glidic}, Kayli and {Strampelli}, Giovanni Maria and {Millar-Blanchaer}, Max},
        title = "{spaceKLIP: JWST coronagraphy data data reduction and analysis pipeline}",
 howpublished = {Astrophysics Source Code Library, record ascl:2502.014},
         year = 2025,
        month = feb,
          eid = {ascl:2502.014},
       adsurl = {https://ui.adsabs.harvard.edu/abs/2025ascl.soft02014C}
}

@ARTICLE{2022ApJ...935L..25L,
       author = {{Lawson}, Kellen and {Currie}, Thayne and {Wisniewski}, John P. and {Groff}, Tyler D. and {McElwain}, Michael W. and {Schlieder}, Joshua E.},
        title = "{Constrained Reference Star Differential Imaging: Enabling High-fidelity Imagery of Highly Structured Circumstellar Disks}",
      journal = {\apjl},
         year = 2022,
        month = aug,
       volume = {935},
       number = {2},
          eid = {L25},
        pages = {L25},
          doi = {10.3847/2041-8213/ac853b},
archivePrefix = {arXiv},
       eprint = {2208.01606},
 primaryClass = {astro-ph.EP},
       adsurl = {https://ui.adsabs.harvard.edu/abs/2022ApJ...935L..25L}
}

@ARTICLE{2023AJ....166..150L,
       author = {{Lawson}, Kellen and {Schlieder}, Joshua E. and {Leisenring}, Jarron M. and {Bogat}, Ell and {Beichman}, Charles A. and {Bryden}, Geoffrey and {G{\'a}sp{\'a}r}, Andr{\'a}s and {Groff}, Tyler D. and {McElwain}, Michael W. and {Meyer}, Michael R. and {Barclay}, Thomas and {Calissendorff}, Per and {De Furio}, Matthew and {Ygouf}, Marie and {Boccaletti}, Anthony and {Greene}, Thomas P. and {Krist}, John and {Plavchan}, Peter and {Rieke}, Marcia J. and {Roellig}, Thomas L. and {Stansberry}, John and {Wisniewski}, John P. and {Young}, Erick T.},
        title = "{JWST/NIRCam Coronagraphy of the Young Planet-hosting Debris Disk AU Microscopii}",
      journal = {\aj},
         year = 2023,
        month = oct,
       volume = {166},
       number = {4},
          eid = {150},
        pages = {150},
          doi = {10.3847/1538-3881/aced08},
archivePrefix = {arXiv},
       eprint = {2308.02486},
 primaryClass = {astro-ph.EP},
       adsurl = {https://ui.adsabs.harvard.edu/abs/2023AJ....166..150L}
}

@ARTICLE{2019SAAS...45.....A,
       author = {{Armitage}, Philip J. and {Kley}, Wilhelm},
        title = "{From Protoplanetary Disks to Planet Formation}",
      journal = {From Protoplanetary Disks to Planet Formation: Saas-Fee Advanced Course 45. Swiss Society for Astrophysics and Astronomy},
         year = 2019,
        month = jan,
          doi = {10.1007/978-3-662-58687-7},
       adsurl = {https://ui.adsabs.harvard.edu/abs/2019SAAS...45.....A}
}

@ARTICLE{2019A&A...623A..85L,
       author = {{Linder}, Esther F. and {Mordasini}, Christoph and {Molli{\`e}re}, Paul and {Marleau}, Gabriel-Dominique and {Malik}, Matej and {Quanz}, Sascha P. and {Meyer}, Michael R.},
        title = "{Evolutionary models of cold and low-mass planets: cooling curves, magnitudes, and detectability}",
      journal = {\aap},
         year = 2019,
        month = mar,
       volume = {623},
          eid = {A85},
        pages = {A85},
          doi = {10.1051/0004-6361/201833873},
archivePrefix = {arXiv},
       eprint = {1812.02027},
 primaryClass = {astro-ph.EP},
       adsurl = {https://ui.adsabs.harvard.edu/abs/2019A&A...623A..85L}
}

@misc{bushouse_2022_7229890,
  author       = {Bushouse, Howard and
                  Eisenhamer, Jonathan and
                  Dencheva, Nadia and
                  Davies, James and
                  Greenfield, Perry and
                  Morrison, Jane and
                  Hodge, Phil and
                  Simon, Bernie and
                  Grumm, David and
                  Droettboom, Michael and
                  Slavich, Edward and
                  Sosey, Megan and
                  Pauly, Tyler and
                  Miller, Todd and
                  Jedrzejewski, Robert and
                  Hack, Warren and
                  Davis, David and
                  Crawford, Steven and
                  Law, David and
                  Gordon, Karl and
                  Regan, Michael and
                  Cara, Mihai and
                  MacDonald, Ken and
                  Bradley, Larry and
                  Shanahan, Clare and
                  Jamieson, William and
                  Teodoro, Mairan and
                  Williams, Thomas},
  title        = {JWST Calibration Pipeline},
  month        = oct,
  year         = 2022,
  publisher    = {Zenodo},
  version      = {1.8.2},
  doi          = {10.5281/zenodo.7229890},
  url          = {https://doi.org/10.5281/zenodo.7229890},
}

@INPROCEEDINGS{2022SPIE12180E..3NK,
       author = {{Kammerer}, Jens and {Girard}, Julien and {Carter}, Aarynn L. and {Perrin}, Marshall D. and {Cooper}, Rachel and {Thatte}, Deepashri and {Vandal}, Thomas and {Leisenring}, Jarron and {Wang}, Jason and {Balmer}, William O. and {Sivaramakrishnan}, Anand and {Pueyo}, Laurent and {Ward-Duong}, Kimberly and {Sunnquist}, Ben and {Adams Redai}, J{\'e}a.},
        title = "{Performance of near-infrared high-contrast imaging methods with JWST from commissioning}",
    booktitle = {Space Telescopes and Instrumentation 2022: Optical, Infrared, and Millimeter Wave},
         year = 2022,
       editor = {{Coyle}, Laura E. and {Matsuura}, Shuji and {Perrin}, Marshall D.},
       series = {Society of Photo-Optical Instrumentation Engineers (SPIE) Conference Series},
       volume = {12180},
        month = aug,
          eid = {121803N},
        pages = {121803N},
          doi = {10.1117/12.2628865},
archivePrefix = {arXiv},
       eprint = {2208.00996},
 primaryClass = {astro-ph.EP},
       adsurl = {https://ui.adsabs.harvard.edu/abs/2022SPIE12180E..3NK}
}

@ARTICLE{2023ApJ...951L..20C,
       author = {{Carter}, Aarynn L. and {Hinkley}, Sasha and {Kammerer}, Jens and {Skemer}, Andrew and {Biller}, Beth A. and {Leisenring}, Jarron M. and {Millar-Blanchaer}, Maxwell A. and {Petrus}, Simon and {Stone}, Jordan M. and {Ward-Duong}, Kimberly and {Wang}, Jason J. and {Girard}, Julien H. and {Hines}, Dean C. and {Perrin}, Marshall D. and {Pueyo}, Laurent and {Balmer}, William O. and {Bonavita}, Mariangela and {Bonnefoy}, Mickael and {Chauvin}, Gael and {Choquet}, Elodie and {Christiaens}, Valentin and {Danielski}, Camilla and {Kennedy}, Grant M. and {Matthews}, Elisabeth C. and {Miles}, Brittany E. and {Patapis}, Polychronis and {Ray}, Shrishmoy and {Rickman}, Emily and {Sallum}, Steph and {Stapelfeldt}, Karl R. and {Whiteford}, Niall and {Zhou}, Yifan and {Absil}, Olivier and {Boccaletti}, Anthony and {Booth}, Mark and {Bowler}, Brendan P. and {Chen}, Christine H. and {Currie}, Thayne and {Fortney}, Jonathan J. and {Grady}, Carol A. and {Greebaum}, Alexandra Z. and {Henning}, Thomas and {Hoch}, Kielan K.~W. and {Janson}, Markus and {Kalas}, Paul and {Kenworthy}, Matthew A. and {Kervella}, Pierre and {Kraus}, Adam L. and {Lagage}, Pierre-Olivier and {Liu}, Michael C. and {Macintosh}, Bruce and {Marino}, Sebastian and {Marley}, Mark S. and {Marois}, Christian and {Matthews}, Brenda C. and {Mawet}, Dimitri and {McElwain}, Michael W. and {Metchev}, Stanimir and {Meyer}, Michael R. and {Molliere}, Paul and {Moran}, Sarah E. and {Morley}, Caroline V. and {Mukherjee}, Sagnick and {Pantin}, Eric and {Quirrenbach}, Andreas and {Rebollido}, Isabel and {Ren}, Bin B. and {Schneider}, Glenn and {Vasist}, Malavika and {Worthen}, Kadin and {Wyatt}, Mark C. and {Briesemeister}, Zackery W. and {Bryan}, Marta L. and {Calissendorff}, Per and {Cantalloube}, Faustine and {Cugno}, Gabriele and {De Furio}, Matthew and {Dupuy}, Trent J. and {Factor}, Samuel M. and {Faherty}, Jacqueline K. and {Fitzgerald}, Michael P. and {Franson}, Kyle and {Gonzales}, Eileen C. and {Hood}, Callie E. and {Howe}, Alex R. and {Kuzuhara}, Masayuki and {Lagrange}, Anne-Marie and {Lawson}, Kellen and {Lazzoni}, Cecilia and {Lew}, Ben W.~P. and {Liu}, Pengyu and {Llop-Sayson}, Jorge and {Lloyd}, James P. and {Martinez}, Raquel A. and {Mazoyer}, Johan and {Palma-Bifani}, Paulina and {Quanz}, Sascha P. and {Redai}, Jea Adams and {Samland}, Matthias and {Schlieder}, Joshua E. and {Tamura}, Motohide and {Tan}, Xianyu and {Uyama}, Taichi and {Vigan}, Arthur and {Vos}, Johanna M. and {Wagner}, Kevin and {Wolff}, Schuyler G. and {Ygouf}, Marie and {Zhang}, Xi and {Zhang}, Keming and {Zhang}, Zhoujian},
        title = "{The JWST Early Release Science Program for Direct Observations of Exoplanetary Systems I: High-contrast Imaging of the Exoplanet HIP 65426 b from 2 to 16 {\ensuremath{\mu}}m}",
      journal = {\apjl},
         year = 2023,
        month = jul,
       volume = {951},
       number = {1},
          eid = {L20},
        pages = {L20},
          doi = {10.3847/2041-8213/acd93e},
archivePrefix = {arXiv},
       eprint = {2208.14990},
 primaryClass = {astro-ph.EP},
       adsurl = {https://ui.adsabs.harvard.edu/abs/2023ApJ...951L..20C}
}

@ARTICLE{2016ApJ...821...82O,
       author = {{Okuzumi}, Satoshi and {Momose}, Munetake and {Sirono}, Sin-iti and {Kobayashi}, Hiroshi and {Tanaka}, Hidekazu},
        title = "{Sintering-induced Dust Ring Formation in Protoplanetary Disks: Application to the HL Tau Disk}",
      journal = {\apj},
         year = 2016,
        month = apr,
       volume = {821},
       number = {2},
          eid = {82},
        pages = {82},
          doi = {10.3847/0004-637X/821/2/82},
archivePrefix = {arXiv},
       eprint = {1510.03556},
 primaryClass = {astro-ph.SR},
       adsurl = {https://ui.adsabs.harvard.edu/abs/2016ApJ...821...82O}
}

@ARTICLE{2022A&A...666A..32X,
       author = {{Xie}, Chen and {Choquet}, Elodie and {Vigan}, Arthur and {Cantalloube}, Faustine and {Benisty}, Myriam and {Boccaletti}, Anthony and {Bonnefoy}, Mickael and {Desgrange}, Celia and {Garufi}, Antonio and {Girard}, Julien and {Hagelberg}, Janis and {Janson}, Markus and {Kenworthy}, Matthew and {Lagrange}, Anne-Marie and {Langlois}, Maud and {Menard}, Fran{\c{c}}ois and {Zurlo}, Alice},
        title = "{Reference-star differential imaging on SPHERE/IRDIS}",
      journal = {\aap},
         year = 2022,
        month = oct,
       volume = {666},
          eid = {A32},
        pages = {A32},
          doi = {10.1051/0004-6361/202243379},
archivePrefix = {arXiv},
       eprint = {2208.07915},
 primaryClass = {astro-ph.EP},
       adsurl = {https://ui.adsabs.harvard.edu/abs/2022A&A...666A..32X}
}

@ARTICLE{2022AJ....163..119S,
       author = {{Sanghi}, Aniket and {Zhou}, Yifan and {Bowler}, Brendan P.},
        title = "{Efficiently Imaging Accreting Protoplanets from Space: Reference Star Differential Imaging of the PDS 70 Planetary System Using the HST/WFC3 Archival PSF Library}",
      journal = {\aj},
         year = 2022,
        month = mar,
       volume = {163},
       number = {3},
          eid = {119},
        pages = {119},
          doi = {10.3847/1538-3881/ac477e},
archivePrefix = {arXiv},
       eprint = {2112.10777},
 primaryClass = {astro-ph.EP},
       adsurl = {https://ui.adsabs.harvard.edu/abs/2022AJ....163..119S}
}

@ARTICLE{2024AJ....168..215S,
       author = {{Sanghi}, Aniket and {Xuan}, Jerry W. and {Wang}, Jason J. and {Mawet}, Dimitri and {Bowler}, Brendan P. and {Ngo}, Henry and {Bryan}, Marta L. and {Ruane}, Garreth and {Absil}, Olivier and {Huby}, Elsa},
        title = "{Efficiently Searching for Close-in Companions Around Young M Dwarfs Using a Multiyear PSF Library}",
      journal = {\aj},
         year = 2024,
        month = nov,
       volume = {168},
       number = {5},
          eid = {215},
        pages = {215},
          doi = {10.3847/1538-3881/ad769f},
archivePrefix = {arXiv},
       eprint = {2408.14268},
 primaryClass = {astro-ph.EP},
       adsurl = {https://ui.adsabs.harvard.edu/abs/2024AJ....168..215S}
}

@ARTICLE{2021MNRAS.501.1999C,
       author = {{Carter}, Aarynn L. and {Hinkley}, Sasha and {Bonavita}, Mariangela and {Phillips}, Mark W. and {Girard}, Julien H. and {Perrin}, Marshall and {Pueyo}, Laurent and {Vigan}, Arthur and {Gagn{\'e}}, Jonathan and {Skemer}, Andrew J.~I.},
        title = "{Direct imaging of sub-Jupiter mass exoplanets with James Webb Space Telescope coronagraphy}",
      journal = {\mnras},
         year = 2021,
        month = feb,
       volume = {501},
       number = {2},
        pages = {1999-2016},
          doi = {10.1093/mnras/staa3579},
archivePrefix = {arXiv},
       eprint = {2011.07075},
 primaryClass = {astro-ph.EP},
       adsurl = {https://ui.adsabs.harvard.edu/abs/2021MNRAS.501.1999C}
}

@ARTICLE{2019A&A...631A.155B,
       author = {{Beuzit}, J. -L. and {Vigan}, A. and {Mouillet}, D. and {Dohlen}, K. and {Gratton}, R. and {Boccaletti}, A. and {Sauvage}, J. -F. and {Schmid}, H.~M. and {Langlois}, M. and {Petit}, C. and {Baruffolo}, A. and {Feldt}, M. and {Milli}, J. and {Wahhaj}, Z. and {Abe}, L. and {Anselmi}, U. and {Antichi}, J. and {Barette}, R. and {Baudrand}, J. and {Baudoz}, P. and {Bazzon}, A. and {Bernardi}, P. and {Blanchard}, P. and {Brast}, R. and {Bruno}, P. and {Buey}, T. and {Carbillet}, M. and {Carle}, M. and {Cascone}, E. and {Chapron}, F. and {Charton}, J. and {Chauvin}, G. and {Claudi}, R. and {Costille}, A. and {De Caprio}, V. and {de Boer}, J. and {Delboulb{\'e}}, A. and {Desidera}, S. and {Dominik}, C. and {Downing}, M. and {Dupuis}, O. and {Fabron}, C. and {Fantinel}, D. and {Farisato}, G. and {Feautrier}, P. and {Fedrigo}, E. and {Fusco}, T. and {Gigan}, P. and {Ginski}, C. and {Girard}, J. and {Giro}, E. and {Gisler}, D. and {Gluck}, L. and {Gry}, C. and {Henning}, T. and {Hubin}, N. and {Hugot}, E. and {Incorvaia}, S. and {Jaquet}, M. and {Kasper}, M. and {Lagadec}, E. and {Lagrange}, A. -M. and {Le Coroller}, H. and {Le Mignant}, D. and {Le Ruyet}, B. and {Lessio}, G. and {Lizon}, J. -L. and {Llored}, M. and {Lundin}, L. and {Madec}, F. and {Magnard}, Y. and {Marteaud}, M. and {Martinez}, P. and {Maurel}, D. and {M{\'e}nard}, F. and {Mesa}, D. and {M{\"o}ller-Nilsson}, O. and {Moulin}, T. and {Moutou}, C. and {Orign{\'e}}, A. and {Parisot}, J. and {Pavlov}, A. and {Perret}, D. and {Pragt}, J. and {Puget}, P. and {Rabou}, P. and {Ramos}, J. and {Reess}, J. -M. and {Rigal}, F. and {Rochat}, S. and {Roelfsema}, R. and {Rousset}, G. and {Roux}, A. and {Saisse}, M. and {Salasnich}, B. and {Santambrogio}, E. and {Scuderi}, S. and {Segransan}, D. and {Sevin}, A. and {Siebenmorgen}, R. and {Soenke}, C. and {Stadler}, E. and {Suarez}, M. and {Tiph{\`e}ne}, D. and {Turatto}, M. and {Udry}, S. and {Vakili}, F. and {Waters}, L.~B.~F.~M. and {Weber}, L. and {Wildi}, F. and {Zins}, G. and {Zurlo}, A.},
        title = "{SPHERE: the exoplanet imager for the Very Large Telescope}",
      journal = {\aap},
         year = 2019,
        month = nov,
       volume = {631},
          eid = {A155},
        pages = {A155},
          doi = {10.1051/0004-6361/201935251},
archivePrefix = {arXiv},
       eprint = {1902.04080},
 primaryClass = {astro-ph.IM},
       adsurl = {https://ui.adsabs.harvard.edu/abs/2019A&A...631A.155B}
}

@ARTICLE{2023PASP..135d8003W,
       author = {{Wright}, Gillian S. and {Rieke}, George H. and {Glasse}, Alistair and {Ressler}, Michael and {Garc{\'\i}a Mar{\'\i}n}, Macarena and {Aguilar}, Jonathan and {Alberts}, Stacey and {{\'A}lvarez-M{\'a}rquez}, Javier and {Argyriou}, Ioannis and {Banks}, Kimberly and {Baudoz}, Pierre and {Boccaletti}, Anthony and {Bouchet}, Patrice and {Bouwman}, Jeroen and {Brandl}, Bernard R. and {Breda}, David and {Bright}, Stacey and {Cale}, Steven and {Colina}, Luis and {Cossou}, Christophe and {Coulais}, Alain and {Cracraft}, Misty and {De Meester}, Wim and {Dicken}, Daniel and {Engesser}, Michael and {Etxaluze}, Mireya and {Fox}, Ori D. and {Friedman}, Scott and {Fu}, Henry and {Gasman}, Danny and {G{\'a}sp{\'a}r}, Andr{\'a}s and {Gastaud}, Ren{\'e} and {Geers}, Vincent and {Glauser}, Adrian Michael and {Gordon}, Karl D. and {Greene}, Thomas and {Greve}, Thomas R. and {Grundy}, Timothy and {G{\"u}del}, Manuel and {Guillard}, Pierre and {Haderlein}, Peter and {Hashimoto}, Ryan and {Henning}, Thomas and {Hines}, Dean and {Holler}, Bryan and {Detre}, {\"O}rs Hunor and {Jahromi}, Amir and {James}, Bryan and {Jones}, Olivia C. and {Justtanont}, Kay and {Kavanagh}, Patrick and {Kendrew}, Sarah and {Klaassen}, Pamela and {Krause}, Oliver and {Labiano}, Alvaro and {Lagage}, Pierre-Olivier and {Lambros}, Scott and {Larson}, Kirsten and {Law}, David and {Lee}, David and {Libralato}, Mattia and {Lorenzo Alverez}, Jose and {Meixner}, Margaret and {Morrison}, Jane and {Mueller}, Migo and {Murray}, Katherine and {Mycroft}, Matthew and {Myers}, Richard and {Nayak}, Omnarayani and {Naylor}, Bret and {Nickson}, Bryony and {Noriega-Crespo}, Alberto and {{\"O}stlin}, G{\"o}ran and {O'Sullivan}, Brian and {Ottens}, Richard and {Patapis}, Polychronis and {Penanen}, Konstantin and {Pietraszkiewicz}, Martin and {Ray}, Tom and {Regan}, Michael and {Roteliuk}, Anthony and {Royer}, Pierre and {Samara-Ratna}, Piyal and {Samuelson}, Bridget and {Sargent}, Beth A. and {Scheithauer}, Silvia and {Schneider}, Analyn and {Schreiber}, J{\"u}rgen and {Shaughnessy}, Bryan and {Sheehan}, Evan and {Shivaei}, Irene and {Sloan}, G.~C. and {Tamas}, Laszlo and {Teague}, Kelly and {Temim}, Tea and {Tikkanen}, Tuomo and {Tustain}, Samuel and {van Dishoeck}, Ewine F. and {Vandenbussche}, Bart and {Weilert}, Mark and {Whitehouse}, Paul and {Wolff}, Schuyler},
        title = "{The Mid-infrared Instrument for JWST and Its In-flight Performance}",
      journal = {\pasp},
         year = 2023,
        month = apr,
       volume = {135},
       number = {1046},
          eid = {048003},
        pages = {048003},
          doi = {10.1088/1538-3873/acbe66},
       adsurl = {https://ui.adsabs.harvard.edu/abs/2023PASP..135d8003W}
}

@ARTICLE{2024A&A...689A.104D,
       author = {{Das}, S. and {Kurtovic}, N.~T. and {Flock}, M.},
        title = "{From traffic jams to roadblocks: The outer regions of TW Hya with ALMA Band 8}",
      journal = {\aap},
         year = 2024,
        month = sep,
       volume = {689},
          eid = {A104},
        pages = {A104},
          doi = {10.1051/0004-6361/202450278},
archivePrefix = {arXiv},
       eprint = {2407.07649},
 primaryClass = {astro-ph.EP},
       adsurl = {https://ui.adsabs.harvard.edu/abs/2024A&A...689A.104D}
}

@ARTICLE{2021A&A...648A..33M,
       author = {{Mac{\'\i}as}, E. and {Guerra-Alvarado}, O. and {Carrasco-Gonz{\'a}lez}, C. and {Ribas}, {\'A}. and {Espaillat}, C.~C. and {Huang}, J. and {Andrews}, S.~M.},
        title = "{Characterizing the dust content of disk substructures in TW Hydrae}",
      journal = {\aap},
         year = 2021,
        month = apr,
       volume = {648},
          eid = {A33},
        pages = {A33},
          doi = {10.1051/0004-6361/202039812},
archivePrefix = {arXiv},
       eprint = {2102.04648},
 primaryClass = {astro-ph.EP},
       adsurl = {https://ui.adsabs.harvard.edu/abs/2021A&A...648A..33M}
}

@ARTICLE{2023ApJ...948...36D,
       author = {{Debes}, John and {Nealon}, Rebecca and {Alexander}, Richard and {Weinberger}, Alycia J. and {Wolff}, Schuyler Grace and {Hines}, Dean and {Kastner}, Joel and {Jang-Condell}, Hannah and {Pinte}, Christophe and {Plavchan}, Peter and {Pueyo}, Laurent},
        title = "{The Surprising Evolution of the Shadow on the TW Hya Disk}",
      journal = {\apj},
         year = 2023,
        month = may,
       volume = {948},
       number = {1},
          eid = {36},
        pages = {36},
          doi = {10.3847/1538-4357/acbdf1},
archivePrefix = {arXiv},
       eprint = {2305.03611},
 primaryClass = {astro-ph.SR},
       adsurl = {https://ui.adsabs.harvard.edu/abs/2023ApJ...948...36D}
}

@ARTICLE{2025Natur.642..905L,
       author = {{Lagrange}, A. -M. and {Wilkinson}, C. and {M{\^a}lin}, M. and {Boccaletti}, A. and {Perrot}, C. and {Matr{\`a}}, L. and {Combes}, F. and {Beust}, H. and {Rouan}, D. and {Chomez}, A. and {Milli}, J. and {Charnay}, B. and {Mazevet}, S. and {Flasseur}, O. and {Olofsson}, J. and {Bayo}, A. and {Kral}, Q. and {Carter}, A. and {Crotts}, K.~A. and {Delorme}, P. and {Chauvin}, G. and {Thebault}, P. and {Rubini}, P. and {Kiefer}, F. and {Radcliffe}, A. and {Mazoyer}, J. and {Bodrito}, T. and {Stasevic}, S. and {Langlois}, M.},
        title = "{Evidence for a sub-Jovian planet in the young TWA 7 disk}",
      journal = {\nat},
         year = 2025,
        month = jun,
       volume = {642},
       number = {8069},
        pages = {905-908},
          doi = {10.1038/s41586-025-09150-4},
archivePrefix = {arXiv},
       eprint = {2502.15081},
 primaryClass = {astro-ph.EP},
       adsurl = {https://ui.adsabs.harvard.edu/abs/2025Natur.642..905L}
}

@ARTICLE{2025ApJ...987L..41C,
       author = {{Crotts}, Katie A. and {Carter}, Aarynn L. and {Lawson}, Kellen and {Mang}, James and {Biller}, Beth and {Booth}, Mark and {Ferrer-Chavez}, Rodrigo and {Girard}, Julien H. and {Lagrange}, Anne-Marie and {Liu}, Michael C. and {Marino}, Sebastian and {Millar-Blanchaer}, Maxwell A. and {Skemer}, Andy and {Strampelli}, Giovanni M. and {Wang}, Jason and {Absil}, Olivier and {Balmer}, William O. and {Bendahan-West}, Rapha{\"e}l and {Bogat}, Ellis and {Bowens-Rubin}, Rachel and {Chauvin}, Ga{\"e}l and {Fontanive}, Cl{\'e}mence and {Franson}, Kyle and {Kammerer}, Jens and {Leisenring}, Jarron and {Morley}, Caroline V. and {Rebollido}, Isabel and {Skaf}, Nour and {Sutlieff}, Ben J. and {Bruinsma}, Evelyn L. and {Hinkley}, Sasha and {Hoch}, Kielan and {James}, Andrew D. and {Kane}, Rohan and {Mawet}, Dimitri and {Meyer}, Michael R. and {Palatnick}, Skyler and {Perrin}, Marshall D. and {Ray}, Shrishmoy and {Rickman}, Emily and {Sanghi}, Aniket and {Stephenson}, Klaus Subbotina},
        title = "{Follow-up Exploration of the TWA 7 Planet{\textendash}Disk System with JWST NIRCam}",
      journal = {\apjl},
         year = 2025,
        month = jul,
       volume = {987},
       number = {2},
          eid = {L41},
        pages = {L41},
          doi = {10.3847/2041-8213/ade798},
archivePrefix = {arXiv},
       eprint = {2506.19932},
 primaryClass = {astro-ph.EP},
       adsurl = {https://ui.adsabs.harvard.edu/abs/2025ApJ...987L..41C}
}

@ARTICLE{2024AJ....167..183M,
       author = {{Mullin}, Camryn and {Dong}, Ruobing and {Leisenring}, Jarron and {Cugno}, Gabriele and {Greene}, Thomas and {Johnstone}, Doug and {Meyer}, Michael R. and {Wagner}, Kevin R. and {Wolff}, Schuyler G. and {Boyer}, Martha and {Horner}, Scott and {Hodapp}, Klaus and {McCarthy}, Don and {Rieke}, George and {Rieke}, Marcia and {Young}, Erick},
        title = "{JWST/NIRCam Imaging of Young Stellar Objects. III. Detailed Imaging of the Nebular Environment around the HL Tau Disk}",
      journal = {\aj},
         year = 2024,
        month = apr,
       volume = {167},
       number = {4},
          eid = {183},
        pages = {183},
          doi = {10.3847/1538-3881/ad2de9},
archivePrefix = {arXiv},
       eprint = {2403.00908},
 primaryClass = {astro-ph.EP},
       adsurl = {https://ui.adsabs.harvard.edu/abs/2024AJ....167..183M}
}

@ARTICLE{2017A&A...605A..34P,
       author = {{Pohl}, A. and {Sissa}, E. and {Langlois}, M. and {M{\"u}ller}, A. and {Ginski}, C. and {van Holstein}, R.~G. and {Vigan}, A. and {Mesa}, D. and {Maire}, A.-L. and {Henning}, Th. and {Gratton}, R. and {Olofsson}, J. and {van Boekel}, R. and {Benisty}, M. and {Biller}, B. and {Boccaletti}, A. and {Chauvin}, G. and {Daemgen}, S. and {de Boer}, J. and {Desidera}, S. and {Dominik}, C. and {Garufi}, A. and {Janson}, M. and {Kral}, Q. and {M{\'e}nard}, F. and {Pinte}, C. and {Stolker}, T. and {Szul{\'a}gyi}, J. and {Zurlo}, A. and {Bonnefoy}, M. and {Cheetham}, A. and {Cudel}, M. and {Feldt}, M. and {Kasper}, M. and {Lagrange}, A.-M. and {Perrot}, C. and {Wildi}, F.},
        title = "{New constraints on the disk characteristics and companion candidates around T Chamaeleontis with VLT/SPHERE}",
      journal = {\aap},
         year = 2017,
        month = sep,
       volume = {605},
          eid = {A34},
        pages = {A34},
          doi = {10.1051/0004-6361/201630234},
archivePrefix = {arXiv},
       eprint = {1705.03477},
 primaryClass = {astro-ph.EP},
       adsurl = {https://ui.adsabs.harvard.edu/abs/2017A&A...605A..34P}
}

@MISC{vizier,
author = { {Ochsenbein}, F. et al.},
title = "{ The VizieR database of astronomical catalogues }",
year = 2019,
month = Mar,
doi = {10.26093/cds/vizier}, 
}

@ARTICLE{2013ApJ...771...45D,
       author = {{Debes}, John H. and {Jang-Condell}, Hannah and {Weinberger}, Alycia J. and {Roberge}, Aki and {Schneider}, Glenn},
        title = "{The 0.5-2.22 {\ensuremath{\mu}}m Scattered Light Spectrum of the Disk around TW Hya: Detection of a Partially Filled Disk Gap at 80 AU}",
      journal = {\apj},
         year = 2013,
        month = jul,
       volume = {771},
       number = {1},
          eid = {45},
        pages = {45},
          doi = {10.1088/0004-637X/771/1/45},
archivePrefix = {arXiv},
       eprint = {1306.2969},
 primaryClass = {astro-ph.EP},
       adsurl = {https://ui.adsabs.harvard.edu/abs/2013ApJ...771...45D}
}

@INPROCEEDINGS{2022SPIE12180E..3QG,
       author = {{Girard}, Julien H. and {Leisenring}, Jarron and {Kammerer}, Jens and {Gennaro}, Mario and {Rieke}, Marcia and {Stansberry}, John and {Rest}, Armin and {Egami}, Eiichi and {Sunnquist}, Ben and {Boyer}, Martha and {Canipe}, Alicia and {Correnti}, Matteo and {Hilbert}, Bryan and {Perrin}, Marshall D. and {Pueyo}, Laurent and {Soummer}, Remi and {Allen}, Marsha and {Bushouse}, Howard and {Aguilar}, Jonathan and {Brooks}, Brian and {Coe}, Dan and {DiFelice}, Audrey and {Golimowski}, David and {Hartig}, George and {Hines}, Dean C. and {Koekemoer}, Anton and {Nickson}, Bryony and {Nikolov}, Nikolay and {Kozhurina-Platais}, Vera and {Pirzkal}, Nor and {Robberto}, Massimo and {Sivaramakrishnan}, Anand and {Sohn}, Sangmo Tony and {Telfer}, Randal and {Wu}, Chi Rai and {Beatty}, Thomas and {Florian}, Michael and {Hainline}, Kevin and {Kelly}, Doug and {Misselt}, Karl and {Schlawin}, Everett and {Sun}, Fengwu and {Williams}, Christina and {Willmer}, Christopher and {Stark}, Christopher and {Ygouf}, Marie and {Carter}, Aarynn and {Beichman}, Charles and {Greene}, Thomas P. and {Roellig}, Thomas and {Krist}, John and {Adams Redai}, J{\'e}a. and {Wang}, Jason and {Clark}, Charles R. and {Lewis}, Dan and {Ferry}, Malcolm},
        title = "{JWST/NIRCam coronagraphy: commissioning and first on-sky results}",
    booktitle = {Space Telescopes and Instrumentation 2022: Optical, Infrared, and Millimeter Wave},
         year = 2022,
       editor = {{Coyle}, Laura E. and {Matsuura}, Shuji and {Perrin}, Marshall D.},
       series = {Society of Photo-Optical Instrumentation Engineers (SPIE) Conference Series},
       volume = {12180},
        month = aug,
          eid = {121803Q},
        pages = {121803Q},
          doi = {10.1117/12.2629636},
archivePrefix = {arXiv},
       eprint = {2208.00998},
 primaryClass = {astro-ph.IM},
       adsurl = {https://ui.adsabs.harvard.edu/abs/2022SPIE12180E..3QG}
}

@ARTICLE{1998ApJ...492L..95T,
       author = {{Thompson}, Rodger I. and {Rieke}, Marcia and {Schneider}, Glenn and {Hines}, Dean C. and {Corbin}, Michael R.},
        title = "{Initial On-Orbit Performance of NICMOS}",
      journal = {\apjl},
         year = 1998,
        month = jan,
       volume = {492},
       number = {2},
        pages = {L95-L97},
          doi = {10.1086/311095},
       adsurl = {https://ui.adsabs.harvard.edu/abs/1998ApJ...492L..95T}
}

@ARTICLE{2018MNRAS.478..758S,
       author = {{Siwak}, Michal and {Ogloza}, Waldemar and {Moffat}, Anthony F.~J. and {Matthews}, Jaymie M. and {Rucinski}, Slavek M. and {Kallinger}, Thomas and {Kuschnig}, Rainer and {Cameron}, Chris and {Weiss}, Werner W. and {Rowe}, Jason F. and {Guenther}, David B. and {Sasselov}, Dimitar},
        title = "{Photometric variability of TW Hya from seconds to years as seen from space and the ground during 2013-2017}",
      journal = {\mnras},
         year = 2018,
        month = jul,
       volume = {478},
       number = {1},
        pages = {758-783},
          doi = {10.1093/mnras/sty1220},
archivePrefix = {arXiv},
       eprint = {1805.04547},
 primaryClass = {astro-ph.SR},
       adsurl = {https://ui.adsabs.harvard.edu/abs/2018MNRAS.478..758S}
}

@ARTICLE{2012ApJ...755L..28S,
       author = {{Soummer}, R{\'e}mi and {Pueyo}, Laurent and {Larkin}, James},
        title = "{Detection and Characterization of Exoplanets and Disks Using Projections on Karhunen-Lo{\`e}ve Eigenimages}",
      journal = {\apjl},
         year = 2012,
        month = aug,
       volume = {755},
       number = {2},
          eid = {L28},
        pages = {L28},
          doi = {10.1088/2041-8205/755/2/L28},
archivePrefix = {arXiv},
       eprint = {1207.4197},
 primaryClass = {astro-ph.IM},
       adsurl = {https://ui.adsabs.harvard.edu/abs/2012ApJ...755L..28S}
}

@ARTICLE{2024A&A...688A.197M,
       author = {{M{\'e}sz{\'a}ros}, Szabolcs and {Bohlin}, Ralph and {Allende Prieto}, Carlos and {Cseh}, Borb{\'a}la and {Kov{\'a}cs}, J{\'o}zsef and {Fleming}, Scott W. and {Dencs}, Zolt{\'a}n and {Deustua}, Susana and {Gordon}, Karl D. and {Hubeny}, Ivan and {Mez{\H{o}}}, Gy{\"o}rgy and {Truszek}, M{\'a}rton},
        title = "{The updated BOSZ synthetic stellar spectral library}",
      journal = {\aap},
         year = 2024,
        month = aug,
       volume = {688},
          eid = {A197},
        pages = {A197},
          doi = {10.1051/0004-6361/202449306},
archivePrefix = {arXiv},
       eprint = {2407.10872},
 primaryClass = {astro-ph.SR},
       adsurl = {https://ui.adsabs.harvard.edu/abs/2024A&A...688A.197M}
}

@ARTICLE{2017AJ....153..234B,
       author = {{Bohlin}, Ralph C. and {M{\'e}sz{\'a}ros}, Szabolcs and {Fleming}, Scott W. and {Gordon}, Karl D. and {Koekemoer}, Anton M. and {Kov{\'a}cs}, J{\'o}zsef},
        title = "{A New Stellar Atmosphere Grid and Comparisons with HST/STIS CALSPEC Flux Distributions}",
      journal = {\aj},
         year = 2017,
        month = may,
       volume = {153},
       number = {5},
          eid = {234},
        pages = {234},
          doi = {10.3847/1538-3881/aa6ba9},
archivePrefix = {arXiv},
       eprint = {1704.00653},
 primaryClass = {astro-ph.SR},
       adsurl = {https://ui.adsabs.harvard.edu/abs/2017AJ....153..234B}
}

@ARTICLE{2018ApJ...867L..14V,
       author = {{van der Marel}, Nienke and {Williams}, Jonathan P. and {Bruderer}, Simon},
        title = "{Rings and Gaps in Protoplanetary Disks: Planets or Snowlines?}",
      journal = {\apjl},
         year = 2018,
        month = nov,
       volume = {867},
       number = {1},
          eid = {L14},
        pages = {L14},
          doi = {10.3847/2041-8213/aae88e},
archivePrefix = {arXiv},
       eprint = {1810.05614},
 primaryClass = {astro-ph.EP},
       adsurl = {https://ui.adsabs.harvard.edu/abs/2018ApJ...867L..14V}
}

@ARTICLE{2013Sci...341..630Q,
       author = {{Qi}, Chunhua and {{\"O}berg}, Karin I. and {Wilner}, David J. and {D'Alessio}, Paola and {Bergin}, Edwin and {Andrews}, Sean M. and {Blake}, Geoffrey A. and {Hogerheijde}, Michiel R. and {van Dishoeck}, Ewine F.},
        title = "{Imaging of the CO Snow Line in a Solar Nebula Analog}",
      journal = {Science},
         year = 2013,
        month = aug,
       volume = {341},
       number = {6146},
        pages = {630-632},
          doi = {10.1126/science.1239560},
archivePrefix = {arXiv},
       eprint = {1307.7439},
 primaryClass = {astro-ph.SR},
       adsurl = {https://ui.adsabs.harvard.edu/abs/2013Sci...341..630Q}
}

@ARTICLE{2017ApJ...835..146D,
       author = {{Dong}, Ruobing and {Fung}, Jeffrey},
        title = "{What is the Mass of a Gap-opening Planet?}",
      journal = {\apj},
         year = 2017,
        month = feb,
       volume = {835},
       number = {2},
          eid = {146},
        pages = {146},
          doi = {10.3847/1538-4357/835/2/146},
archivePrefix = {arXiv},
       eprint = {1612.04821},
 primaryClass = {astro-ph.EP},
       adsurl = {https://ui.adsabs.harvard.edu/abs/2017ApJ...835..146D}
}

@ARTICLE{2016MNRAS.459.2790R,
       author = {{Rosotti}, Giovanni P. and {Juhasz}, Attila and {Booth}, Richard A. and {Clarke}, Cathie J.},
        title = "{The minimum mass of detectable planets in protoplanetary discs and the derivation of planetary masses from high-resolution observations}",
      journal = {\mnras},
         year = 2016,
        month = jul,
       volume = {459},
       number = {3},
        pages = {2790-2805},
          doi = {10.1093/mnras/stw691},
archivePrefix = {arXiv},
       eprint = {1603.02141},
 primaryClass = {astro-ph.EP},
       adsurl = {https://ui.adsabs.harvard.edu/abs/2016MNRAS.459.2790R}
}

@ARTICLE{2015ApJ...809...93D,
       author = {{Dong}, Ruobing and {Zhu}, Zhaohuan and {Whitney}, Barbara},
        title = "{Observational Signatures of Planets in Protoplanetary Disks I. Gaps Opened by Single and Multiple Young Planets in Disks}",
      journal = {\apj},
         year = 2015,
        month = aug,
       volume = {809},
       number = {1},
          eid = {93},
        pages = {93},
          doi = {10.1088/0004-637X/809/1/93},
archivePrefix = {arXiv},
       eprint = {1411.6063},
 primaryClass = {astro-ph.EP},
       adsurl = {https://ui.adsabs.harvard.edu/abs/2015ApJ...809...93D}
}

@ARTICLE{1998ApJ...509..802C,
       author = {{Calvet}, Nuria and {Gullbring}, Erik},
        title = "{The Structure and Emission of the Accretion Shock in T Tauri Stars}",
      journal = {\apj},
         year = 1998,
        month = dec,
       volume = {509},
       number = {2},
        pages = {802-818},
          doi = {10.1086/306527},
       adsurl = {https://ui.adsabs.harvard.edu/abs/1998ApJ...509..802C}
}

@ARTICLE{1987ApJ...323..714K,
       author = {{Kenyon}, S.~J. and {Hartmann}, L.},
        title = "{Spectral Energy Distributions of T Tauri Stars: Disk Flaring and Limits on Accretion}",
      journal = {\apj},
         year = 1987,
        month = dec,
       volume = {323},
        pages = {714},
          doi = {10.1086/165866},
       adsurl = {https://ui.adsabs.harvard.edu/abs/1987ApJ...323..714K}
}

@misc{2025zndo..15747364P,
       author = {{Perrin}, Marshall and {Long}, Joseph and {Osborne}, Shannon and {Geda}, Robel and {Sappington}, Bradley and {Mel{\'e}ndez}, Marcio and {Lajoie}, Charles-Philippe and {Leisenring}, Jarron and {Zimmerman}, Neil and {Brooks}, Keira and {Otor}, O. Justin and {Kulp}, Trey and {Chambers}, Lauren and {Jurling}, Alden},
        title = "{STPSF}",
         year = 2025,
        month = jun,
          eid = {10.5281/zenodo.15747364},
          doi = {10.5281/zenodo.15747364},
      version = {2.1.0},
    publisher = {Zenodo},
       adsurl = {https://ui.adsabs.harvard.edu/abs/2025zndo..15747364P}
}

@ARTICLE{2000ApJ...538..793K,
       author = {{Krist}, John E. and {Stapelfeldt}, Karl R. and {M{\'e}nard}, Fran{\c{c}}ois and {Padgett}, Deborah L. and {Burrows}, Christopher J.},
        title = "{WFPC2 Images of a Face-on Disk Surrounding TW Hydrae}",
      journal = {\apj},
         year = 2000,
        month = aug,
       volume = {538},
       number = {2},
        pages = {793-800},
          doi = {10.1086/309170},
       adsurl = {https://ui.adsabs.harvard.edu/abs/2000ApJ...538..793K}
}

@ARTICLE{2023ApJ...956..102H,
       author = {{Herczeg}, Gregory J. and {Chen}, Yuguang and {Donati}, Jean-Francois and {Dupree}, Andrea K. and {Walter}, Frederick M. and {Hillenbrand}, Lynne A. and {Johns-Krull}, Christopher M. and {Manara}, Carlo F. and {G{\"u}nther}, Hans Moritz and {Fang}, Min and {Schneider}, P. Christian and {Valenti}, Jeff A. and {Alencar}, Silvia H.~P. and {Venuti}, Laura and {Alcal{\'a}}, Juan Manuel and {Frasca}, Antonio and {Arulanantham}, Nicole and {Linsky}, Jeffrey L. and {Bouvier}, Jerome and {Brickhouse}, Nancy S. and {Calvet}, Nuria and {Espaillat}, Catherine C. and {Campbell-White}, Justyn and {Carpenter}, John M. and {Chang}, Seok-Jun and {Cruz}, Kelle L. and {Dahm}, S.~E. and {Eisl{\"o}ffel}, Jochen and {Edwards}, Suzan and {Fischer}, William J. and {Guo}, Zhen and {Henning}, Thomas and {Ji}, Tao and {Jose}, Jessy and {Kastner}, Joel H. and {Launhardt}, Ralf and {Principe}, David A. and {Robinson}, Connor E. and {Serna}, Javier and {Siwak}, Michal and {Sterzik}, Michael F. and {Takasao}, Shinsuke},
        title = "{Twenty-five Years of Accretion onto the Classical T Tauri Star TW Hya}",
      journal = {\apj},
         year = 2023,
        month = oct,
       volume = {956},
       number = {2},
          eid = {102},
        pages = {102},
          doi = {10.3847/1538-4357/acf468},
archivePrefix = {arXiv},
       eprint = {2308.14590},
 primaryClass = {astro-ph.SR},
       adsurl = {https://ui.adsabs.harvard.edu/abs/2023ApJ...956..102H}
}

@ARTICLE{2014A&A...564A..93M,
       author = {{Menu}, J. and {van Boekel}, R. and {Henning}, Th. and {Chandler}, C.~J. and {Linz}, H. and {Benisty}, M. and {Lacour}, S. and {Min}, M. and {Waelkens}, C. and {Andrews}, S.~M. and {Calvet}, N. and {Carpenter}, J.~M. and {Corder}, S.~A. and {Deller}, A.~T. and {Greaves}, J.~S. and {Harris}, R.~J. and {Isella}, A. and {Kwon}, W. and {Lazio}, J. and {Le Bouquin}, J.-B. and {M{\'e}nard}, F. and {Mundy}, L.~G. and {P{\'e}rez}, L.~M. and {Ricci}, L. and {Sargent}, A.~I. and {Storm}, S. and {Testi}, L. and {Wilner}, D.~J.},
        title = "{On the structure of the transition disk around TW Hydrae}",
      journal = {\aap},
         year = 2014,
        month = apr,
       volume = {564},
          eid = {A93},
        pages = {A93},
          doi = {10.1051/0004-6361/201322961},
archivePrefix = {arXiv},
       eprint = {1402.6597},
 primaryClass = {astro-ph.SR},
       adsurl = {https://ui.adsabs.harvard.edu/abs/2014A&A...564A..93M}
}

@ARTICLE{2017AJ....154...73R,
       author = {{Ruane}, G. and {Mawet}, D. and {Kastner}, J. and {Meshkat}, T. and {Bottom}, M. and {Femen{\'\i}a Castell{\'a}}, B. and {Absil}, O. and {Gomez Gonzalez}, C. and {Huby}, E. and {Zhu}, Z. and {Jensen-Clem}, R. and {Choquet}, {\'E}. and {Serabyn}, E.},
        title = "{Deep Imaging Search for Planets Forming in the TW Hya Protoplanetary Disk with the Keck/NIRC2 Vortex Coronagraph}",
      journal = {\aj},
         year = 2017,
        month = aug,
       volume = {154},
       number = {2},
          eid = {73},
        pages = {73},
          doi = {10.3847/1538-3881/aa7b81},
archivePrefix = {arXiv},
       eprint = {1706.07489},
 primaryClass = {astro-ph.EP},
       adsurl = {https://ui.adsabs.harvard.edu/abs/2017AJ....154...73R}
}

@ARTICLE{2015ApJ...815L..26R,
       author = {{Rapson}, Valerie A. and {Kastner}, Joel H. and {Millar-Blanchaer}, Maxwell A. and {Dong}, Ruobing},
        title = "{Peering into the Giant-planet-forming Region of the TW Hydrae Disk with the Gemini Planet Imager}",
      journal = {\apjl},
         year = 2015,
        month = dec,
       volume = {815},
       number = {2},
          eid = {L26},
        pages = {L26},
          doi = {10.1088/2041-8205/815/2/L26},
archivePrefix = {arXiv},
       eprint = {1512.01865},
 primaryClass = {astro-ph.SR},
       adsurl = {https://ui.adsabs.harvard.edu/abs/2015ApJ...815L..26R}
}

@ARTICLE{2016ApJ...819L...7N,
       author = {{Nomura}, Hideko and {Tsukagoshi}, Takashi and {Kawabe}, Ryohei and {Ishimoto}, Daiki and {Okuzumi}, Satoshi and {Muto}, Takayuki and {Kanagawa}, Kazuhiro D. and {Ida}, Shigeru and {Walsh}, Catherine and {Millar}, T.~J. and {Bai}, Xue-Ning},
        title = "{ALMA Observations of a Gap and a Ring in the Protoplanetary Disk around TW Hya}",
      journal = {\apjl},
         year = 2016,
        month = mar,
       volume = {819},
       number = {1},
          eid = {L7},
        pages = {L7},
          doi = {10.3847/2041-8205/819/1/L7},
archivePrefix = {arXiv},
       eprint = {1512.05440},
 primaryClass = {astro-ph.EP},
       adsurl = {https://ui.adsabs.harvard.edu/abs/2016ApJ...819L...7N}
}

@ARTICLE{2018ApJ...866..110D,
       author = {{Dong}, Ruobing and {Li}, Shengtai and {Chiang}, Eugene and {Li}, Hui},
        title = "{Multiple Disk Gaps and Rings Generated by a Single Super-Earth. II. Spacings, Depths, and Number of Gaps, with Application to Real Systems}",
      journal = {\apj},
         year = 2018,
        month = oct,
       volume = {866},
       number = {2},
          eid = {110},
        pages = {110},
          doi = {10.3847/1538-4357/aadadd},
archivePrefix = {arXiv},
       eprint = {1808.06613},
 primaryClass = {astro-ph.EP},
       adsurl = {https://ui.adsabs.harvard.edu/abs/2018ApJ...866..110D}
}

@ARTICLE{2019ApJ...878L...8T,
       author = {{Tsukagoshi}, Takashi and {Muto}, Takayuki and {Nomura}, Hideko and {Kawabe}, Ryohei and {Kanagawa}, Kazuhiro D. and {Okuzumi}, Satoshi and {Ida}, Shigeru and {Walsh}, Catherine and {Millar}, Tom J. and {Takahashi}, Sanemichi Z. and {Hashimoto}, Jun and {Uyama}, Taichi and {Tamura}, Motohide},
        title = "{Discovery of An au-scale Excess in Millimeter Emission from the Protoplanetary Disk around TW Hya}",
      journal = {\apjl},
         year = 2019,
        month = jun,
       volume = {878},
       number = {1},
          eid = {L8},
        pages = {L8},
          doi = {10.3847/2041-8213/ab224c},
archivePrefix = {arXiv},
       eprint = {1905.07891},
 primaryClass = {astro-ph.SR},
       adsurl = {https://ui.adsabs.harvard.edu/abs/2019ApJ...878L...8T}
}

@ARTICLE{2022ApJ...936..163T,
       author = {{Teague}, Richard and {Bae}, Jaehan and {Andrews}, Sean M. and {Benisty}, Myriam and {Bergin}, Edwin A. and {Facchini}, Stefano and {Huang}, Jane and {Longarini}, Cristiano and {Wilner}, David},
        title = "{Mapping the Complex Kinematic Substructure in the TW Hya Disk}",
      journal = {\apj},
         year = 2022,
        month = sep,
       volume = {936},
       number = {2},
          eid = {163},
        pages = {163},
          doi = {10.3847/1538-4357/ac88ca},
archivePrefix = {arXiv},
       eprint = {2208.04837},
 primaryClass = {astro-ph.EP},
       adsurl = {https://ui.adsabs.harvard.edu/abs/2022ApJ...936..163T}
}

@ARTICLE{2025AJ....170..317C,
       author = {{Cugno}, Gabriele and {Facchini}, Stefano and {Alarcon}, Felipe and {Bae}, Jaehan and {Benisty}, Myriam and {Eilers}, Anna-Christina and {Leung}, Gene C.~K. and {Meyer}, Michael and {Pueyo}, Laurent and {Teague}, Richard and {Bergin}, Edwin and {Girard}, Julien and {Helled}, Ravit and {Huang}, Jane and {Leisenring}, Jarron},
        title = "{Direct Measurement of Extinction in a Planet-hosting Gap}",
      journal = {\aj},
         year = 2025,
        month = dec,
       volume = {170},
       number = {6},
          eid = {317},
        pages = {317},
          doi = {10.3847/1538-3881/ae0acd},
archivePrefix = {arXiv},
       eprint = {2509.26617},
 primaryClass = {astro-ph.EP},
       adsurl = {https://ui.adsabs.harvard.edu/abs/2025AJ....170..317C}
}

@ARTICLE{2020A&A...633A..63D,
       author = {{de Boer}, J. and {Langlois}, M. and {van Holstein}, R.~G. and {Girard}, J.~H. and {Mouillet}, D. and {Vigan}, A. and {Dohlen}, K. and {Snik}, F. and {Keller}, C.~U. and {Ginski}, C. and {Stam}, D.~M. and {Milli}, J. and {Wahhaj}, Z. and {Kasper}, M. and {Schmid}, H.~M. and {Rabou}, P. and {Gluck}, L. and {Hugot}, E. and {Perret}, D. and {Martinez}, P. and {Weber}, L. and {Pragt}, J. and {Sauvage}, J.-F. and {Boccaletti}, A. and {Le Coroller}, H. and {Dominik}, C. and {Henning}, T. and {Lagadec}, E. and {M{\'e}nard}, F. and {Turatto}, M. and {Udry}, S. and {Chauvin}, G. and {Feldt}, M. and {Beuzit}, J.-L.},
        title = "{Polarimetric imaging mode of VLT/SPHERE/IRDIS. I. Description, data reduction, and observing strategy}",
      journal = {\aap},
         year = 2020,
        month = jan,
       volume = {633},
          eid = {A63},
        pages = {A63},
          doi = {10.1051/0004-6361/201834989},
archivePrefix = {arXiv},
       eprint = {1909.13107},
 primaryClass = {astro-ph.IM},
       adsurl = {https://ui.adsabs.harvard.edu/abs/2020A&A...633A..63D}
}

@ARTICLE{2017ApJ...843..127D,
       author = {{Dong}, Ruobing and {Li}, Shengtai and {Chiang}, Eugene and {Li}, Hui},
        title = "{Multiple Disk Gaps and Rings Generated by a Single Super-Earth}",
      journal = {\apj},
         year = 2017,
        month = jul,
       volume = {843},
       number = {2},
          eid = {127},
        pages = {127},
          doi = {10.3847/1538-4357/aa72f2},
archivePrefix = {arXiv},
       eprint = {1705.04687},
 primaryClass = {astro-ph.EP},
       adsurl = {https://ui.adsabs.harvard.edu/abs/2017ApJ...843..127D}
}

@ARTICLE{2006ApJ...637L.133E,
       author = {{Eisner}, J.~A. and {Chiang}, E.~I. and {Hillenbrand}, L.~A.},
        title = "{Spatially Resolving the Inner Disk of TW Hydrae}",
      journal = {\apjl},
         year = 2006,
        month = feb,
       volume = {637},
       number = {2},
        pages = {L133-L136},
          doi = {10.1086/500689},
archivePrefix = {arXiv},
       eprint = {astro-ph/0601034},
 primaryClass = {astro-ph},
       adsurl = {https://ui.adsabs.harvard.edu/abs/2006ApJ...637L.133E}
}

@ARTICLE{2023PASP..135b8001R,
       author = {{Rieke}, Marcia J. and {Kelly}, Douglas M. and {Misselt}, Karl and {Stansberry}, John and {Boyer}, Martha and {Beatty}, Thomas and {Egami}, Eiichi and {Florian}, Michael and {Greene}, Thomas P. and {Hainline}, Kevin and {Leisenring}, Jarron and {Roellig}, Thomas and {Schlawin}, Everett and {Sun}, Fengwu and {Tinnin}, Lee and {Williams}, Christina C. and {Willmer}, Christopher N.~A. and {Wilson}, Debra and {Clark}, Charles R. and {Rohrbach}, Scott and {Brooks}, Brian and {Canipe}, Alicia and {Correnti}, Matteo and {DiFelice}, Audrey and {Gennaro}, Mario and {Girard}, Julien H. and {Hartig}, George and {Hilbert}, Bryan and {Koekemoer}, Anton M. and {Nikolov}, Nikolay K. and {Pirzkal}, Norbert and {Rest}, Armin and {Robberto}, Massimo and {Sunnquist}, Ben and {Telfer}, Randal and {Wu}, Chi Rai and {Ferry}, Malcolm and {Lewis}, Dan and {Baum}, Stefi and {Beichman}, Charles and {Doyon}, Ren{\'e} and {Dressler}, Alan and {Eisenstein}, Daniel J. and {Ferrarese}, Laura and {Hodapp}, Klaus and {Horner}, Scott and {Jaffe}, Daniel T. and {Johnstone}, Doug and {Krist}, John and {Martin}, Peter and {McCarthy}, Donald W. and {Meyer}, Michael and {Rieke}, George H. and {Trauger}, John and {Young}, Erick T.},
        title = "{Performance of NIRCam on JWST in Flight}",
      journal = {\pasp},
         year = 2023,
        month = feb,
       volume = {135},
       number = {1044},
          eid = {028001},
        pages = {028001},
          doi = {10.1088/1538-3873/acac53},
archivePrefix = {arXiv},
       eprint = {2212.12069},
 primaryClass = {astro-ph.IM},
       adsurl = {https://ui.adsabs.harvard.edu/abs/2023PASP..135b8001R}
}

@ARTICLE{2018PASA...35...10W,
       author = {{Wolf}, Christian and {Onken}, Christopher A. and {Luvaul}, Lance C. and {Schmidt}, Brian P. and {Bessell}, Michael S. and {Chang}, Seo-Won and {Da Costa}, Gary S. and {Mackey}, Dougal and {Martin-Jones}, Tony and {Murphy}, Simon J. and {Preston}, Tim and {Scalzo}, Richard A. and {Shao}, Li and {Smillie}, Jon and {Tisserand}, Patrick and {White}, Marc C. and {Yuan}, Fang},
        title = "{SkyMapper Southern Survey: First Data Release (DR1)}",
      journal = {\pasa},
         year = 2018,
        month = feb,
       volume = {35},
          eid = {e010},
        pages = {e010},
          doi = {10.1017/pasa.2018.5},
archivePrefix = {arXiv},
       eprint = {1801.07834},
 primaryClass = {astro-ph.IM},
       adsurl = {https://ui.adsabs.harvard.edu/abs/2018PASA...35...10W}
}

@ARTICLE{2019MNRAS.484L.130M,
       author = {{Mentiplay}, Daniel and {Price}, Daniel J. and {Pinte}, Christophe},
        title = "{Super-Earths in the TW Hya disc}",
      journal = {\mnras},
         year = 2019,
        month = mar,
       volume = {484},
       number = {1},
        pages = {L130-L135},
          doi = {10.1093/mnrasl/sly209},
archivePrefix = {arXiv},
       eprint = {1811.03636},
 primaryClass = {astro-ph.EP},
       adsurl = {https://ui.adsabs.harvard.edu/abs/2019MNRAS.484L.130M}
}

@ARTICLE{2021AJ....161...38O,
       author = {{{\"O}berg}, Karin I. and {Cleeves}, L. Ilsedore and {Bergner}, Jennifer B. and {Cavanaro}, Joseph and {Teague}, Richard and {Huang}, Jane and {Loomis}, Ryan A. and {Bergin}, Edwin A. and {Blake}, Geoffrey A. and {Calahan}, Jenny and {Cazzoletti}, Paolo and {Guzm{\'a}n}, Viviana Veloso and {Hogerheijde}, Michiel R. and {Kama}, Mihkel and {Terwisscha van Scheltinga}, Jeroen and {Qi}, Chunhua and {van Dishoeck}, Ewine and {Walsh}, Catherine and {Wilner}, David J.},
        title = "{The TW Hya Rosetta Stone Project. I. Radial and Vertical Distributions of DCN and DCO$^{+}$}",
      journal = {\aj},
         year = 2021,
        month = jan,
       volume = {161},
       number = {1},
          eid = {38},
        pages = {38},
          doi = {10.3847/1538-3881/abc74d},
archivePrefix = {arXiv},
       eprint = {2011.06774},
 primaryClass = {astro-ph.EP},
       adsurl = {https://ui.adsabs.harvard.edu/abs/2021AJ....161...38O}
}

@ARTICLE{2021ApJ...908....8C,
       author = {{Calahan}, Jenny K. and {Bergin}, Edwin and {Zhang}, Ke and {Teague}, Richard and {Cleeves}, Ilsedore and {Bergner}, Jennifer and {Blake}, Geoffrey A. and {Cazzoletti}, Paolo and {Guzm{\'a}n}, Viviana and {Hogerheijde}, Michiel R. and {Huang}, Jane and {Kama}, Mihkel and {Loomis}, Ryan and {{\"O}berg}, Karin and {Qi}, Charlie and {van Dishoeck}, Ewine F. and {Terwisscha van Scheltinga}, Jeroen and {Walsh}, Catherine and {Wilner}, David},
        title = "{The TW Hya Rosetta Stone Project. III. Resolving the Gaseous Thermal Profile of the Disk}",
      journal = {\apj},
         year = 2021,
        month = feb,
       volume = {908},
       number = {1},
          eid = {8},
        pages = {8},
          doi = {10.3847/1538-4357/abd255},
archivePrefix = {arXiv},
       eprint = {2012.05927},
 primaryClass = {astro-ph.EP},
       adsurl = {https://ui.adsabs.harvard.edu/abs/2021ApJ...908....8C}
}

@ARTICLE{2002ApJ...569..997R,
       author = {{Rafikov}, R.~R.},
        title = "{Nonlinear Propagation of Planet-generated Tidal Waves}",
      journal = {\apj},
         year = 2002,
        month = apr,
       volume = {569},
       number = {2},
        pages = {997-1008},
          doi = {10.1086/339399},
archivePrefix = {arXiv},
       eprint = {astro-ph/0110496},
 primaryClass = {astro-ph},
       adsurl = {https://ui.adsabs.harvard.edu/abs/2002ApJ...569..997R}
}

@ARTICLE{2006ApJ...638..314D,
       author = {{D'Alessio}, Paola and {Calvet}, Nuria and {Hartmann}, Lee and {Franco-Hern{\'a}ndez}, Ramiro and {Serv{\'\i}n}, Hermelinda},
        title = "{Effects of Dust Growth and Settling in T Tauri Disks}",
      journal = {\apj},
         year = 2006,
        month = feb,
       volume = {638},
       number = {1},
        pages = {314-335},
          doi = {10.1086/498861},
archivePrefix = {arXiv},
       eprint = {astro-ph/0511564},
 primaryClass = {astro-ph},
       adsurl = {https://ui.adsabs.harvard.edu/abs/2006ApJ...638..314D}
}

@ARTICLE{2006MNRAS.373.1619R,
       author = {{Rice}, W.~K.~M. and {Armitage}, Philip J. and {Wood}, Kenneth and {Lodato}, G.},
        title = "{Dust filtration at gap edges: implications for the spectral energy distributions of discs with embedded planets}",
      journal = {\mnras},
         year = 2006,
        month = dec,
       volume = {373},
       number = {4},
        pages = {1619-1626},
          doi = {10.1111/j.1365-2966.2006.11113.x},
archivePrefix = {arXiv},
       eprint = {astro-ph/0609808},
 primaryClass = {astro-ph},
       adsurl = {https://ui.adsabs.harvard.edu/abs/2006MNRAS.373.1619R}
}

@ARTICLE{2012ApJ...755....6Z,
       author = {{Zhu}, Zhaohuan and {Nelson}, Richard P. and {Dong}, Ruobing and {Espaillat}, Catherine and {Hartmann}, Lee},
        title = "{Dust Filtration by Planet-induced Gap Edges: Implications for Transitional Disks}",
      journal = {\apj},
         year = 2012,
        month = aug,
       volume = {755},
       number = {1},
          eid = {6},
        pages = {6},
          doi = {10.1088/0004-637X/755/1/6},
archivePrefix = {arXiv},
       eprint = {1205.5042},
 primaryClass = {astro-ph.SR},
       adsurl = {https://ui.adsabs.harvard.edu/abs/2012ApJ...755....6Z}
}

@ARTICLE{2003ARA&A..41..241D,
       author = {{Draine}, B.~T.},
        title = "{Interstellar Dust Grains}",
      journal = {\araa},
         year = 2003,
        month = jan,
       volume = {41},
        pages = {241-289},
          doi = {10.1146/annurev.astro.41.011802.094840},
archivePrefix = {arXiv},
       eprint = {astro-ph/0304489},
 primaryClass = {astro-ph},
       adsurl = {https://ui.adsabs.harvard.edu/abs/2003ARA&A..41..241D}
}
\bibliographystyle{aasjournalv7}

\end{document}